\documentclass[twoside]{article}
\usepackage[a4paper]{geometry}
\usepackage[latin1]{inputenc} 
\usepackage[T1]{fontenc} 
\usepackage{RR}
\usepackage{hyperref}
\usepackage{perlazaRR}

\newcommand{\squeezeequ}{\medmuskip=2mu \thinmuskip=1mu \thickmuskip=3mu}

\newcommand{\Tsupersqueezeequ}{\medmuskip=0.1mu \thinmuskip=0mu \thickmuskip=0.1mu \nulldelimiterspace=-1pt \scriptspace=0pt}

\renewcommand{\eqref}[1]{(\ref{#1})}

\usepackage{array, makecell}

\usepackage{subequation}
\RRNo{9492}
\RRdate{November 2022}
\RRauthor{
Ke~Sun 
}
\authorhead{Ke~Sun }
\RRtitle{Quelques propri\'{e}t\'{e}s des \'{e}quilibres de Nash des jeux $2 × 2$ \`{a} somme nulle}
\RRetitle{Some Properties of the Nash Equilibrium in 2~$\boldsymbol{\times}$~2 Zero-Sum Games}
\titlehead{Some Properties of the NE in $2 \times 2$ Zero-Sum Games}
\RRnote{This work is supported by the Inria Exploratory Action -- Information and Decision Making (IDEM).}
\RRnote{The author thanks Samir M. Perlaza (NEO) and Alain Jean-Marie (NEO) for insightful discussions.}
\RRnote{Ke Sun is with the Inria centre at Université Côte d'Azur, Sophia Antipolis, France.}

\RRresume{Dans ce rapport, certaines propri\'{e}t\'{e}s de l'ensemble des \'{e}quilibres de Nash (EN) des jeux $2 \times 2$ à somme nulle  sont étudiés. 
En particulier, la cardinalité de l'ensemble des EN est donnée en termes des entrées de la matrice du jeu. De plus, des expressions explicites pour les stratégies d'équilibre et la valeur du jeu sont présentées. Ces résultats ne sont pas nécessairement des connaissances nouvelles, car ils découlent de la définition de l'EN après quelques calculs.
Néanmoins, cette présentation synthétique est originale dans la littérature.
}
\RRabstract{In this report, some properties of the set of Nash equilibria (NEs) of $2 \times 2$ zero-sum games  are reviewed. 
In particular, the cardinality of the set of NEs is given in terms of the entries of the payoff matrix.  Moreover, closed-form expressions for the NE strategies and  the payoff at the NE (the value of the game) are provided in terms of the entries of the payoff matrix. 
The results presented in this report are not necessarily new knowledge, as they follow from the definition of the NE after some tedious calculations. 
 Nevertheless this synthetic presentation is original in the literature. 
}
\RRmotcle{jeux $2\times 2$ à somme nulle,  jeux en forme normale, et équilibre de Nash.}
\RRkeyword{ $2 \times 2$ Zero Sum Games, Normal-form Game, Nash Equilibrium}
\RRprojet{NEO}  
\RCSophia 

\begin{document}
\pagenumbering{roman}
\makeRR   
\tableofcontents

\newpage 

\pagenumbering{arabic}
\setcounter{page}{1}

\section{Zero-Sum Games in Normal Form}\label{Sec:ZS}
The notations that are used in this report are listed in the following. 
Given a finite set $\set{X}$, the notation $2^{\set{X}}$ represents 
the power set of $\set{X}$. The notation $\simplex{\set{X}}$ represents the set of all probability measures that can be defined on the measurable space $\left( \set{X}, 2^{\set{X}}\right)$. The set of all subsets of $\simplex{\set{X}}$ is denoted by $2^{\simplex{\set{X}}}$.

Consider a two-player two-action zero-sum game in normal form with payoff matrix 
\begin{IEEEeqnarray}{rcl}
\label{Equ:u}
\matx{u} & = & 
\begin{pmatrix}
u_{1,1} & u_{1,2}\\
u_{2,1} & u_{2,2}
\end{pmatrix}.
\end{IEEEeqnarray}
Let the elements of the set $\set{K} \triangleq \{1,2\}$ represent the indices of the players; 
and let the elements of the set $\set{A}_1 = \set{A}_2 \triangleq \{a_1, a_2\}$ represent the actions of the players.
Hence, for all $(i,j) \inCountTwo^2$, when \Pone plays $a_i$ and \Ptwo plays $a_j$, the outcome of the game is $u_{i, j}$.
In the following, such a game is represented by the tuple 
\begin{IEEEeqnarray}{rcl}
\label{EqTheGame}
\game{G} & \triangleq & \left(\set{K}, \set{A}_1 , \set{A}_2 , \matx{u} \right).
\end{IEEEeqnarray} 

For all $k \in \set{K}$, the strategy of \Pkth  is a probability measure denoted by $P_{A_k} \in \simplex{\set{A}_k}$.  
At each repetition of the game, players choose their actions by sampling their probability measures (strategies). 
Let the average payoff be represented by the function $u: \simplex{\set{A}_1} \times \simplex{\set{A}_2} \to \reals$ such that, given the strategies $P_{A_1}$ and $P_{A_2}$,  
\begin{IEEEeqnarray}{rcl}
\label{Eqv}
u\left(P_{A_1}, P_{A_2} \right) & = &  \sum_{(i,j) \in \{1,2\}^2} P_{A_1} \left( a_i \right)  P_{A_2}\left(a_j\right) u_{i,j}. \squeezeequ
\end{IEEEeqnarray}
\Pone chooses its strategy $P_{A_1}$ aiming to maximize the expected payoff $u\left(P_{A_1}, P_{A_2} \right)$, whereas \Ptwo  chooses the strategy $P_{A_2}$ to minimize it. 

The set of best responses of \Pone to a given strategy of \Ptwo is determined by the correspondence $\BR_1: \simplex{\set{A}_{2}} \to 2^{\simplex{\set{A}_1}}$,  such that given the strategy $P$ for \Ptwo, it holds that
\begin{IEEEeqnarray}{rcl}\label{Equ:BR_1}
\BR_1\left( P \right) =  \arg\max_{Q \in \simplex{\set{A}_1}} u(Q, P), 
\end{IEEEeqnarray}
where the function $u$ is defined in \eqref{Eqv}.
Similarly, the set of best responses of \Ptwo to a given strategy of \Pone is determined by the correspondence $\BR_2: \simplex{\set{A}_{1}} \to 2^{\simplex{\set{A}_2}}$,  such that given the strategy $Q$ for \Pone, it holds that
\begin{IEEEeqnarray}{rcl}\label{Equ:BR_2}
\BR_2\left( Q \right) =  \arg\min_{P \in \simplex{\set{A}_2}} u(Q, P). 
\end{IEEEeqnarray}

When both players choose their actions simultaneously, a relevant outcome of the game is the Nash equilibrium (NE). 
\begin{definition}[Nash Equilibrium]\label{Def:Equ_1}
A pair of strategies $\left( P_{A_1}^{\star}, P_{A_2}^{\star} \right) \in \simplex{\set{A}_1}\times\simplex{\set{A}_2}$ forms an NE in the game $\game{G}$ in \eqref{EqTheGame}, if and only if the following conditions simultaneously hold:
\begin{enumerate}
\item[(i)] For all $Q \in \simplex{\set{A}_1}$, 
\begin{IEEEeqnarray}{l}\label{EqNECon1}
u\left(P_{A_1}^{\star}, P_{A_2}^{\star}\right) \geqslant u\left(Q, P_{A_2}^{\star}\right); \mbox{ and }
\end{IEEEeqnarray}
\item[(ii)] For all $Q \in \simplex{\set{A}_2}$, 
\begin{IEEEeqnarray}{l}\label{EqNECon2}
u\left(P_{A_1}^{\star}, P_{A_2}^{\star}\right) \leqslant u\left(P_{A_1}^{\star}, Q\right),
\end{IEEEeqnarray}
\end{enumerate}
where the function $u$ is defined in \eqref{Eqv}.
\end{definition}

\subsection{Multiplicity of the Nash Equilibrium}\label{SubSec:NE}

The following lemma shows that if the game $\game{G}$ in \eqref{EqTheGame} possesses an NE in which one of the players uses a pure strategy, while the other uses a strictly mixed strategy, then the game $\game{G}$ possesses infinitely many NEs.

\begin{lemma}\label{Lemma:infinite}
If a pair of strategies $\left(P_{A_1}^\star, P_{A_2}^\star\right)\in \simplex{\set{A}_1}\times\simplex{\set{A}_2} $ forms an NE in the game $\game{G}$ in \eqref{EqTheGame}, and one of the following statements holds: 
\begin{enumerate}
\item $P_{A_1}^\star (a_1) \in \{0,1\}$ and $P_{A_2}^\star (a_1) \in (0,1)$; \quad  \textnormal{or} 
\item $P_{A_1}^\star (a_1) \in (0,1) $ and $P_{A_2}^\star (a_1) \in \{0,1\}$, 
\end{enumerate}
then, the game $\game{G}$ possesses infinitely many NEs. 
\end{lemma}
\begin{IEEEproof}
The proof of Lemma \ref{Lemma:infinite} is provided in Appendix \ref{Sec:App1}.

\end{IEEEproof}

The following lemma shows that if the number of NEs in the game $\game{G}$ in \eqref{EqTheGame} is finite, then an NE in which both players use pure strategies and an NE in which both players use strictly mixed strategies cannot simultaneously exist.

\begin{lemma}\label{Lemma:twoNE_unique}
Let $\set{P} \subset \simplex{\set{A}_1} \times \simplex{\set{A}_2}$ be the set of NEs of the game $\game{G}$ in \eqref{EqTheGame}.
Assume that  $|\set{P}| < \infty$. 
The following statements hold: 
\begin{enumerate}
\item If there exists a pair of strategies $\left(P_{A_1}^\star, P_{A_2}^\star\right) \in \set{P}$ such that for all $i \in \{1,2 \}$, $P_{A_i}^\star(a_1) \in \{0,1\}$, then, a pair $\left(Q_{A_1}, Q_{A_2}\right) \in \simplex{\set{A}_1} \times \simplex{\set{A}_2}$ such that for all $j \in \{1,2 \}$, $Q_{A_j}(a_1) \in (0,1)$, satisfies $(Q_{A_1}, Q_{A_2})  \not\in \set{P}$. 
\item If there exists a pair of strategies $\left(P_{A_1}^\star, P_{A_2}^\star\right) \in \set{P}$ such that for all $i \in \{1,2 \}$, $P_{A_i}^\star(a_1) \in (0,1)$, then, a pair $\left(Q_{A_1}, Q_{A_2}\right) \in \simplex{\set{A}_1} \times \simplex{\set{A}_2}$ such that for all $j \in \{1,2 \}$, $Q_{A_j}(a_1) \in \{0,1\}$, satisfies $(Q_{A_1}, Q_{A_2})  \not\in \set{P}$. 
\end{enumerate}
\end{lemma}

\begin{IEEEproof}
The proof of Lemma \ref{Lemma:twoNE_unique} is provided in Appendix \ref{Sec:App2}.

\end{IEEEproof}

The following lemma shows that if there exists a finite number of NEs in the game $\game{G}$ in \eqref{EqTheGame} and an NE in strictly mixed strategies exists, then such equilibrium is unique. 

\begin{lemma}\label{Lemma:strictlymixed_unique}
Let $\set{P} \subset \simplex{\set{A}_1} \times \simplex{\set{A}_2}$ be the set of NEs of the game $\game{G}$ in \eqref{EqTheGame}.
If $|\set{P}| < \infty$ and there exists  a pair of strategies $\left(P_{A_1}^\star, P_{A_2}^\star\right) \in \set{P}$ be such that for all $i \in \{1,2\}$,
\begin{IEEEeqnarray}{c}
P_{A_i}^\star(a_1) \in (0,1), 
\end{IEEEeqnarray}
then, the pair $\left(P_{A_1}^\star, P_{A_2}^\star\right)$ forms the unique NE. 
\end{lemma}
\begin{IEEEproof}
The proof of Lemma \ref{Lemma:strictlymixed_unique} is provided in Appendix \ref{Sec:App3}.

\end{IEEEproof}

The following lemma shows that if there exists a finite number of NEs in the game $\game{G}$ in \eqref{EqTheGame} and an  NE in pure strategies exists, then such NE is unique. 

\begin{lemma}\label{Lemma:pure_unique}
Let $\set{P} \subset \simplex{\set{A}_1} \times \simplex{\set{A}_2}$ be the set of NEs of the game $\game{G}$ in \eqref{EqTheGame}. 
If $|\set{P}| < \infty$ and there exists a pair of strategies $\left(P_{A_1}^\star, P_{A_2}^\star\right) \in \set{P}$ such that for all $i \in \{1,2\}$,
\begin{IEEEeqnarray}{c}
P_{A_i}^\star(a_1) \in \{0,1\}, 
\end{IEEEeqnarray}
then, the pair $\left(P_{A_1}^\star, P_{A_2}^\star\right)$ forms the unique NE. 
\end{lemma}
\begin{IEEEproof}
The proof of Lemma \ref{Lemma:pure_unique} is provided in Appendix \ref{Sec:App4}. 

\end{IEEEproof}


The following theorem specifies the number of NEs that can be observed in a $2 \times 2$ zero-sum game in normal form. 

 \begin{theorem}[Multiplicity of the Nash Equilibria]\label{TheoNumberNE}
 Let $\set{P} \subset \simplex{\set{A}_1} \times \simplex{\set{A}_2}$ be the set of NEs of the game $\game{G}$ in \eqref{EqTheGame}. 
Then either $|\set{P}| =1$ or $|\set{P}| =\infty$. 
In particular, if it holds that $\set{P} = \{ \left(P_{A_1}^\star, P_{A_2}^\star\right) \}$, then only one of the following statement is true: 
\begin{enumerate}
\item for all $i \in \{1,2\}$, $P_{A_i}^\star(a_1) \in \{0,1 \}$; \quad \textnormal{or}
\item for all $i \in \{1,2\}$, $P_{A_i}^\star(a_1) \in (0,1 )$.  
\end{enumerate}
 \end{theorem}
\begin{IEEEproof} 
The proof is provided in Appendix \ref{Sec:App5}.

\end{IEEEproof}

\subsection{Closed-form Expressions for the Nash Equilibrium Strategies}

Before showing the closed-form expression of the NE, closed-form expressions for the best responses in \eqref{Equ:BR_1} and \eqref{Equ:BR_2} are presented. 

\begin{lemma}\label{LemmaBR1}
Let $\delta \triangleq u_{1,1} - u_{1,2} -u_{2,1}+u_{2,2} $. 
For a given $P_{A_2} \in \simplex{\set{A}_2}$, the best response $\BR_1(P_{A_2})$  in \eqref{Equ:BR_1}
satisfies: 

(i) if $\delta \neq 0$, then 
\begin{IEEEeqnarray}{rl}
&\BR_1\left( P_{A_2} \right)= \left\{ 
 \begin{array}{cl}
 \{ P_{A_1} \in \Delta(\set{A}_1): P_{A_1}(a_1) = 0 \}, & \makecell[t]{ \textnormal{if }\delta  > 0 \textnormal{ and } P_{A_2}(a_1) <p^{(2)}, \textnormal{ or  } \\
\hspace{-1em}\delta  < 0  \textnormal{ and }  P_{A_2}(a_1) > p^{(2)} } \\
\{ P_{A_1} \in \Delta(\set{A}_1): P_{A_1}(a_1) = 1 \}, &  \makecell[t]{ \textnormal{if }\delta  > 0 \textnormal{ and } P_{A_2}(a_1) > p^{(2)}, \textnormal{ or  } \\
\hspace{-1em} \delta  < 0  \textnormal{ and }  P_{A_2}(a_1) < p^{(2)}} \\
\{ P_{A_1} \in \Delta(\set{A}_1): P_{A_1}(a_1) = \beta, \beta  \in [0,1] \}, & 
\makecell[t]{ \textnormal{if } P_{A_2}(a_1) = p^{(2)}, }
 \end{array}
 \right. \IEEEeqnarraynumspace \squeezeequ \label{Equ:BR_11}
\end{IEEEeqnarray}
where
\begin{IEEEeqnarray}{c}
p^{(2)} \triangleq \frac{1}{\delta }\left(u_{2,2} - u_{1,2} \right)
\label{Equ:P_{A_2}}; 
\end{IEEEeqnarray}

(ii) if $\delta  = 0$, then 
\begin{IEEEeqnarray}{rl}
\BR_1\left( P_{A_2} \right)= 
&\left\{ 
 \begin{array}{cl}
 \{ P_{A_1} \in \Delta(\set{A}_1): P_{A_1}(a_1) = 0 \}, &   \textnormal{if } u_{1,2} < u_{2,2} \\
\{ P_{A_1} \in \Delta(\set{A}_1): P_{A_1}(a_1) = 1 \}, &   \textnormal{if } u_{1,2} > u_{2,2} \\
\{ P_{A_1} \in \Delta(\set{A}_1): P_{A_1}(a_1) = \beta, \beta  \in [0,1] \}, & 
\textnormal{if } u_{1,2} = u_{2,2}
 \end{array}.  \label{Equ:BR_12}
 \right. \IEEEeqnarraynumspace \squeezeequ
\end{IEEEeqnarray}
\end{lemma}

\begin{IEEEproof}
The proof of Lemma \ref{LemmaBR1} is provided in Appendix \ref{AppLemmaBR1}.
\end{IEEEproof}

The best response of \Ptwo in  \eqref{Equ:BR_2} follows directly from Lemma \ref{LemmaBR1}  by exchanging the roles of the players, i.e. \Pone is the column player and \Ptwo is the row player, and reverting the payoff matrix from $\matx{u}$ to $- \matx{u}$, which is stated in the following lemma.

\begin{lemma}\label{LemmaBR2}
Let $\delta  \triangleq u_{1,1} - u_{1,2} -u_{2,1}+u_{2,2} $. 
For a given $P_{A_1} \in \simplex{\set{A}_1}$, the best response $\BR_2(P_{A_1})$  in \eqref{Equ:BR_2}
satisfies: 

(i) if $\delta  \neq 0$, then 
\begin{IEEEeqnarray}{rl}
&\BR_2\left( P_{A_1} \right)= \left\{ 
 \begin{array}{cl}
 \{ P_{A_2} \in \Delta(\set{A}_2): P_{A_2}(a_1) = 0 \}, & \makecell[t]{ \textnormal{if } \delta  > 0 \textnormal{ and } P_{A_1}(a_1) >p^{(1)}, \textnormal{ or  } \\
\hspace{-1em} \delta  < 0  \textnormal{ and }  P_{A_1}(a_1) <p^{(1)} } \\
\{ P_{A_2} \in \Delta(\set{A}_2): P_{A_2}(a_1) = 1 \}, &  \makecell[t]{ \textnormal{if } \delta  > 0 \textnormal{ and } P_{A_1}(a_1) < p^{(1)}, \textnormal{ or  } \\
\hspace{-1em} \delta  < 0  \textnormal{ and }  P_{A_1}(a_1) >p^{(1)}} \\
\{ P_{A_2} \in \Delta(\set{A}_2): P_{A_2}(a_1) = \beta, \beta  \in [0,1] \}, & 
\makecell[t]{ \textnormal{if } P_{A_1}(a_1) = p^{(1)}, }
 \end{array}
 \right. \IEEEeqnarraynumspace \squeezeequ \label{Equ:BR_21}
\end{IEEEeqnarray}
where
\begin{IEEEeqnarray}{c}
p^{(1)} \triangleq  \frac{1}{\delta }\left(u_{2,2} - u_{2,1} \right); \label{Equ:P_{A_1}} 
\end{IEEEeqnarray}

(ii) if $\delta = 0$, then 
\begin{IEEEeqnarray}{rl}
\BR_2\left( P_{A_1} \right)= 
&\left\{ 
 \begin{array}{cl}
 \{ P_{A_2} \in \Delta(\set{A}_2): P_{A_2}(a_1) = 0 \}, & \textnormal{if } u_{2,1} > u_{2,2}\\
\{ P_{A_2} \in \Delta(\set{A}_2): P_{A_2}(a_1) = 1 \}, &  \textnormal{if } u_{2,1} < u_{2,2}\\
\{ P_{A_2} \in \Delta(\set{A}_2): P_{A_2}(a_1) = \beta, \beta  \in [0,1] \}, & 
\textnormal{if } u_{2,1} = u_{2,2}
 \end{array}.
 \right.   \IEEEeqnarraynumspace \squeezeequ \label{Equ:BR_22}
\end{IEEEeqnarray}
\end{lemma}

\begin{IEEEproof}
The proof of Lemma \ref{LemmaBR2} is provided in Appendix \ref{AppLemmaBR2}.
\end{IEEEproof}

In the case in which the zero-sum game $\game{G}$ in \eqref{EqTheGame} exhibits a unique NE, the strategies of each player have the following closed-form expressions.

\begin{theorem}[Unique Nash Equilibrium]\label{TheoNE}
Let $\set{P} \subset \simplex{\set{A}_1} \times \simplex{\set{A}_2}$ be the set of NEs of the game $\game{G}$ in \eqref{EqTheGame} such that $|\set{P}|=1$. 
Let the pair of strategies $\left( P_{A_1}^\star, P_{A_2}^\star \right) \in \set{P}$ form an NE in $\game{G}$ in \eqref{EqTheGame}. 

\noindent
(i) The pair of strategies $\left( P_{A_1}^\star, P_{A_2}^\star \right)$ is the unique NE with 
\begin{IEEEeqnarray}{c}\label{EqUniqueMixCon}
P_{A_1}^\star(a_{1})  \in (0,1) \quad \textnormal{and} \quad P_{A_2}^\star(a_{1})  \in (0,1), 
\end{IEEEeqnarray}
if and only if, the entries in the matrix $\matx{u}$ in (\ref{Equ:u}) satisfy
\begin{IEEEeqnarray}{rl}\label{Equ:SMNE_condition}
&(u_{1,1} -u_{1,2})(u_{2,2} -u_{2,1}) > 0 \quad \textnormal{and} \ \\
& (u_{1,1} -u_{2,1})(u_{2,2} -u_{1,2}) > 0. \label{Equ:SMNE_condition_1}
\end{IEEEeqnarray}
Furthermore, the unique NE satisfies 
\begin{IEEEeqnarray}{rcl}
P_{A_1}^\star(a_{1})  &=&  
\frac{ u_{2,2} - u_{2,1}}{u_{1,1} - u_{1,2} - u_{2,1} + u_{2,2}} \in (0,1) \ \textnormal{and} \ \label{Equ:uniqueMNE_1}\\
P_{A_2}^\star(a_{1})    &=&  
\frac{ u_{2,2} - u_{1,2}}{u_{1,1} - u_{1,2} - u_{2,1} + u_{2,2}} \in (0,1) \label{Equ:uniqueMNE_2}
\end{IEEEeqnarray} 
and 
 \begin{IEEEeqnarray}{c}
u(P_{A_1}^\star, P_{A_2}^\star) = \frac{u_{1,1} u_{2,2} - u_{1,2} u_{2,1}}{u_{1,1} - u_{1,2} - u_{2,1} + u_{2,2}}.
 \end{IEEEeqnarray}

\noindent 
(ii) The pair of strategies $\left( P_{A_1}^\star, P_{A_2}^\star \right)$ is the unique NE with 
\begin{IEEEeqnarray}{c}
P_{A_1}^\star(a_{1})  =   1 \quad \textnormal{and} \quad 
P_{A_2}^\star(a_{1})    =  1,
\end{IEEEeqnarray} 
if and only if, the entries in the matrix $\matx{u}$ in (\ref{Equ:u}) satisfy
\begin{IEEEeqnarray}{rl}\label{Equ:pen01}
u_{1,2} > u_{1,1} > u_{2,1}.
\end{IEEEeqnarray}
Furthermore, the payoff at the NE is
 \begin{IEEEeqnarray}{c}
 u(P_{A_1}^\star, P_{A_2}^\star) = u_{1,1}.
 \end{IEEEeqnarray}

\noindent 
(iii) The pair of strategies $\left( P_{A_1}^\star, P_{A_2}^\star \right)$ is the unique NE with 
\begin{IEEEeqnarray}{c}
P_{A_1}^\star(a_{1})  =   1  \quad  \textnormal{and} \quad 
P_{A_2}^\star(a_{1})    =  0, 
\end{IEEEeqnarray} 
if and only if, the entries in the matrix $\matx{u}$ in (\ref{Equ:u}) satisfy
\begin{IEEEeqnarray}{rl}\label{Equ:pen02}
u_{1,1} > u_{1,2} > u_{2,2}. 
\end{IEEEeqnarray}
Furthermore, the payoff at the NE is
 \begin{IEEEeqnarray}{c}
 u(P_{A_1}^\star, P_{A_2}^\star) = u_{1,2}.
 \end{IEEEeqnarray}

\noindent 
(iv) The pair of strategies $\left( P_{A_1}^\star, P_{A_2}^\star \right)$ is the unique NE with 
\begin{IEEEeqnarray}{c}
P_{A_1}^\star(a_{1})    = 0 \quad  \textnormal{and} \quad 
P_{A_2}^\star(a_{1})   =1, 
\end{IEEEeqnarray} 
if and only if, the entries in the matrix $\matx{u}$ in (\ref{Equ:u}) satisfy
\begin{IEEEeqnarray}{rl}\label{Equ:pen03}
u_{2,2} > u_{2,1} > u_{1,1}. 
\end{IEEEeqnarray}
Furthermore, the payoff at the NE is
 \begin{IEEEeqnarray}{c}
u(P_{A_1}^\star, P_{A_2}^\star) = u_{2,1}.
 \end{IEEEeqnarray}

\noindent 
(v) The pair of strategies $\left( P_{A_1}^\star, P_{A_2}^\star \right)$ is the unique NE with 
\begin{IEEEeqnarray}{c}
P_{A_1}^\star(a_{1})    = 0  \quad  \textnormal{and} \quad 
P_{A_2}^\star(a_{1})     =0,
\end{IEEEeqnarray} 
if and only if, the entries in the matrix $\matx{u}$ in (\ref{Equ:u}) satisfy
\begin{IEEEeqnarray}{rl}\label{Equ:pen04}
u_{2,1} > u_{2,2} > u_{1,2}. 
\end{IEEEeqnarray}
Furthermore, the payoff at the NE is
 \begin{IEEEeqnarray}{c}
 u(P_{A_1}^\star, P_{A_2}^\star) = u_{2,2}.  
 \end{IEEEeqnarray}

\end{theorem}
\begin{IEEEproof}
The proof of Theorem \ref{TheoNE} is provided in Appendix \ref{Sec:App6}.

\end{IEEEproof}

In the case in which the zero-sum game $\game{G}$ in \eqref{EqTheGame} exhibits an infinite number of NEs, the following lemma shows the closed-form expressions for the NE strategies when there exists an $i \in \{1,2\}$ such that $P_{A_i}^\star(a_1) \in [0,1]$. 

\begin{theorem}[Infinite Nash Equilibrium-I]\label{TheInfNE1}
Let $\set{P} \subset \simplex{\set{A}_1} \times \simplex{\set{A}_2}$ be the set of NEs of the game $\game{G}$ in \eqref{EqTheGame}.

\noindent
(i) The set $\set{P}$ satisfies 
\begin{IEEEeqnarray}{c}
\set{P} = \left\{ \left( P_{1}, P_{2}\right) \in \simplex{\set{A}_1} \times \simplex{\set{A}_2}: P_{1}(a_{1})    = 1 \ \textnormal{and} \ P_{2} \in \simplex{\set{A}_2} \right\},
\end{IEEEeqnarray}
if and only if, the entries of the matrix $\matx{u}$ in  (\ref{Equ:u}) satisfy 
\begin{IEEEeqnarray}{rll}\label{Equ:mouse01}
& u_{1,1} = u_{1,2} > u_{2,1} = u_{2,2} &\qquad \textnormal{or} \\
&u_{1,1} = u_{1,2} \geq \textnormal{max} \{ u_{2,1}, u_{2,2}\} > \textnormal{min} \{ u_{2,1}, u_{2,2}\}&. \label{Equ:mouse011}
\end{IEEEeqnarray}
Furthermore, the payoff at the NEs is 
\begin{IEEEeqnarray}{c}
u(P_{A_1}^\star, P_{A_2}^\star) = u_{1,1} = u_{1,2}.
\end{IEEEeqnarray}

\noindent 
(ii) The set $\set{P}$ satisfies 
\begin{IEEEeqnarray}{c}
\set{P} = \left\{ \left( P_{1}, P_{2}\right) \in \simplex{\set{A}_1} \times \simplex{\set{A}_2}: P_{1}(a_{1})    = 0 \ \textnormal{and} \ P_{2} \in \simplex{\set{A}_2} \right\},  
\end{IEEEeqnarray}
if and only if, the entries of the matrix $\matx{u}$ in  (\ref{Equ:u}) satisfy 
\begin{IEEEeqnarray}{rll}\label{Equ:mouse02}
&u_{2,1} = u_{2,2} > u_{1,1} = u_{1,2} & \qquad \textnormal{or}\\
&u_{2,1} = u_{2,2} \geq \textnormal{max} \{ u_{1,1}, u_{1,2}\} > \textnormal{min} \{ u_{1,1}, u_{1,2}\}.\label{Equ:mouse021}
\end{IEEEeqnarray}
Furthermore, the payoff at the NEs is
\begin{IEEEeqnarray}{c}
u(P_{A_1}^\star, P_{A_2}^\star) = u_{2,1} = u_{2,2}.
\end{IEEEeqnarray}

\noindent 
(iii) The set $\set{P}$ satisfies 
\begin{IEEEeqnarray}{c}
\set{P} = \left\{ \left( P_{1}, P_{2}\right) \in \simplex{\set{A}_1} \times \simplex{\set{A}_2}: P_1 \in \simplex{\set{A}_1} \textnormal{and} \ P_{2}(a_{1})    = 1  \right\},  
\end{IEEEeqnarray}
if and only if, the entries of the matrix $\matx{u}$ in  (\ref{Equ:u})  satisfy 
\begin{IEEEeqnarray}{rll}\label{Equ:mouse03}
&u_{1,1} = u_{2,1} < u_{1,2} = u_{2,2} &\qquad \textnormal{or} \\
&u_{1,1} = u_{2,1} \leq \textnormal{min} \{ u_{1,2}, u_{2,2}\} < \textnormal{max} \{ u_{1,2}, u_{2,2}\}&,\label{Equ:mouse031}
\end{IEEEeqnarray}
Furthermore, the payoff at the NEs is
\begin{IEEEeqnarray}{c}
u(P_{A_1}^\star, P_{A_2}^\star) = u_{1,1} = u_{2,1}.
\end{IEEEeqnarray}

\noindent 
(iv) The set $\set{P}$ satisfies 
\begin{IEEEeqnarray}{c}
\set{P} = \left\{ \left( P_{1}, P_{2}\right) \in \simplex{\set{A}_1} \times \simplex{\set{A}_2}: P_1 \in \simplex{\set{A}_1} \textnormal{and} \ P_{2}(a_{1})    = 0  \right\},  
\end{IEEEeqnarray}
if and only if, the entries of the matrix $\matx{u}$ in  (\ref{Equ:u})  satisfy 
\begin{IEEEeqnarray}{rll}\label{Equ:mouse04}
&u_{1,2} = u_{2,2} < u_{1,1} = u_{2,1} & \qquad \textnormal{or}\\
&u_{1,2} = u_{2,2} \leq \textnormal{min} \{ u_{1,1}, u_{2,1}\} < \textnormal{max} \{ u_{1,1}, u_{2,1}\}.\label{Equ:mouse041}
\end{IEEEeqnarray}
Furthermore, the payoff at the NEs is 
\begin{IEEEeqnarray}{c}
u(P_{A_1}^\star, P_{A_2}^\star) = u_{1,2} = u_{2,2}.
\end{IEEEeqnarray}

\noindent
(v) The set $\set{P}$ satisfies 
\begin{IEEEeqnarray}{c}
\set{P} = \left\{ \left( P_{1}, P_{2}\right) \in \simplex{\set{A}_1} \times \simplex{\set{A}_2}: P_1 \in \simplex{\set{A}_1} \textnormal{and} \ P_2 \in \simplex{\set{A}_2}  \right\},  
\end{IEEEeqnarray}
if and only if, the entries of the matrix $\matx{u}$ in  (\ref{Equ:u})  satisfy 
\begin{IEEEeqnarray}{c}\label{Equ:mouse05}
u_{1,1} = u_{1,2} = u_{2,1} = u_{2,2}. 
\end{IEEEeqnarray}
Furthermore, the payoff at the NEs is
\begin{IEEEeqnarray}{c}
u(P_{A_1}^\star, P_{A_2}^\star) = u_{1,1}= u_{1,2}= u_{2,1} = u_{2,2}. 
\end{IEEEeqnarray}
\end{theorem}

\begin{IEEEproof}
The proof of Theorem \ref{TheInfNE1} is provided in Appendix \ref{Sec:App7}.

\end{IEEEproof}

In the case in which the zero-sum game $\game{G}$ in \eqref{EqTheGame} exhibits an infinite number of NEs, the following lemma shows the closed-form expressions for the NE strategies when there exists an $i \in \{1,2\}$ such that $P_{A_i}^\star(a_1) \in [0,a]$ or $P_{A_i}^\star(a_1) \in [b,1]$ for some $a \in (0,1)$ and $b\in (0,1)$. 

\begin{theorem}[Infinite Nash Equilibrium-II]\label{TheInfNE2}
Let $\set{P} \subset \simplex{\set{A}_1} \times \simplex{\set{A}_2}$ be the set of NEs of the game $\game{G}$ in \eqref{EqTheGame}.

\noindent 
(i) The set $\set{P}$ satisfies 
\begin{IEEEeqnarray}{c}
\set{P} = \left\{ \left( P_{1}, P_{2}\right) \in \simplex{\set{A}_1} \times \simplex{\set{A}_2}: P_{1}(a_{1})    = 0 \ \textnormal{and} \ P_{2}(a_1) \in \left[0, \frac{u_{2,2} -u_{1,2}}{u_{1,1} - u_{1,2} -u_{2,1}+u_{2,2}} \right] \right\}  \IEEEeqnarraynumspace  \squeezeequ  \label{Equ:mous0112}
\end{IEEEeqnarray}
with $0 <\frac{u_{2,2} -u_{1,2}}{u_{1,1} - u_{1,2} -u_{2,1}+u_{2,2}} < 1$, 
if and only if, the entries of the matrix $\matx{u}$ in  (\ref{Equ:u}) satisfy 
\begin{IEEEeqnarray}{rll}\label{Equ:mous011}
u_{1,1} >u_{2,1} = u_{2,2}>u_{1,2}. 
\end{IEEEeqnarray}
Furthermore, the payoff at the NEs satisfies 
\begin{IEEEeqnarray}{c}
u(P_{A_1}^\star, P_{A_2}^\star) = u_{2,1} = u_{2,2}. 
\end{IEEEeqnarray}

\noindent 
(ii) The set $\set{P}$ satisfies 
\begin{IEEEeqnarray}{c}
\set{P} = \left\{ \left( P_{1}, P_{2}\right) \in \simplex{\set{A}_1} \times \simplex{\set{A}_2}: P_{1}(a_{1})    = 0 \ \textnormal{and} \ P_{2}(a_1) \in \left[\frac{u_{2,2} -u_{1,2}}{u_{1,1} - u_{1,2} -u_{2,1}+u_{2,2}} , 1\right] \right\}  \IEEEeqnarraynumspace  \squeezeequ  \label{Equ:mous0122}
\end{IEEEeqnarray}
with $0 <\frac{u_{2,2} -u_{1,2}}{u_{1,1} - u_{1,2} -u_{2,1}+u_{2,2}} < 1$,
if and only if, the entries of the matrix $\matx{u}$ in  (\ref{Equ:u}) satisfy 
\begin{IEEEeqnarray}{rll}\label{Equ:mous012}
u_{1,1} <u_{2,1} = u_{2,2}<u_{1,2}. 
\end{IEEEeqnarray}
Furthermore, the payoff at the NEs satisfies 
\begin{IEEEeqnarray}{c}
u(P_{A_1}^\star, P_{A_2}^\star) = u_{2,1} =u_{2,2}. 
\end{IEEEeqnarray}

\noindent 
(iii) The set $\set{P}$ satisfies 
\begin{IEEEeqnarray}{c}
\set{P} = \left\{ \left( P_{1}, P_{2}\right) \in \simplex{\set{A}_1} \times \simplex{\set{A}_2}: P_{1}(a_{1})    = 1 \ \textnormal{and} \ P_{2}(a_1) \in \left[\frac{u_{2,2} -u_{1,2}}{u_{1,1} - u_{1,2} -u_{2,1}+u_{2,2}} , 1\right] \right\}  \IEEEeqnarraynumspace  \squeezeequ  \label{Equ:mous0212}
\end{IEEEeqnarray}
with $0 <\frac{u_{2,2} -u_{1,2}}{u_{1,1} - u_{1,2} -u_{2,1}+u_{2,2}} < 1$,
if and only if, the entries of the matrix $\matx{u}$ in  (\ref{Equ:u}) satisfy 
\begin{IEEEeqnarray}{rll}\label{Equ:mous021}
u_{2,2} > u_{1,2} = u_{1,1} >u_{2,1}. 
\end{IEEEeqnarray}
Furthermore, the payoff at the NEs satisfies 
\begin{IEEEeqnarray}{c}
u(P_{A_1}^\star, P_{A_2}^\star) = u_{1,1} =u_{1,2}. 
\end{IEEEeqnarray}

\noindent 
(iv) The set $\set{P}$ satisfies 
\begin{IEEEeqnarray}{c}
\set{P} = \left\{ \left( P_{1}, P_{2}\right) \in \simplex{\set{A}_1} \times \simplex{\set{A}_2}: P_{1}(a_{1})    = 1 \ \textnormal{and} \ P_{2}(a_1) \in \left[0, \frac{u_{2,2} -u_{1,2}}{u_{1,1} - u_{1,2} -u_{2,1}+u_{2,2}} \right] \right\}  \IEEEeqnarraynumspace  \squeezeequ  \label{Equ:mous0222}
\end{IEEEeqnarray}
with $0 <\frac{u_{2,2} -u_{1,2}}{u_{1,1} - u_{1,2} -u_{2,1}+u_{2,2}} < 1$,
if and only if, the entries of the matrix $\matx{u}$ in  (\ref{Equ:u}) satisfy 
\begin{IEEEeqnarray}{rll}\label{Equ:mous022}
u_{2,2} < u_{1,2} = u_{1,1} <u_{2,1}.
\end{IEEEeqnarray}
Furthermore, the payoff at the NEs satisfies 
\begin{IEEEeqnarray}{c}
u(P_{A_1}^\star, P_{A_2}^\star) = u_{1,1} =u_{1,2}. 
\end{IEEEeqnarray}

\noindent 
(v) The set $\set{P}$ satisfies 
\begin{IEEEeqnarray}{c}
\set{P} = \left\{ \left( P_{1}, P_{2}\right) \in \simplex{\set{A}_1} \times \simplex{\set{A}_2}: P_{1}(a_{1})   \in \left[\frac{u_{2,2} -u_{2,1}}{u_{1,1} - u_{1,2} -u_{2,1}+u_{2,2}} , 1\right]  \ \textnormal{and} \ P_{2}(a_1)=0  \right\}   \IEEEeqnarraynumspace  \squeezeequ  \label{Equ:mous0312}
\end{IEEEeqnarray}
with $0 <\frac{u_{2,2} -u_{2,1}}{u_{1,1} - u_{1,2} -u_{2,1}+u_{2,2}} < 1$,
if and only if, the entries of the matrix $\matx{u}$ in  (\ref{Equ:u})  satisfy 
\begin{IEEEeqnarray}{rll}\label{Equ:mous031}
u_{1,1} > u_{1,2} = u_{2,2}  >u_{2,1}.
\end{IEEEeqnarray}
Furthermore, the payoff at the NEs satisfies 
\begin{IEEEeqnarray}{c}
u(P_{A_1}^\star, P_{A_2}^\star) = u_{1,2} = u_{2,2}. 
\end{IEEEeqnarray}

\noindent 
(vi) The set $\set{P}$ satisfies 
\begin{IEEEeqnarray}{c}
\set{P} = \left\{ \left( P_{1}, P_{2}\right) \in \simplex{\set{A}_1} \times \simplex{\set{A}_2}: P_{1}(a_{1})   \in \left[0, \frac{u_{2,2} -u_{2,1}}{u_{1,1} - u_{1,2} -u_{2,1}+u_{2,2}} \right]  \ \textnormal{and} \ P_{2}(a_1)=0  \right\}   \IEEEeqnarraynumspace  \squeezeequ \label{Equ:mouse0322}
\end{IEEEeqnarray}
with $0 <\frac{u_{2,2} -u_{2,1}}{u_{1,1} - u_{1,2} -u_{2,1}+u_{2,2}} < 1$,
if and only if, the entries of the matrix $\matx{u}$ in  (\ref{Equ:u})  satisfy 
\begin{IEEEeqnarray}{rll}\label{Equ:mouse032}
u_{1,1} < u_{1,2} = u_{2,2}  <u_{2,1}. 
\end{IEEEeqnarray}
Furthermore, the payoff at the NEs satisfies 
\begin{IEEEeqnarray}{c}
u(P_{A_1}^\star, P_{A_2}^\star) = u_{1,2} = u_{2,2}. 
\end{IEEEeqnarray}

\noindent 
(vii)  The set $\set{P}$ satisfies 
\begin{IEEEeqnarray}{c}
\set{P} = \left\{ \left( P_{1}, P_{2}\right) \in \simplex{\set{A}_1} \times \simplex{\set{A}_2}: P_{1}(a_{1})   \in \left[0, \frac{u_{2,2} -u_{2,1}}{u_{1,1} - u_{1,2} -u_{2,1}+u_{2,2}} \right]  \ \textnormal{and} \ P_{2}(a_1)=1  \right\}   \IEEEeqnarraynumspace  \squeezeequ \label{Equ:mouse0411}
\end{IEEEeqnarray}
with $0 <\frac{u_{2,2} -u_{2,1}}{u_{1,1} - u_{1,2} -u_{2,1}+u_{2,2}} < 1$,
 if and only if, the entries of the matrix $\matx{u}$ in  (\ref{Equ:u})  satisfy
\begin{IEEEeqnarray}{rll}\label{Equ:mouse0410}
u_{2,2} >u_{2,1} = u_{1,1}  > u_{1,2}. 
\end{IEEEeqnarray}
Furthermore, the payoff at the NEs satisfies 
\begin{IEEEeqnarray}{c}
u(P_{A_1}^\star, P_{A_2}^\star) = u_{1,1}=u_{2,1}. 
\end{IEEEeqnarray}

\noindent 
(viii)  The set $\set{P}$ satisfies 
\begin{IEEEeqnarray}{c}
\set{P} = \left\{ \left( P_{1}, P_{2}\right) \in \simplex{\set{A}_1} \times \simplex{\set{A}_2}: P_{1}(a_{1})   \in \left[\frac{u_{2,2} -u_{2,1}}{u_{1,1} - u_{1,2} -u_{2,1}+u_{2,2}},1 \right]  \ \textnormal{and} \ P_{2}(a_1)=1  \right\}   \IEEEeqnarraynumspace  \squeezeequ \label{Equ:mouse0421}
\end{IEEEeqnarray}
with $0 <\frac{u_{2,2} -u_{2,1}}{u_{1,1} - u_{1,2} -u_{2,1}+u_{2,2}} < 1$,
if and only if, the entries of the matrix $\matx{u}$ in  (\ref{Equ:u})  satisfy 
\begin{IEEEeqnarray}{rll}\label{Equ:mouse042}
u_{2,2} <u_{2,1} = u_{1,1}  < u_{1,2}. 
\end{IEEEeqnarray}
Furthermore, the payoff at the NEs satisfies 
\begin{IEEEeqnarray}{c}
u(P_{A_1}^\star, P_{A_2}^\star) = u_{1,1} =u_{2,1}. 
\end{IEEEeqnarray}

\end{theorem}

\begin{IEEEproof}
The proof of Theorem \ref{TheInfNE2} is provided in Appendix \ref{Sec:App7_1}.

\end{IEEEproof}

\subsection{Payoff at the Nash Equilibrium}
The payoff at the NE (the value of the game) is characterized by the following theorem. 

\begin{theorem}[Value of the Game]\label{TheoremNE}
Let the probability measures $P^{\star}_{A_1} \in \simplex{\set{A}_1}$ and $P^{\star}_{A_2} \in \simplex{\set{A}_2}$ form an NE of the game $\game{G}$ in~\eqref{EqTheGame}. 
If the entries of the matrix $\matx{u}$ in~\eqref{Equ:u} satisfy
\begin{subequations}
\begin{IEEEeqnarray}{lcl}
\left( u_{1,1} - u_{1,2} \right) \left( u_{2,2} - u_{2,1} \right)   >   0 & \mbox{ and } & \\
\left( u_{1,1} - u_{2,1} \right) \left( u_{2,2} - u_{1,2} \right)   >   0,
\end{IEEEeqnarray}
\label{EqMixedAssumption}
\end{subequations}
then, the NE of the game $\game{G}$ in~\eqref{EqTheGame} is unique, with 
\begin{subequations}\label{EqNEStratExample}
\begin{IEEEeqnarray}{rcl}
\label{EqPA1StarExample}
P^{\star}_{A_1}(a_1) & = &  \frac{u_{2,2}-u_{2,1}}{u_{1,1} - u_{1,2} - u_{2,1}+u_{2,2}} \in (0,1)\mbox{  and  } \IEEEeqnarraynumspace\\
\label{EqPA2StarExample}
P^{\star}_{A_2}(a_1) & = &\frac{u_{2,2}-u_{1,2}}{u_{1,1} - u_{1,2} - u_{2,1}+u_{2,2}}\in (0,1).
\end{IEEEeqnarray} 
\end{subequations}
Moreover,  the expected payoff at the NE is
\begin{IEEEeqnarray}{rcl}
u(P_{A_1}^{\star},P_{A_2}^{\star}) & = & \frac{u_{1,1}u_{2,2} - u_{1,2}u_{2,1}}{u_{1,1} - u_{1,2} - u_{2,1}+u_{2,2}}.
\end{IEEEeqnarray}
If the entries of the matrix $\matx{u}$ in~\eqref{Equ:u}  satisfy
\begin{subequations}
\begin{IEEEeqnarray}{lcl}
\left( u_{1,1} - u_{1,2} \right) \left( u_{2,2} - u_{2,1} \right)   \leqslant   0 & \mbox{ or } & \\
\left( u_{1,1} - u_{2,1} \right) \left( u_{2,2} - u_{1,2} \right)   \leqslant   0,
\end{IEEEeqnarray}
\label{EqNotMixedAssumption}
\end{subequations}
then, there exists either a unique NE or infinitely many NEs. Moreover, all NE strategies lead to the same payoff,
\begin{subequations}
\begin{IEEEeqnarray}{rcl}\label{Equ:f_3}
u(P_{A_1}^{\star},P_{A_2}^{\star}) & = & \min \lbrace \max\lbrace u_{1,1}, u_{2,1}\rbrace,  \max\lbrace u_{1,2}, u_{2,2}\rbrace \rbrace \squeezeequ\\
& = & \max \lbrace \min\lbrace u_{1,1}, u_{1,2}\rbrace,  \min\lbrace u_{2,1}, u_{2,2}\rbrace \rbrace. \IEEEeqnarraynumspace \squeezeequ \label{Equ:f_4}
\end{IEEEeqnarray}
\end{subequations}
\end{theorem}

\begin{IEEEproof}
The proof of Theorem \ref{TheoremNE} is provided in Appendix \ref{AppTheoremNE}. 

\end{IEEEproof}

A key observation of the proof of Theorem \ref{TheoremNE} is that if the entries of the payoff matrix $\matx{u}$ satisfy \eqref{EqNotMixedAssumption}, 
the value of $2 \times 2$ zero sum game can be searched exclusively in pure strategies.
The following lemma shows that if the entries of the payoff matrix $\matx{u}$ satisfy \eqref{EqMixedAssumption}, it is not true. 
\begin{lemma}\label{LemmaMix}
If the entries of the payoff matrix $\matx{u}$ in \eqref{Equ:u} satisfy \eqref{EqMixedAssumption}, then 
\begin{IEEEeqnarray}{rl}
\min \left\{ \max \left\{ u_{1,1},  u_{2,1} \right\}, \max \left\{ u_{1,2},  u_{2,2} \right\}\right\} &> \frac{u_{1,1}u_{2,2} - u_{1,2} u_{2,1}}{u_{1,1} - u_{1,2} -u_{2,1} +u_{2,2}} \\
&> \max \left\{ \min \left\{ u_{1,1},  u_{1,2} \right\}, \min \left\{ u_{2,2},  u_{2,1} \right\}\right\}. 
\end{IEEEeqnarray}
\end{lemma}
\begin{IEEEproof} 
The proof of Lemma \ref{LemmaMix} is provided in Appendix \ref{AppLemmaMix}. 
\end{IEEEproof}

\subsection{Leadership of \Ptwo}

In this subsection, the case in which the players do not choose the strategies simultaneously is considered. 
Without loss of generality, it is assumed that \Ptwo chooses the strategy first. 
\Pone observes the strategy of \Ptwo and chooses a strategy to maximize the payoff. 
Hence, the payoff under this case is fully characterized by the function $\hat{u}: \simplex{\set{A}_2} \to \reals$ such that for all $P \in \simplex{\set{A}_2}$, 
\begin{IEEEeqnarray}{rcl}
\label{EqHatu}
\hat{u}\left( P \right) & = & \max_{Q \in \simplex{\set{A}_1}} u\left(Q,  P\right),
\end{IEEEeqnarray}
with the function $u$ defined in~\eqref{Eqv}. 

Note that the best response of \Pone in Lemma \ref{LemmaBR1} is either a pure strategy or the simplex $\simplex{\set{A}_1}$.
Hence, it follows that 
\begin{IEEEeqnarray}{rl}\label{EqUp1}
\hat{u}\left( P \right) & = \max_{Q \in \{ T \in \simplex{\set{A}_1}: T(a_1) \in \{ 0,1\} \} } u(Q,P) \\
&= \max \left\{ u(Q', P), u(Q'', P)\right\}, 
\end{IEEEeqnarray}
where $Q' \in \simplex{\set{A}_1}$ satisfies $Q'(a_1)=1$ and  $Q'' \in \simplex{\set{A}_1}$ satisfies $Q''(a_1)=0$.
Plugging \eqref{Eqv} into \eqref{EqUp1} yields 
\begin{IEEEeqnarray}{c}\label{EqUp2}
\hat{u}\left( P \right) = \max \left\{ u_{1,1}P(a_1) + u_{1,2}P(a_2),  u_{2,1}P(a_1) + u_{2,2}P(a_2)\right\}.
\end{IEEEeqnarray}

The following lemma shows a closed-form expression when the entries of the payoff matrix $\matx{u}$ in \eqref{Equ:u} satisfy \eqref{EqMixedAssumption}. 

\begin{lemma}\label{LemmaHatu}
Assume that the matrix $\matx{u}$ in~\eqref{Equ:u} satisfies~\eqref{EqMixedAssumption}. 
Let $P_{A_1}^{\star} \in \simplex{\set{A}_1}$ and $P_{A_2}^{\star} \in \simplex{\set{A}_2}$ form an NE in the game $\game{g}(\matx{u})$ in~\eqref{EqTheGame}. 
And let $\delta \triangleq u_{1,1} -u_{1,2}-u_{2,1}+u_{2,2}$.
Then for all $P \in \simplex{\set{A}_2}$, the function $\hat{u}$ in~\eqref{EqHatu} satisfies: 
\begin{IEEEeqnarray}{c}\label{Equ:g_1}
\hat{u}(P)= 
\quad  \left\{ 
\begin{array}{cl}
u_{2,1} P(a_1)+u_{2,2}P(a_2), &  \ \ \makecell[t]{ \textnormal{ if } P(a_1) < P^{\star}_{A_2}(a_1) \ \textnormal{and} \ \delta> 0 \quad \textnormal{or } \\ \!\!\!\! P(a_1) > P^{\star}_{A_2}(a_1) \ \textnormal{and} \ \delta \leq 0,} \\
u_{1,1}P(a_1)  +u_{1,2}P(a_2), & \ \ \makecell[t]{\textnormal{ if }  P(a_1) > P^{\star}_{A_2}(a_1) \ \textnormal{and} \ \delta > 0 \quad \textnormal{or } \\ \!\!\!\!  P(a_1) < P^{\star}_{A_2}(a_1) \ \textnormal{and} \ \delta \leq 0,} \\
u(P_{A_1}^{\star}, P_{A_2}^{\star}), &\ \   \makecell[t]{\textnormal{ if }   P(a_1) = P^{\star}_{A_2}(a_1), }
\end{array}
\right.  \IEEEeqnarraynumspace 
\end{IEEEeqnarray}
with $P^{\star}_{A_2}(a_1)$ in~\eqref{EqPA2StarExample}. 
\end{lemma}

\begin{IEEEproof}
The proof of Lemma \ref{LemmaHatu} is provided in Appendix \ref{AppLemmaHatu}.

\end{IEEEproof}

The following lemma shows the monotonicity of $\hat{u}(P_{A_2})$ in \eqref{EqHatu} in the interval $P_{A_2}(a_1) \in \left[0, P_{A_2}^\star (a_1) \right]$ and $P_{A_2}(a_1) \in \left[ P_{A_2}^\star (a_1), 1\right]$. 

\begin{lemma}\label{LemmaMono}
Assume that the entries of the matrix $\matx{u}$ in~\eqref{Equ:u} satisfy \eqref{EqMixedAssumption}. 
For all tuples $\left(P, Q \right) \in \simplex{\set{A}_2} \times \simplex{\set{A}_2}$, if $0 \leq P(a_1) < Q(a_1) \leq P_{A_2}^\star(a_1)$, then it holds that 
\begin{IEEEeqnarray}{c}
\hat{u}(P) > \hat{v}(Q).
\end{IEEEeqnarray}
Alternatively, if $ P_{A_2}^\star(a_1) \leq P(a_1) < Q(a_1) \leq 1$, then it holds that 
\begin{IEEEeqnarray}{c}
\hat{u}(P) <\hat{v}(Q). 
\end{IEEEeqnarray}
\end{lemma}

\begin{IEEEproof}
The proof of Lemma \ref{LemmaMono} is provided in Appendix \ref{AppLemmaMono}.

\end{IEEEproof}

The following lemma shows that the minimum value of $\hat{u}$ in \eqref{EqHatu} equals the value of the game. 

\begin{lemma}\label{LemmaNESE}
Let the probability measures $P^{\star}_{A_1} \in \simplex{\set{A}_1}$ and $P^{\star}_{A_2} \in \simplex{\set{A}_2}$ be one of the NEs of the game $\game{g}$ in \eqref{EqTheGame}. Then, 
\begin{IEEEeqnarray}{rcl}
\min_{P \in \simplex{\set{A}_2}} \hat{u}(P) & = & u(P_{A_1}^{\star},P_{A_2}^{\star}), 
\end{IEEEeqnarray}
where  $\hat{u}$ is in \eqref{EqHatu}. 
\end{lemma}

\begin{IEEEproof}
The proof of Lemma \ref{LemmaNESE} is provided in Appendix \ref{AppLemmaNESE}. 

\end{IEEEproof}
\newpage

\appendix

%
%
%
%

\section{Proof of Lemma \ref{Lemma:infinite}}\label{Sec:App1}
The proof consists in studying the following cases.

\noindent Case I: 
\begin{IEEEeqnarray}{rl}
 &P_{A_1}^\star(a_1) = 1 - P_{A_1}^\star(a_2) =0  \quad \textnormal{and} \label{Equ:P_A1}\\
 & P_{A_2}^\star(a_1) = 1 - P_{A_2}^\star(a_2) = \alpha, \label{Equ:P_A2}
\end{IEEEeqnarray}
with $\alpha \in (0,1)$. 

\noindent Case II: 
\begin{IEEEeqnarray}{rl}
 &P_{A_1}^\star(a_1) = 1 - P_{A_1}^\star(a_2) =1  \quad \textnormal{and} \label{Equ:P_A1_1}\\
 & P_{A_2}^\star(a_1) = 1 - P_{A_2}^\star(a_2) = \alpha,  \label{Equ:P_A2_1}
\end{IEEEeqnarray}
with $\alpha \in (0,1)$. 

\noindent Case III: 
\begin{IEEEeqnarray}{rl}
 &P_{A_1}^\star(a_1) = 1 - P_{A_1}^\star(a_2) = \alpha  \quad \textnormal{and} \label{Equ:P_A1_2}\\
 & P_{A_2}^\star(a_1) = 1 - P_{A_2}^\star(a_2) = 0, \label{Equ:P_A2_2}
\end{IEEEeqnarray}
with $\alpha \in (0,1)$. 

\noindent Case IV: 
\begin{IEEEeqnarray}{rl}
 &P_{A_1}^\star(a_1) = 1 - P_{A_1}^\star(a_2) = \alpha  \quad \textnormal{and} \label{Equ:P_A1_3}\\
 & P_{A_2}^\star(a_1) = 1 - P_{A_2}^\star(a_2) = 1, \label{Equ:P_A2_3}
\end{IEEEeqnarray}
with $\alpha \in (0,1)$. 

Note that in the proof, the set of all the NEs in the game $\game{G}$ in \eqref{EqTheGame} is denoted by $\set{P} \subset \simplex{\set{A}_1} \times  \simplex{\set{A}_2}$. 

The proof of Case I is as follows. 
Assume that the pair of strategies $\left( P_{A_1}^\star, P_{A_2}^\star \right)$ in (\ref{Equ:P_A1}) and (\ref{Equ:P_A2}) forms an NE in the game $\game{G}$ in \eqref{EqTheGame}.
Then, from the indifference principle \cite[Theorem 5.18]{Maschler_2013_game}, it follows that Player 2 is indifferent to using either action $a_1$ or $a_2$. 
That is, 
\begin{IEEEeqnarray}{c}\label{Equ:Lemma1_1}
 u(P_{A_1}^\star, P_{A_2}^\star) = \left(0, \  1\right) \matx{u} \left(1, \  0\right)^{\sf{T}} =  \left(0, \  1\right) \matx{u} \left(0, \  1\right)^{\sf{T}}, 
\end{IEEEeqnarray}
which implies that 
\begin{IEEEeqnarray}{c}\label{Equ:Lemma1_2}
u_{2,1} = u_{2,2}. 
\end{IEEEeqnarray}
Under the assumption that $\left( P_{A_1}^\star, P_{A_2}^\star \right)$ forms an NE,
it follows from \eqref{EqNECon1} that  for all $ \beta \in \left[0,1\right]$, it holds that 
\begin{IEEEeqnarray}{c}
 u(P_{A_1}^\star, P_{A_2}^\star) = (0,1) \matx{u} \left( \alpha, 1-\alpha\right)^{\sf T} \geq (\beta,1 -\beta) \matx{u} \left( \alpha, 1-\alpha\right)^{\sf T}, 
\end{IEEEeqnarray} 
which implies that 
\begin{IEEEeqnarray}{c}
u_{2,1} \alpha +u_{2,2} (1-\alpha) \geq \alpha \beta u_{1,1} + \beta (1-\alpha) u_{1,2} + (1-\beta)\alpha u_{2,1} + (1-\beta)(1-\alpha) u_{2,2}. \label{EqCof1}
\end{IEEEeqnarray}
Note that taking the equality in \eqref{Equ:Lemma1_2} into \eqref{EqCof1} yields that
\begin{IEEEeqnarray}{c}\label{EqBott1}
 u_{2,1} \geq  \alpha u_{1,1} + \left( 1-\alpha\right)u_{1,2}. 
\end{IEEEeqnarray}

The following proof proves that either 
\begin{IEEEeqnarray}{l}
\left\{(P_1,P_2) \in \simplex{\set{A}_1} \times \simplex{\set{A}_2}: P_1(a_1) = 1 \ \textnormal{and} \ P_2(a_1) \in [0,\alpha]\right\} \subset \set{P} \quad \textnormal{or} \\
\left\{(P_1,P_2) \in \simplex{\set{A}_1} \times \simplex{\set{A}_2}: P_1(a_1) = 1 \ \textnormal{and} \ P_2(a_1) \in [\alpha, 1]\right\} \subset \set{P} 
\end{IEEEeqnarray}
holds, which implies that $\set{P}$ is a set of cardinality infinite. 

First, consider the case in which $u_{1,1} \geq u_{1,2}$. 
From (\ref{Equ:Lemma1_2}), for all $P\in \Delta (\set{A}_2)$ such that $P(a_1) \in [0,\alpha]$, and for all $\alpha' \in [0,1]$, it holds that 
\begin{IEEEeqnarray}{c}
\left(0, \  1\right) \matx{u} \bigl(P(a_1), \  1- P(a_1)\bigr)^{\sf{T}} = u_{2,1} = \left(0, \  1\right) \matx{u} \bigl(\alpha', \  1-\alpha'\bigr)^{\sf{T}} = u_{2,1}. \IEEEeqnarraynumspace
\end{IEEEeqnarray}
Furthermore, from (\ref{Equ:Lemma1_2}) and \eqref{EqBott1}, for all $P\in \Delta (\set{A}_2)$ such that $P(a_1) \in [0,\alpha]$ and all $\beta \in [0,1]$, 
\begin{IEEEeqnarray}{rl}
& \left(0, \  1\right) \matx{u} \bigl(P(a_1), \  1- P(a_1)\bigr)^{\sf{T}} \IEEEnonumber \\
&= u_{2,1}  \\
& \geq  \beta \bigl(P(a_1)u_{1,1} + (1-P(a_1)) u_{1,2}  \bigr) + (1-\beta)u_{2,1} \\
& = P(a_1) \beta u_{1,1} + \beta (1-P(a_1)) u_{1,2} + (1-\beta)P(a_1) u_{2,1} + (1-\beta)(1-P(a_1)) u_{2,2} \IEEEeqnarraynumspace  \\
& = \left(\beta, \  1 -\beta \right) \matx{u} \bigl(P(a_1), \  1- P(a_1)\bigr)^{\sf{T}},\IEEEeqnarraynumspace
\end{IEEEeqnarray}
where the inequality follows from the fact that if $u_{1,1} \geq u_{1,2}$ and $P(a_1)\leq \alpha$, from \eqref{EqBott1}, it holds that 
\begin{IEEEeqnarray}{c}
 u_{2,1} \geq  \alpha u_{1,1} + \left( 1-\alpha\right)u_{1,2} \geq P(a_1)u_{1,1} + (1-P(a_1)) u_{1,2}. 
\end{IEEEeqnarray}
Hence, it holds that 
\begin{IEEEeqnarray}{c}
\left\{(P_1,P_2) \in \simplex{\set{A}_1} \times \simplex{\set{A}_2}: P_1(a_1) = 0 \ \textnormal{and} \ P_2(a_1) \in [0,\alpha]\right\} \subset \set{P}. 
\end{IEEEeqnarray}

Alternatively, consider the case in which $u_{1,1} < u_{1,2}$. 
From (\ref{Equ:Lemma1_2}), for all $P\in \Delta (\set{A}_2)$ such that $P(a_1) \in [\alpha, 1]$, and for all $\alpha' \in [0,1]$, it holds that
\begin{IEEEeqnarray}{c}
\left(0, \  1\right) \matx{u} \bigl(P(a_1), \  1- P(a_1)\bigr)^{\sf{T}} = u_{2,1} = \left(0, \  1\right) \matx{u} \bigl(\alpha', \  1-\alpha'\bigr)^{\sf{T}} = u_{2,1}. \IEEEeqnarraynumspace
\end{IEEEeqnarray}
Furthermore, from (\ref{Equ:Lemma1_2}) and \eqref{EqBott1}, for all $P\in \Delta (\set{A}_2)$ such that $P(a_1) \in [\alpha, 1]$ and all $\beta \in [0,1]$, 
\begin{IEEEeqnarray}{rl}
& \left(0, \  1\right) \matx{u} \bigl(P(a_1), \  1- P(a_1)\bigr)^{\sf{T}} \IEEEnonumber \\
&= u_{2,1}  \\
& \geq   \beta \bigl(P(a_1)u_{1,1} + (1-P(a_1)\bigr) u_{1,2}  ) + (1-\beta)u_{2,1} \\
& = P(a_1) \beta u_{1,1} + \beta (1-P(a_1)) u_{1,2} + (1-\beta)P(a_1) u_{2,1} + (1-\beta)(1-P(a_1)) u_{2,2} \IEEEeqnarraynumspace  \\
& = \left(\beta, \  1 -\beta \right) \matx{u} \bigl(P(a_1), \  1- P(a_1)\bigr)^{\sf{T}},\IEEEeqnarraynumspace
\end{IEEEeqnarray}
where the inequality follows from the fact that if $u_{1,1} < u_{1,2}$ and $P(a_1)\geq \alpha$, from \eqref{EqBott1}, it holds that 
\begin{IEEEeqnarray}{c}
 u_{2,1} \geq  \alpha u_{1,1} + \left( 1-\alpha\right)u_{1,2} \geq P(a_1)u_{1,1} + (1-P(a_1)) u_{1,2}. 
\end{IEEEeqnarray}
Hence, it holds that 
\begin{IEEEeqnarray}{c}
\left\{(P_1,P_2) \in \simplex{\set{A}_1} \times \simplex{\set{A}_2}: P_1(a_1) = 0 \ \textnormal{and} \ P_2(a_1) \in [\alpha, 1]\right\} \subset \set{P}. 
\end{IEEEeqnarray}

This completes the proof.

The proof of Case II is as follows. 
Assume that the pair of strategies $\left( P_{A_1}^\star, P_{A_2}^\star \right)$ in (\ref{Equ:P_A1_1}) and (\ref{Equ:P_A2_1}) forms an NE in the game $\game{G}$ in \eqref{EqTheGame}.
Then from the indifference principle \cite[Theorem 5.18]{Maschler_2013_game}, it follows that Player 2 is indifferent to using either action $a_1$ or $a_2$. 
That is, 
\begin{IEEEeqnarray}{c}\label{Equ:Lemma1_3}
 \left(1, \  0\right) \matx{u} \left(1, \  0\right)^{\sf{T}} =  \left(1, \  0\right) \matx{u} \left(0, \  1\right)^{\sf{T}}, 
\end{IEEEeqnarray}
which implies that 
\begin{IEEEeqnarray}{c}\label{Equ:Lemma1_4}
u_{1,1} = u_{1,2}. 
\end{IEEEeqnarray}
Under the assumption that $\left( P_{A_1}^\star, P_{A_2}^\star \right)$ forms an NE, from \eqref{EqNECon1}, for all $ \beta \in \left[0,1\right]$, it holds that
\begin{IEEEeqnarray}{c}
(1,0) \matx{u} \left( \alpha, 1-\alpha\right)^{\sf T} \geq (\beta,1 -\beta) \matx{u} \left( \alpha, 1-\alpha\right)^{\sf T}, 
\end{IEEEeqnarray} 
which implies that 
\begin{IEEEeqnarray}{c}
u_{1,1} \alpha +u_{1,2} (1-\alpha) \geq \alpha \beta u_{1,1} + \beta (1-\alpha) u_{1,2} + (1-\beta)\alpha u_{2,1} + (1-\beta)(1-\alpha) u_{2,2}. \label{EqCof2}
\end{IEEEeqnarray}
Note that taking the equality in \eqref{Equ:Lemma1_4} into \eqref{EqCof2} yields that for all $\beta \in \left[ 0,1\right]$, it holds that 
\begin{IEEEeqnarray}{c}\label{EqBott2}
 u_{1,1} \geq \alpha u_{2,1} + \left( 1-\alpha\right)u_{2,2}.
\end{IEEEeqnarray}

The following proof proves that either 
\begin{IEEEeqnarray}{l}
\left\{(P_1,P_2) \in \simplex{\set{A}_1} \times \simplex{\set{A}_2}: P_1(a_1) = 1 \ \textnormal{and} \ P_2(a_1) \in [0,\alpha]\right\} \subset \set{P} \quad \textnormal{or} \\
\left\{(P_1,P_2) \in \simplex{\set{A}_1} \times \simplex{\set{A}_2}: P_1(a_1) = 1 \ \textnormal{and} \ P_2(a_1) \in [\alpha, 1]\right\} \subset \set{P} 
\end{IEEEeqnarray}
holds, which implies that $\set{P}$ is a set of cardinality infinite. 

First, consider the case in which $u_{2,1} \geq u_{2,2}$. 
From (\ref{Equ:Lemma1_2}), for all $P\in \Delta (\set{A}_2)$ such that $P(a_1) \in [0,\alpha]$, and for all $\alpha' \in [0,1]$, it holds that 
\begin{IEEEeqnarray}{c}
\left(1, \  0\right) \matx{u} \bigl(P(a_1), \  1- P(a_1)\bigr)^{\sf{T}} = u_{1,1} = \left(1, \  0\right) \matx{u} \bigl(\alpha', \  1-\alpha'\bigr)^{\sf{T}} = u_{1,1}. \IEEEeqnarraynumspace
\end{IEEEeqnarray}
Furthermore, from (\ref{Equ:Lemma1_4}) and \eqref{EqBott2}, for all $P\in \Delta (\set{A}_2)$ such that $P(a_1) \in [0,\alpha]$ and all $\beta \in [0,1]$, 
\begin{IEEEeqnarray}{rl}
& \left(1, \  0\right) \matx{u} \bigl(P(a_1), \  1- P(a_1)\bigr)^{\sf{T}} \IEEEnonumber \\
&= u_{1,1}  \\
& \geq  \beta u_{1,1} + (1-\beta) \bigl(P(a_1)u_{2,1} + (1-P(a_1)) u_{2,2}  \bigr) \\
& = P(a_1) \beta u_{1,1} + \beta (1-P(a_1)) u_{1,2} + (1-\beta)P(a_1) u_{2,1} + (1-\beta)(1-P(a_1)) u_{2,2} \IEEEeqnarraynumspace  \\
& = \left(\beta, \  1 -\beta \right) \matx{u} \bigl(P(a_1), \  1- P(a_1)\bigr)^{\sf{T}},\IEEEeqnarraynumspace
\end{IEEEeqnarray}
where the inequality follows from the fact that if $u_{2,1} \geq u_{2,2}$ and $P(a_1)\leq \alpha$, from \eqref{EqBott2}, it holds that 
\begin{IEEEeqnarray}{c}
 u_{1,1} \geq  \alpha u_{2,1} + \left( 1-\alpha\right)u_{2,2} \geq P(a_1)u_{1,1} + (1-P(a_1)) u_{1,2}. 
\end{IEEEeqnarray}
Hence, it holds that 
\begin{IEEEeqnarray}{c}
\left\{(P_1,P_2) \in \simplex{\set{A}_1} \times \simplex{\set{A}_2}: P_1(a_1) = 1 \ \textnormal{and} \ P_2(a_1) \in [0,\alpha]\right\} \subset \set{P}. 
\end{IEEEeqnarray}

Alternatively, consider the case in which $u_{2,1} < u_{2,2}$. 
From (\ref{Equ:Lemma1_4}), for all $P\in \Delta (\set{A}_2)$ such that $P(a_1) \in [\alpha, 1]$, and for all $\alpha' \in [0,1]$, it holds that
\begin{IEEEeqnarray}{c}
\left(1, \  0\right) \matx{u} \bigl(P(a_1), \  1- P(a_1)\bigr)^{\sf{T}} = u_{1,2} = \left(1, \  0\right) \matx{u} \bigl(\alpha', \  1-\alpha'\bigr)^{\sf{T}} = u_{1,2}. \IEEEeqnarraynumspace
\end{IEEEeqnarray}
Furthermore, from (\ref{Equ:Lemma1_2}) and \eqref{EqBott1}, for all $P\in \Delta (\set{A}_2)$ such that $P(a_1) \in [\alpha, 1]$ and all $\beta \in [0,1]$, 
\begin{IEEEeqnarray}{rl}
& \left(1, \  0\right) \matx{u} \bigl(P(a_1), \  1- P(a_1)\bigr)^{\sf{T}} \IEEEnonumber \\
&= u_{1,1}  \\
& \geq  \beta u_{1,1} + (1-\beta) \bigl(P(a_1)u_{2,1} + (1-P(a_1)) u_{2,2}  \bigr) \\
& = P(a_1) \beta u_{1,1} + \beta (1-P(a_1)) u_{1,2} + (1-\beta)P(a_1) u_{2,1} + (1-\beta)(1-P(a_1)) u_{2,2} \IEEEeqnarraynumspace  \\
& = \left(\beta, \  1 -\beta \right) \matx{u} \bigl(P(a_1), \  1- P(a_1)\bigr)^{\sf{T}},\IEEEeqnarraynumspace
\end{IEEEeqnarray}
where the inequality follows from the fact that if $u_{2,1} < u_{2,2}$ and $P(a_1)\geq \alpha$, from \eqref{EqBott2}, it holds that 
\begin{IEEEeqnarray}{c}
 u_{1,1} \geq  \alpha u_{2,1} + \left( 1-\alpha\right)u_{2,2} \geq  P(a_1)u_{2,1} + (1-P(a_1)) u_{2,2}. 
\end{IEEEeqnarray}
Hence, it holds that 
\begin{IEEEeqnarray}{c}
\left\{(P_1,P_2) \in \simplex{\set{A}_1} \times \simplex{\set{A}_2}: P_1(a_1) = 1 \ \textnormal{and} \ P_2(a_1) \in [\alpha, 1]\right\} \subset \set{P}. 
\end{IEEEeqnarray}

This completes the proof.

The proof of Case III is as follows. 
Assume that the pair of strategies $\left( P_{A_1}^\star, P_{A_2}^\star \right)$ in (\ref{Equ:P_A1_2}) and (\ref{Equ:P_A2_2}) forms an NE in the game $\game{G}$ in \eqref{EqTheGame}.
Then from the indifference principle \cite[Theorem 5.18]{Maschler_2013_game}, it follows that Player 1 is indifferent to using either action $a_1$ or $a_2$. 
That is, 
\begin{IEEEeqnarray}{c}\label{Equ:Lemma1_5}
 \left(1, \  0\right) \matx{u} \left(0, \  1\right)^{\sf{T}} =  \left(0, \  1\right) \matx{u} \left(0, \  1\right)^{\sf{T}}, 
\end{IEEEeqnarray}
which implies that 
\begin{IEEEeqnarray}{c}\label{Equ:Lemma1_6}
u_{1,2} = u_{2,2}. 
\end{IEEEeqnarray}
Under the assumption that $\left( P_{A_1}^\star, P_{A_2}^\star \right)$ forms an NE, from \eqref{EqNECon2}, for all $\beta \in \left[0,1\right]$, it holds that
\begin{IEEEeqnarray}{c}
\left( \alpha, 1-\alpha\right)  \matx{u} (0,1)^{\sf T} \leq \left( \alpha, 1-\alpha\right)  \matx{u} (\beta,1 -\beta)^{\sf T}, 
\end{IEEEeqnarray} 
which implies that 
\begin{IEEEeqnarray}{c}
u_{1,2} \alpha +u_{2,2} (1-\alpha) \leq \alpha \beta u_{1,1} + \alpha (1-\beta) u_{1,2} + (1-\alpha)\beta u_{2,1} + (1-\beta)(1-\alpha) u_{2,2}. \label{EqCof3}
\end{IEEEeqnarray}
Note that taking the equality in \eqref{Equ:Lemma1_6} into \eqref{EqCof3} yields that for all $\beta \in \left[ 0,1\right]$, it holds that 
\begin{IEEEeqnarray}{c}\label{EqBott3}
 u_{2,2} \leq   \alpha u_{1,1} + \left( 1-\alpha\right)u_{2,1}. 
\end{IEEEeqnarray}

The following proof proves that either 
\begin{IEEEeqnarray}{l}
\left\{(P_1,P_2) \in \simplex{\set{A}_1} \times \simplex{\set{A}_2}: P_1(a_1) \in [0, \alpha] \ \textnormal{and} \ P_2(a_1)= 0 \right\} \subset \set{P} \quad \textnormal{or} \\
\left\{(P_1,P_2) \in \simplex{\set{A}_1} \times \simplex{\set{A}_2}: P_1(a_1) \in [\alpha, 1] \ \textnormal{and} \ P_2(a_1) =0 \right\} \subset \set{P} 
\end{IEEEeqnarray}
holds, which implies that $\set{P}$ is a set of cardinality infinite. 

First, consider the case in which $u_{2,1} \geq u_{1,1}$. 
Given the fact that the equality in (\ref{Equ:Lemma1_6}) holds, for all $P\in \Delta (\set{A}_1)$ such that $P(a_1) \in [0,\alpha]$, and for all $\alpha' \in [0,1]$, 
\begin{IEEEeqnarray}{c}
\left(P(a_1), \  1-P(a_1)\right) \matx{u} \bigl(0, \  1\bigr)^{\sf{T}} = u_{1,2} = \left(\alpha', \  1-\alpha' \right) \matx{u} \bigl(0, \  1\bigr)^{\sf{T}} = u_{1,2}. \IEEEeqnarraynumspace
\end{IEEEeqnarray}
Furthermore, from (\ref{Equ:Lemma1_6}) and \eqref{EqBott3}, for all $P\in \Delta (\set{A}_1)$ such that $P(a_1) \in [0,\alpha]$ and all $\beta \in [0,1]$, 
\begin{IEEEeqnarray}{rl}
& \left(P(a_1), \  1- P(a_1)\right) \matx{u} (0, \  1)^{\sf{T}} \IEEEnonumber \\
&= u_{2,2}  \\
& \leq  (1-\beta) u_{2,2} + \beta \bigl(P(a_1)u_{1,1} + (1-P(a_1)) u_{2,1}  \bigr) \\
& = P(a_1) \beta u_{1,1} + (1-\beta)P(a_1) u_{1,2} +  \beta (1-P(a_1)) u_{2,1} + (1-\beta)(1-P(a_1)) u_{2,2} \IEEEeqnarraynumspace  \\
& = \left(P(a_1), \  1 -P(a_1) \right) \matx{u} \bigl(\beta, \  1- \beta\bigr)^{\sf{T}},\IEEEeqnarraynumspace
\end{IEEEeqnarray}
where the inequality follows from the fact that if $u_{2,1} \geq u_{1,1}$ and $P(a_1)\leq \alpha$, from \eqref{EqBott3}, it holds that 
\begin{IEEEeqnarray}{c}
 u_{2,2} \leq  \alpha u_{1,1} + \left( 1-\alpha\right)u_{2,1} \leq P(a_1)u_{1,1} + (1-P(a_1)) u_{1,2}. 
\end{IEEEeqnarray}
Hence, it holds that 
\begin{IEEEeqnarray}{c}
\left\{(P_1,P_2) \in \simplex{\set{A}_1} \times \simplex{\set{A}_2}: P_1(a_1) \in [0, \alpha] \ \textnormal{and} \ P_2(a_1)= 0 \right\} \subset \set{P}. 
\end{IEEEeqnarray}

Alternatively, consider the case in which $u_{2,1} < u_{1,1}$. 
From (\ref{Equ:Lemma1_6}), for all $P\in \Delta (\set{A}_1)$ such that $P(a_1) \in [\alpha, 1]$, and for all $\alpha' \in [0,1]$, it holds that 
\begin{IEEEeqnarray}{c}
\left(P(a_1), \  1-P(a_1)\right) \matx{u} \bigl(0, \  1\bigr)^{\sf{T}} = u_{1,2} = \left(\alpha', \  1-\alpha' \right) \matx{u} \bigl(0, \  1\bigr)^{\sf{T}} = u_{1,2}. \IEEEeqnarraynumspace
\end{IEEEeqnarray}
Furthermore, from (\ref{Equ:Lemma1_6}) and \eqref{EqBott3}, for all $P\in \Delta (\set{A}_1)$ such that $P(a_1) \in [0,\alpha]$ and all $\beta \in [0,1]$, 
\begin{IEEEeqnarray}{rl}
& \left(P(a_1), \  1- P(a_1)\right) \matx{u} (0, \  1)^{\sf{T}} \IEEEnonumber \\
&= u_{2,2}  \\
& \leq (1-\beta) u_{2,2} + \beta (P(a_1)u_{1,1} + (1-P(a_1)) u_{2,1}  ) \\
& = P(a_1) \beta u_{1,1} + (1-\beta)P(a_1) u_{1,2} +  \beta (1-P(a_1)) u_{2,1} + (1-\beta)(1-P(a_1)) u_{2,2} \IEEEeqnarraynumspace  \\
& = \left(P(a_1), \  1 -P(a_1) \right) \matx{u} \bigl(\beta, \  1- \beta\bigr)^{\sf{T}},\IEEEeqnarraynumspace
\end{IEEEeqnarray}
where the inequality follows from the fact that if $u_{2,1} < u_{1,1}$ and $P(a_1)\geq \alpha$, from \eqref{EqBott3}, it holds that 
\begin{IEEEeqnarray}{c}
 u_{2,2} \leq  \alpha u_{1,1} + \left( 1-\alpha\right)u_{2,1} \leq P(a_1)u_{1,1} + (1-P(a_1)) u_{1,2}. 
\end{IEEEeqnarray}
Hence, it holds that 
\begin{IEEEeqnarray}{c}
\left\{(P_1,P_2) \in \simplex{\set{A}_1} \times \simplex{\set{A}_2}: P_1(a_1) \in [ \alpha, 1] \ \textnormal{and} \ P_2(a_1)= 0 \right\} \subset \set{P}. 
\end{IEEEeqnarray}

This completes the proof.

The proof of Case VI is as follows. 
Assume that the pair of strategies $\left( P_{A_1}^\star, P_{A_2}^\star \right)$ in (\ref{Equ:P_A1_3}) and (\ref{Equ:P_A2_3}) forms an NE in the game $\game{G}$ in \eqref{EqTheGame}.
Then from the indifference principle \cite[Theorem 5.18]{Maschler_2013_game}, it follows that Player 1 is indifferent to using either action $a_1$ or $a_2$. 
That is, 
\begin{IEEEeqnarray}{c}\label{Equ:Lemma1_7}
 \left(1, \  0\right) \matx{u} \left(1, \  0\right)^{\sf{T}} =  \left(0, \  1\right) \matx{u} \left(1, \  0\right)^{\sf{T}}, 
\end{IEEEeqnarray}
which implies that 
\begin{IEEEeqnarray}{c}\label{Equ:Lemma1_8}
u_{1,1} = u_{2,1}. 
\end{IEEEeqnarray}
Under the assumption that $\left( P_{A_1}^\star, P_{A_2}^\star \right)$ forms an NE, from \eqref{EqNECon2}, for all $\beta \in \left[0,1\right]$, it holds that
\begin{IEEEeqnarray}{c}
\left( \alpha, 1-\alpha\right)  \matx{u} (1,0)^{\sf T} \leq \left( \alpha, 1-\alpha\right)  \matx{u} (\beta,1 -\beta)^{\sf T}, 
\end{IEEEeqnarray} 
which implies that 
\begin{IEEEeqnarray}{c}
u_{1,1} \alpha +u_{2,1} (1-\alpha) \leq \alpha \beta u_{1,1} + \alpha (1-\beta) u_{1,2} + (1-\alpha)\beta u_{2,1} + (1-\beta)(1-\alpha) u_{2,2}. \label{EqCof4}
\end{IEEEeqnarray}
Note that taking the equality in \eqref{Equ:Lemma1_8} into \eqref{EqCof4} yields that for all $\beta \in \left[ 0,1\right]$, it holds that 
\begin{IEEEeqnarray}{c}\label{EqBott4}
 u_{1,1} \leq  \alpha u_{1,2} + \left( 1-\alpha\right)u_{2,2}. 
\end{IEEEeqnarray}

The following proof proves that either 
\begin{IEEEeqnarray}{l}
\left\{(P_1,P_2) \in \simplex{\set{A}_1} \times \simplex{\set{A}_2}: P_1(a_1) \in [0, \alpha] \ \textnormal{and} \ P_2(a_1)= 1 \right\} \subset \set{P} \quad \textnormal{or} \\
\left\{(P_1,P_2) \in \simplex{\set{A}_1} \times \simplex{\set{A}_2}: P_1(a_1) \in [\alpha, 1] \ \textnormal{and} \ P_2(a_1) =1 \right\} \subset \set{P} 
\end{IEEEeqnarray}
holds, which implies that $\set{P}$ is a set of cardinality infinite. 

First, consider the case in which $u_{1,2} \geq u_{2,2}$. 
Given the fact that the equality in (\ref{Equ:Lemma1_8}) holds, for all $P\in \Delta (\set{A}_1)$ such that $P(a_1) \in [0,\alpha]$, and for all $\alpha' \in [0,1]$, 
\begin{IEEEeqnarray}{c}
\left(P(a_1), \  1-P(a_1)\right) \matx{u} \bigl(1, \  0\bigr)^{\sf{T}} = u_{1,1} = \left(\alpha', \  1-\alpha' \right) \matx{u} \bigl(1, \  0\bigr)^{\sf{T}} = u_{1,1}. \IEEEeqnarraynumspace
\end{IEEEeqnarray}
Furthermore, from (\ref{Equ:Lemma1_8}) and \eqref{EqBott4}, for all $P\in \Delta (\set{A}_1)$ such that $P(a_1) \in [0,\alpha]$ and all $\beta \in [0,1]$, 
\begin{IEEEeqnarray}{rl}
& \left(P(a_1), \  1- P(a_1)\right) \matx{u} (1, \  0)^{\sf{T}} \IEEEnonumber \\
&= u_{1,1}  \\
& \leq  \beta u_{1,1} + (1-\beta) \bigl(P(a_1)u_{1,2} + (1-P(a_1)) u_{2,2}  \bigr) \\
& = P(a_1) \beta u_{1,1} + (1-\beta)P(a_1) u_{1,2} +  \beta (1-P(a_1)) u_{2,1} + (1-\beta)(1-P(a_1)) u_{2,2} \IEEEeqnarraynumspace  \\
& = \left(P(a_1), \  1 -P(a_1) \right) \matx{u} \bigl(\beta, \  1- \beta\bigr)^{\sf{T}},\IEEEeqnarraynumspace
\end{IEEEeqnarray}
where the inequality follows from the fact that if $u_{2,2} \geq u_{1,2}$ and $P(a_1)\leq \alpha$, from \eqref{EqBott4}, it holds that 
\begin{IEEEeqnarray}{c}
 u_{1,1} \leq  \alpha u_{1,2} + \left( 1-\alpha\right)u_{2,2} \leq P(a_1)u_{1,2} + (1-P(a_1)) u_{2,2}. 
\end{IEEEeqnarray}
Hence, it holds that 
\begin{IEEEeqnarray}{c}
\left\{(P_1,P_2) \in \simplex{\set{A}_1} \times \simplex{\set{A}_2}: P_1(a_1) \in [0, \alpha] \ \textnormal{and} \ P_2(a_1)= 1 \right\} \subset \set{P}. 
\end{IEEEeqnarray}

Alternatively, consider the case in which $u_{1,2} < u_{2,2}$. 
From (\ref{Equ:Lemma1_6}), for all $P\in \Delta (\set{A}_1)$ such that $P(a_1) \in [\alpha, 1]$, and for all $\alpha' \in [0,1]$, it holds that 
\begin{IEEEeqnarray}{c}
\left(P(a_1), \  1-P(a_1)\right) \matx{u} \bigl(1, \  0\bigr)^{\sf{T}} = u_{1,1} = \left(\alpha', \  1-\alpha' \right) \matx{u} \bigl(0, \  1\bigr)^{\sf{T}} = u_{1,1}. \IEEEeqnarraynumspace
\end{IEEEeqnarray}
Furthermore, from (\ref{Equ:Lemma1_8}) and \eqref{EqBott4}, for all $P\in \Delta (\set{A}_1)$ such that $P(a_1) \in [\alpha, 1]$ and all $\beta \in [0,1]$, 
\begin{IEEEeqnarray}{rl}
& \left(P(a_1), \  1- P(a_1)\right) \matx{u} (1, \  0)^{\sf{T}} \IEEEnonumber \\
&= u_{1,1}  \\
& \leq  \beta u_{1,1} + (1-\beta) \bigl(P(a_1)u_{1,2} + (1-P(a_1)) u_{2,2}  \bigr) \\
& = P(a_1) \beta u_{1,1} + (1-\beta)P(a_1) u_{1,2} +  \beta (1-P(a_1)) u_{2,1} + (1-\beta)(1-P(a_1)) u_{2,2} \IEEEeqnarraynumspace  \\
& = \left(P(a_1), \  1 -P(a_1) \right) \matx{u} \bigl(\beta, \  1- \beta\bigr)^{\sf{T}},\IEEEeqnarraynumspace
\end{IEEEeqnarray}
where the inequality follows from the fact that if $u_{2,2} < u_{1,2}$ and $P(a_1)\geq \alpha$, from \eqref{EqBott4}, it holds that 
\begin{IEEEeqnarray}{c}
 u_{1,1} \leq  \alpha u_{1,2} + \left( 1-\alpha\right)u_{2,2} \leq P(a_1)u_{1,2} + (1-P(a_1)) u_{2,2}. 
\end{IEEEeqnarray}
Hence, it holds that 
\begin{IEEEeqnarray}{c}
\left\{(P_1,P_2) \in \simplex{\set{A}_1} \times \simplex{\set{A}_2}: P_1(a_1) \in [ \alpha, 1] \ \textnormal{and} \ P_2(a_1)= 1 \right\} \subset \set{P}. 
\end{IEEEeqnarray}

This completes the proof.

%

This completes the whole proof.

\section{Proof of Lemma \ref{Lemma:twoNE_unique}}\label{Sec:App2}
The proof of the first statement is as follows. 
 Assume that the pair of strategies $\left( P_{A_1}^\star, P_{A_2}^\star \right) \in \set{P}$ satisfies
\begin{IEEEeqnarray}{c}
 P_{A_1}^\star(a_1) = 1- P_{A_1}^\star (a_2)\in \{0,1\} \quad \textnormal{and} \quad P_{A_2}^\star(a_1) = 1 - P_{A_2}^\star(a_2) \in \{0,1\}.
\end{IEEEeqnarray}
If there exists another pair of strategies $\left( Q_{A_1}^\star, Q_{A_2}^\star \right) \in \set{P}$ satisfying
 \begin{IEEEeqnarray}{c}
 Q_{A_1}^\star (a_1) = 1- Q_{A_1}^\star (a_2)\in (0,1) \quad \textnormal{and} \quad Q_{A_2}^\star(a_1) = 1 - Q_{A_2}^\star(a_2) \in (0,1), 
\end{IEEEeqnarray}
then from  \cite[Corollary 4.46]{Maschler_2013_game}, it holds that the pair of strategies $(P_{A_1}^\star, Q_{A_2}^\star)$ also forms an NE. 
Then, if $(P_{A_1}^\star, Q_{A_2}^\star)$ forms an NE, from Lemma \ref{Lemma:infinite}, it holds that there exist infinitely many NEs, which contradicts the assumption of the lemma. 

The proof of the second statement is as follows. 
 Assume that the pair of strategies $\left( P_{A_1}^\star, P_{A_2}^\star \right) \in \set{P}$ satisfies
\begin{IEEEeqnarray}{c}
 P_{A_1}^\star (a_1) = 1- P_{A_1} ^\star(a_2)\in (0,1) \quad \textnormal{and} \quad P_{A_2}^\star(a_1) = 1 - P_{A_2}^\star(a_2) \in (0,1).
\end{IEEEeqnarray}
If there exists another pair of strategies $\left( Q_{A_1}^\star, Q_{A_2}^\star \right) \in \set{P}$ satisfying
 \begin{IEEEeqnarray}{c}
 Q_{A_1}^\star (a_1) = 1- Q_{A_1}^\star (a_2)\in \{0,1\} \quad \textnormal{and} \quad Q_{A_2}^\star(a_1) = 1 - Q_{A_2}(a_2) \in \{0,1\}, 
\end{IEEEeqnarray}
then from  \cite[Corollary 4.46]{Maschler_2013_game}, it holds that the pair of strategies $(P_{A_1}^\star, Q_{A_2}^\star)$ also forms an NE.
Then, if $(P_{A_1}^\star, Q_{A_2}^\star)$ forms an NE, from Lemma \ref{Lemma:infinite}, it holds that there exist infinitely many NEs, which contradicts the assumption of the lemma. 

This completes the proof.

\section{Proof of Lemma \ref{Lemma:strictlymixed_unique}}\label{Sec:App3}

Assume that the pair of strategies $\left( P_{A_1}^\star, P_{A_2}^\star \right) \in \set{P}$ satisfies 
\begin{IEEEeqnarray}{rl}
&P_{A_1}^\star(a_1) = 1 - P_{A_1}^\star(a_2) \in (0,1) \quad \textnormal{and} \\
&  P_{A_2}^\star(a_1) = 1 - P_{A_2}^\star(a_2) \in (0,1).
\end{IEEEeqnarray}
Assume also that there exists another pair of strategies $\left( Q_{A_1}^\star, Q_{A_2}^\star \right) \in \set{P}$ satisfying
\begin{IEEEeqnarray}{rl}
&Q_{A_1}^\star(a_1) = 1 - Q_{A_1}^\star(a_2) \in [0,1] \quad \textnormal{and} \\
&  Q_{A_2}^\star(a_1) = 1 - Q_{A_2}^\star(a_2) \in [0,1].
\end{IEEEeqnarray}
Note that if the pair of strategies $\left( Q_{A_1}^\star, Q_{A_2}^\star \right)$ satisfies 
\begin{IEEEeqnarray}{rl}
&Q_{A_1}^\star(a_1) = 1 - Q_{A_1}^\star(a_2) \in \{0,1\} \quad \textnormal{and} \\
&  Q_{A_2}^\star(a_1) = 1 - Q_{A_2}^\star(a_2) \in \{0,1\}, 
\end{IEEEeqnarray}
then from Lemma \ref{Lemma:twoNE_unique}, there are infinitely many NEs in the game $\game{G}$, 
which contradicts the assumption of the lemma. 
Hence, the proof continues by considering the following cases.

\noindent Case I: the pair of strategies $\left( Q_{A_1}^\star, Q_{A_2}^\star \right)$ satisfies 
\begin{IEEEeqnarray}{rl}
&Q_{A_1}^\star(a_1) = 1 - Q_{A_1}^\star(a_2) \in \{0,1\}, \quad \textnormal{and} \\
&  Q_{A_2}^\star(a_1) = 1 - Q_{A_2}^\star(a_2) \in (0,1).
\end{IEEEeqnarray}

\noindent Case II: the pair of strategies $\left( Q_{A_1}^\star, Q_{A_2}^\star \right)$ satisfies 
\begin{IEEEeqnarray}{rl}
&Q_{A_1}^\star(a_1) = 1 - Q_{A_1}^\star(a_2) \in (0,1), \quad \textnormal{and} \\
&  Q_{A_2}^\star(a_1) = 1 - Q_{A_2}^\star(a_2) \in \{0,1\}.
\end{IEEEeqnarray}

\noindent Case III: the pair of strategies $\left( Q_{A_1}^\star, Q_{A_2}^\star \right)$ satisfies 
\begin{IEEEeqnarray}{rl}
&Q_{A_1}^\star(a_1) = 1 - Q_{A_1}^\star(a_2) \in (0,1), \quad \textnormal{and} \\
&  Q_{A_2}^\star(a_1) = 1 - Q_{A_2}^\star(a_2) \in (0,1).
\end{IEEEeqnarray}

Proof of Case I: From Lemma \ref{Lemma:infinite}, it holds that there are infinitely many NEs in the game~$\game{G}$, which contradicts the assumption that the game $\game{G}$ possesses a finite number of NEs. 

Proof of Case II: From Lemma \ref{Lemma:infinite}, it holds that there are infinitely many NEs in the game~$\game{G}$, which contradicts the assumption that the game $\game{G}$ possesses a finite number of NEs. 

Proof of Case III: 
From the  indifference principle \cite[Theorem 5.18]{Maschler_2013_game}, it follows that Player 1 is indifferent to use action $a_1$ or action $a_2$ when Player 2 uses strategy $P_{A_2}^\star$. 
That is, 
\begin{IEEEeqnarray}{c}
 \left(1, \  0\right) \matx{u} \left(P_{A_2}^\star(a_1), \  1 - P_{A_2}^\star(a_1)\right)^{\sf{T}} =  \left(0, \  1\right) \matx{u} \left(P_{A_2}^\star(a_1), \  1 - P_{A_2}^\star(a_1)\right)^{\sf{T}}. 
\end{IEEEeqnarray}
The above equality yields
\begin{IEEEeqnarray}{c}
u_{1,1} P_{A_2}^\star(a_1) + u_{1,2} \bigl(1- P_{A_2}^\star(a_1) \bigr) = u_{2,1} P_{A_2}^\star(a_1) + u_{2,2} \bigl(1- P_{A_2}^\star(a_1) \bigr), 
\end{IEEEeqnarray}
which is equivalent to 
\begin{IEEEeqnarray}{c} \label{Equ:P_{A_2}_optimal}
(u_{1,1} - u_{1,2} - u_{2,1} + u_{2,2}) P_{A_2}^\star(a_1) = u_{2,2} - u_{1,2}.
\end{IEEEeqnarray}
Similarly, Player 1 is also indifferent to use action $a_1$ or action $a_2$ when Player 2 uses strategy $Q_{A_2}^\star$. 
As a result, it holds that 
\begin{IEEEeqnarray}{c} \label{Equ:P_{A_2}_optimal_1}
(u_{1,1} - u_{1,2} - u_{2,1} + u_{2,2}) Q_{A_2}^\star(a_1) = u_{2,2} - u_{1,2}.
\end{IEEEeqnarray}

Note that the equalities in \eqref{Equ:P_{A_2}_optimal} and \eqref{Equ:P_{A_2}_optimal_1} are both first-order equations. Hence, there is a unique solution in $(0,1)$ if  $u_{1,1} - u_{1,2} - u_{2,1} + u_{2,2} \neq 0$ and $u_{2,2} - u_{1,2} \neq0$, which implies that
$P_{A_2}^\star(a_1) = Q_{A_2}^\star(a_1)$. Alternatively, there are infinitely many solutions if $u_{1,1} - u_{1,2} - u_{2,1} + u_{2,2} = 0$ and $u_{2,2} - u_{1,2} = 0$, which implies that all strategies $Q_{A_2} \in \Delta (\set{A}_2)$, together with $P_{A_1}^\star$, form an NE.
This contradicts the assumption that there is a finite number of NEs. 
Finally, the case in which there are no solutions to \eqref{Equ:P_{A_2}_optimal} and \eqref{Equ:P_{A_2}_optimal_1} contradicts the initial assumption, as $(P_{A_1}^\star, P_{A_2}^\star)$ forms an NE. 
As a result, if the game $\game{G}$ exhibits a finite number of NEs, it holds that $P_{A_2}^\star = Q_{A_2}^\star$.

For Player 2, the indifferent principle  \cite[Theorem 5.18]{Maschler_2013_game} yields that
\begin{IEEEeqnarray}{c}
\left(P_{A_1}^\star(a_1), \  1 - P_{A_1}^\star(a_1)\right) \matx{u} \left(1, \  0\right)^{\sf{T}} = \left(P_{A_1}^\star(a_1), \  1 - P_{A_1}^\star(a_1)\right) \matx{u} \left(0, \  1\right)^{\sf{T}}, 
\end{IEEEeqnarray}
which is equivalent to 
\begin{IEEEeqnarray}{c} \label{Equ:P_{A_1}_optimal}
(u_{1,1} - u_{1,2} - u_{2,1} + u_{2,2}) P_{A_1}^\star(a_1) = u_{2,2} - u_{2,1}. 
\end{IEEEeqnarray}
Similarly, Player 2 is also indifferent to use action $a_1$ or action $a_2$ when Player 1 uses strategy $Q_{A_1}$. 
As a result, it holds that 
\begin{IEEEeqnarray}{c} \label{Equ:P_{A_1}_optimal_1}
(u_{1,1} - u_{1,2} - u_{2,1} + u_{2,2}) Q_{A_1}^\star(a_1) = u_{2,2} - u_{2,1}.
\end{IEEEeqnarray}

Note that the equalities in \eqref{Equ:P_{A_1}_optimal} and \eqref{Equ:P_{A_1}_optimal_1} are both a first-order equations. Hence, there is a unique solution in $(0,1)$ if  $u_{1,1} - u_{1,2} - u_{2,1} + u_{2,2} \neq 0$ and $u_{2,2} - u_{2,1} \neq 0$, which implies that
$P_{A_1}^{\star}(a_1) = Q_{A_1}^\star(a_1)$. Alternatively, there are infinitely many solutions if $u_{1,1} - u_{1,2} - u_{2,1} + u_{2,2} = 0$ and $u_{2,2} - u_{2,1} = 0$, which implies that all strategies $Q_{A_1} \in \Delta (\set{A}_1)$, together with $P_{A_2}^\star$, form an NE.
This contradicts the assumption that there is  a finite number of NEs. 
Finally, the case in which there are no solutions to \eqref{Equ:P_{A_1}_optimal} and \eqref{Equ:P_{A_1}_optimal_1} also contradicts the initial assumption, as $(P_{A_1}^\star, P_{A_2}^\star)$ forms an NE. 
As a result, if the game $\game{G}$ exhibits a finite number of NEs, it holds that $P_{A_1}^\star = Q_{A_1}^\star$. 


This completes the proof.

\section{Proof of Lemma \ref{Lemma:pure_unique}}\label{Sec:App4}

Note that from Lemma \ref{Lemma:infinite},  if there exist a finite number of NEs in the game $\game{G}$, then an NE in pure strategies and an NE in strictly mixed strategies cannot coexist. 
Furthermore, from  Lemma \ref{Lemma:twoNE_unique}, if there exist a finite number of NEs in the game $\game{G}$, then an NE in which one player uses a pure strategy and the other players uses a strictly mixed strategy does not exists.
 Hence, the proof is conducted by contradiction to prove that {\it if there exist a finite number of NEs in the game $\game{G}$, two NEs in pure strategies cannot coexist}.

Assume that there exists another pair of strategies $(Q_{A_1}^\star, Q_{A_2}^\star)\in \set{P} $ such that 
\begin{IEEEeqnarray}{rl}
& Q_{A_1}^\star (a_1) = 1- Q_{A_1}^\star (a_2)\in \{0,1\}  \quad \textnormal{and}\\
& Q_{A_2}^\star (a_1) = 1- Q_{A_2}^\star (a_2)\in \{0,1\}. 
\end{IEEEeqnarray}
Given the fact that there are at most four NEs in pure strategies in $2 \times 2$ games, 
there are $\left( 
\begin{array}{c}
4 \\
2
\end{array}
\right) =6$ scenarios in which two NEs in pure strategies exist. 
Hence, the proof considers the following cases. 

\noindent Case I: Strategies $P_{A_1}^\star$, $P_{A_2}^\star$, $Q_{A_1}^\star$, and $Q_{A_2}^\star$ satisfy 
\begin{IEEEeqnarray}{rl}
& P_{A_1}^\star(a_2) = 1- P_{A_1}^\star(a_1) = 0 \quad \textnormal{and} \quad P_{A_2}^\star(a_2) = 1- P_{A_2}^\star(a_1) = 0,\\
& Q_{A_1}^\star(a_2) = 1- Q_{A_1}^\star(a_1) = 0 \quad \textnormal{and} \quad Q_{A_2}^\star(a_2) = 1- Q_{A_2}^\star(a_1) = 1.
\end{IEEEeqnarray}

\noindent Case II: Strategies $P_{A_1}^\star$, $P_{A_2}^\star$, $Q_{A_1}^\star$, and $Q_{A_2}^\star$ satisfy 
\begin{IEEEeqnarray}{rl}
& P_{A_1}^\star(a_2) = 1- P_{A_1}^\star(a_1) = 0 \quad \textnormal{and} \quad P_{A_2}^\star(a_2) = 1- P_{A_2}^\star(a_1) = 0,\\
& Q_{A_1}^\star(a_2) = 1- Q_{A_1}^\star(a_1) = 1 \quad \textnormal{and} \quad Q_{A_2}^\star(a_2) = 1- Q_{A_2}^\star(a_1) = 0.
\end{IEEEeqnarray}

\noindent Case III: Strategies $P_{A_1}^\star$, $P_{A_2}^\star$, $Q_{A_1}^\star$, and $Q_{A_2}^\star$ satisfy 
\begin{IEEEeqnarray}{rl}
& P_{A_1}^\star(a_2) = 1- P_{A_1}^\star(a_1) = 0 \quad \textnormal{and} \quad P_{A_2}^\star(a_2) = 1- P_{A_2}^\star(a_1) = 0,\\
& Q_{A_1}^\star(a_2) = 1- Q_{A_1}^\star(a_1) = 1 \quad \textnormal{and} \quad Q_{A_2}^\star(a_2) = 1- Q_{A_2}^\star(a_1) = 1.
\end{IEEEeqnarray}

\noindent Case IV: Strategies $P_{A_1}^\star$, $P_{A_2}^\star$, $Q_{A_1}^\star$, and $Q_{A_2}^\star$ satisfy 
\begin{IEEEeqnarray}{rl}
& P_{A_1}^\star(a_2) = 1- P_{A_1}^\star(a_1) = 1 \quad \textnormal{and} \quad P_{A_2}^\star(a_2) = 1- P_{A_2}^\star(a_1) = 1,\\
& Q_{A_1}^\star(a_2) = 1- Q_{A_1}^\star(a_1) = 0 \quad \textnormal{and} \quad Q_{A_2}^\star(a_2) = 1- Q_{A_2}^\star(a_1) = 1.
\end{IEEEeqnarray}

\noindent Case V: Strategies $P_{A_1}^\star$, $P_{A_2}^\star$, $Q_{A_1}^\star$, and $Q_{A_2}^\star$ satisfy 
\begin{IEEEeqnarray}{rl}
& P_{A_1}^\star(a_2) = 1- P_{A_1}^\star(a_1) = 1 \quad \textnormal{and} \quad P_{A_2}^\star(a_2) = 1- P_{A_2}^\star(a_1) = 1,\\
& Q_{A_1}^\star(a_2) = 1- Q_{A_1}^\star(a_1) = 1 \quad \textnormal{and} \quad Q_{A_2}^\star(a_2) = 1- Q_{A_2}^\star(a_1) = 0.
\end{IEEEeqnarray}

\noindent Case VI: Strategies $P_{A_1}^\star$, $P_{A_2}^\star$, $Q_{A_1}^\star$, and $Q_{A_2}^\star$ satisfy 
\begin{IEEEeqnarray}{rl}
& P_{A_1}^\star(a_2) = 1- P_{A_1}^\star(a_1) = 1 \quad \textnormal{and} \quad P_{A_2}^\star(a_2) = 1- P_{A_2}^\star(a_1) = 0,\\
& Q_{A_1}^\star(a_2) = 1- Q_{A_1}^\star(a_1) = 0 \quad \textnormal{and} \quad Q_{A_2}^\star(a_2) = 1- Q_{A_2}^\star(a_1) = 1.
\end{IEEEeqnarray}

The proof of Case I is as follows. 
If $(P_{A_1}^\star, P_{A_2}^\star)$ forms the NE, then from Definition \ref{Def:Equ_1}, it holds that for all $\alpha \in [0,1]$, 
\begin{IEEEeqnarray}{c}\label{EqBor1}
u(P_{A_1}^\star, P_{A_2}^\star) = u_{1,1} \geq (\alpha, 1- \alpha) \matx{u} (1,0)^{\sf T} ;
\end{IEEEeqnarray}
and for all $Q \in \simplex{\set{A}_2}$ such that $Q(a_1)=\beta$, 
\begin{IEEEeqnarray}{c}\label{EqBor2}
u(P_{A_1}^\star, P_{A_2}^\star) =u_{1,1} \leq (1, 0) \matx{u} (\beta, 1-\beta)^{\sf T}. 
\end{IEEEeqnarray}
Setting $\alpha$ in \eqref{EqBor1} and $\beta$ in \eqref{EqBor2} to zero yields
\begin{IEEEeqnarray}{c}\label{EqCof5}
u_{1,2} \geq u_{1,1} \geq u_{2,1}. 
\end{IEEEeqnarray}
Similarly, if $(Q_{A_1}^\star, Q_{A_2}^\star)$ forms the NE, then from Definition \ref{Def:Equ_1}, it holds that for all $\alpha \in [0,1]$, 
\begin{IEEEeqnarray}{c}\label{EqBor3}
u(Q_{A_1}^\star, Q_{A_2}^\star) = u_{1,2} \geq (\alpha, 1- \alpha) \matx{u} (0,1)^{\sf T};
\end{IEEEeqnarray}
and for all $\beta \in [0,1]$, 
\begin{IEEEeqnarray}{c}\label{EqBor4}
u(Q_{A_1}^\star, Q_{A_2}^\star) = u_{1,2} \leq (1, 0) \matx{u} (\beta, 1-\beta)^{\sf T}.
\end{IEEEeqnarray}
Setting $\alpha$ in \eqref{EqBor3} to zero and $\beta$ in \eqref{EqBor4} to one yields
\begin{IEEEeqnarray}{c}\label{EqCof6}
u_{1,1} \geq u_{1,2} \geq u_{2,2}. 
\end{IEEEeqnarray}
Hence, the inequalities in \eqref{EqCof5} and \eqref{EqCof6} yields 
\begin{IEEEeqnarray}{l}
u_{1,1} = u_{1,2} \geq u_{2,1} \quad \textnormal{and} \label{EqWait1}\\
u_{1,1} = u_{1,2} \geq u_{2,2}.\label{EqWait2}
\end{IEEEeqnarray}
If at most one of the inequalities in \eqref{EqWait1} and  \eqref{EqWait2} holds with equality, then, for \Pone, action $a_1$ dominates action $a_2$.
For this case, all strategies $P_{A_2} \in \simplex{\set{A}_2}$, together with strategy $P_{A_1} \in \simplex{\set{A}_1}$ such that $P_{A_1}(a_1) = 1$, form an NE. 
Then there are infinitely many NEs, which contradicts the assumption of the lemma.
Alternatively, if both the inequalities \eqref{EqWait1} and  \eqref{EqWait2} hold with equality, all pair of strategies $(P_{A_1}, P_{A_2}) \in \simplex{\set{A}_1} \times \simplex{\set{A}_2}$ form an NE. 
Then there are infinitely many NEs, which contradicts the assumption of the lemma and completes the proof. 

The proof of Case II is as follows. 
If $(P_{A_1}^\star, P_{A_2}^\star)$ forms the NE, then from Definition \ref{Def:Equ_1}, it holds that for all $Q \in \simplex{\set{A}_1}$ such that $Q(a_1)=\alpha$, 
\begin{IEEEeqnarray}{c}\label{EqBor5}
u(P_{A_1}^\star, P_{A_2}^\star) = u_{1,1} \geq (\alpha, 1- \alpha) \matx{u} (1,0)^{\sf T};
\end{IEEEeqnarray}
and for all $Q \in \simplex{\set{A}_2}$ such that $Q(a_1)=\beta$, 
\begin{IEEEeqnarray}{c}\label{EqBor6}
u(P_{A_1}^\star, P_{A_2}^\star) = u_{1,1} \leq (1, 0) \matx{u} (\beta, 1-\beta)^{\sf T}.
\end{IEEEeqnarray}
Setting $\alpha$ in \eqref{EqBor5} and $\beta$ in \eqref{EqBor6} to zero yields
\begin{IEEEeqnarray}{c}\label{EqCof7}
u_{1,2} \geq u_{1,1} \geq u_{2,1}. 
\end{IEEEeqnarray}
Similarly, if $(Q_{A_1}^\star, Q_{A_2}^\star)$ forms the NE, then from Definition \ref{Def:Equ_1}, it holds that for all $Q \in \simplex{\set{A}_1}$ such that $Q(a_1)=\alpha$, 
\begin{IEEEeqnarray}{c}\label{EqBor7}
u(Q_{A_1}^\star, Q_{A_2}^\star)  = u_{2,1} \geq (\alpha, 1- \alpha) \matx{u} (1,0)^{\sf T};
\end{IEEEeqnarray}
and for all $Q \in \simplex{\set{A}_2}$ such that $Q(a_1)=\beta$, 
\begin{IEEEeqnarray}{c}\label{EqBor8}
u(Q_{A_1}^\star, Q_{A_2}^\star)  = u_{2,1} \leq (0, 1) \matx{u} (\beta, 1-\beta)^{\sf T}.
\end{IEEEeqnarray}
Setting $\alpha$ in \eqref{EqBor7} to one and $\beta$ in \eqref{EqBor8} to zero yields
\begin{IEEEeqnarray}{c}\label{EqCof8}
u_{2,2} \geq u_{2,1} \geq u_{1,1}. 
\end{IEEEeqnarray}
Hence, the inequalities in \eqref{EqCof7} and \eqref{EqCof8} yields 
\begin{IEEEeqnarray}{l}
u_{1,1} = u_{2,1} \leq u_{1,2} \quad \textnormal{and} \label{EqWait3}\\
u_{1,1} = u_{2,1} \leq u_{2,2}. \label{EqWait4}
\end{IEEEeqnarray}
If at most one of the inequalities in \eqref{EqWait3} and  \eqref{EqWait4} holds with equality, then, for \Ptwo, action $a_1$ dominates action $a_2$.
For this case, all strategies $P_{A_1} \in \simplex{\set{A}_1}$, together with strategy $P_{A_2} \in \simplex{\set{A}_2}$ such that $P_{A_2}(a_1) = 1$, form an NE. 
Then there are infinitely many NEs, which contradicts the assumption of the lemma.
Alternatively, if both the inequalities \eqref{EqWait3} and  \eqref{EqWait4} hold with equality, all pair of strategies $(P_{A_1}, P_{A_2}) \in \simplex{\set{A}_1} \times \simplex{\set{A}_2}$ form an NE. 
Then there are infinitely many NEs, which contradicts the assumption of the lemma and completes the proof.


If Case III holds, from Definition \ref{Def:Equ_1}, the pair of strategies $(P_{A_1}^\star, P_{A_2}^\star)$ forming the NE implies that for all $\beta \in [0,1]$ and $\alpha \in [0,1]$, it holds that 
\begin{IEEEeqnarray}{c}\label{Equ:r1}
(1, 0) \matx{u} (1,0)^{\sf T} \leq (1, 0) \matx{u} (\beta,1-\beta)^{\sf T}; 
\end{IEEEeqnarray}
and 
\begin{IEEEeqnarray}{c}\label{Equ:r2}
(1, 0) \matx{u} (1,0)^{\sf T} \geq (\alpha, 1- \alpha) \matx{u} (1,0)^{\sf T}. 
\end{IEEEeqnarray}
Setting $\beta$ in (\ref{Equ:r1}) and $\alpha$ in (\ref{Equ:r2}) to zero yields 
\begin{IEEEeqnarray}{c}\label{Equ:r21}
u_{2,1} \leq u_{1,1} \leq u_{1,2}. 
\end{IEEEeqnarray}
Similarly, from Definition \ref{Def:Equ_1}, the pair of strategies $(Q_{A_1}^\star, Q_{A_2}^\star)$ forming the NE implies that for all $\beta \in [0,1]$ and $\alpha \in [0,1]$, it holds that 
\begin{IEEEeqnarray}{c}\label{Equ:r3}
( 0, 1) \matx{u} (0, 1)^{\sf T} \leq (0, 1) \matx{u} (\beta,1-\beta)^{\sf T}; 
\end{IEEEeqnarray}
and 
\begin{IEEEeqnarray}{c}\label{Equ:r4}
(0, 1) \matx{u} (0, 1)^{\sf T} \geq (\alpha, 1- \alpha) \matx{u} (0, 1)^{\sf T}. 
\end{IEEEeqnarray}
Setting $\beta$ in (\ref{Equ:r3}) and $\alpha$ in (\ref{Equ:r4}) to one yields 
\begin{IEEEeqnarray}{c}\label{Equ:r5}
u_{1,2} \leq u_{2,2} \leq u_{2,1}. 
\end{IEEEeqnarray}
Combing (\ref{Equ:r21}) with (\ref{Equ:r5}) yields that 
\begin{IEEEeqnarray}{c}
u_{1,1} = u_{1,2} = u_{2,1} = u_{2,2}. 
\end{IEEEeqnarray}
Then, for all strategies $Q \in \Delta (\set{A}_1)$ and  $P \in \Delta (\set{A}_2)$, it holds that
\begin{IEEEeqnarray}{c}
( Q(a_1), 1- Q(a_1))\matx{u} ( P(a_1), 1- P(a_1))^{\sf T} = u_{1,1} = u_{1,2} = u_{2,1} = u_{2,2}, 
\end{IEEEeqnarray}
which implies that there exist infinitely many NEs. 
This contradicts the setting that there exists a finite number of NEs in the game $\game{G}$.

The proof of Case IV is as follows. 
If $(P_{A_1}^\star, P_{A_2}^\star)$ forms the NE, then from Definition \ref{Def:Equ_1}, it holds that for all $Q \in \simplex{\set{A}_1}$ such that $Q(a_1)=\alpha$, 
\begin{IEEEeqnarray}{c}\label{EqBor11}
u_{2,2} \geq (\alpha, 1- \alpha) \matx{u} (0,1)^{\sf T};
\end{IEEEeqnarray}
and for all $Q \in \simplex{\set{A}_2}$ such that $Q(a_1)=\beta$, 
\begin{IEEEeqnarray}{c}\label{EqBor12}
u_{2,2} \leq (0, 1) \matx{u} (\beta, 1-\beta)^{\sf T}. 
\end{IEEEeqnarray}
Setting $\alpha$ in \eqref{EqBor11} and $\beta$ in \eqref{EqBor12} to one yields
\begin{IEEEeqnarray}{c}\label{EqCof15}
u_{2,1} \geq u_{2,2} \geq u_{1,2}. 
\end{IEEEeqnarray}
Similarly, if $(Q_{A_1}^\star, Q_{A_2}^\star)$ forms the NE, then from Definition \ref{Def:Equ_1}, it holds that for all $Q \in \simplex{\set{A}_1}$ such that $Q(a_1)=\alpha$, 
\begin{IEEEeqnarray}{c}\label{EqBor13}
u_{1,2} \geq (\alpha, 1- \alpha) \matx{u} (0,1)^{\sf T};
\end{IEEEeqnarray}
and for all $Q \in \simplex{\set{A}_2}$ such that $Q(a_1)=\beta$, 
\begin{IEEEeqnarray}{c}\label{EqBor14}
u_{1,2} \leq (1, 0) \matx{u} (\beta, 1-\beta)^{\sf T}.
\end{IEEEeqnarray}
Setting $\alpha$ in \eqref{EqBor13} to zero and $\beta$ in \eqref{EqBor14} to one yields
\begin{IEEEeqnarray}{c}\label{EqCof16}
u_{1,1} \geq u_{1,2} \geq u_{2,2}. 
\end{IEEEeqnarray}
Hence, the inequalities in \eqref{EqCof15} and \eqref{EqCof16} yields 
\begin{IEEEeqnarray}{l}
u_{1,2} = u_{2,2} \leq u_{2,1} \quad \textnormal{and} \label{EqWait5}\\
u_{1,2} = u_{2,2} \leq u_{1,1}. \label{EqWait6}
\end{IEEEeqnarray}
If at most one of the inequalities in \eqref{EqWait5} and  \eqref{EqWait6} holds with equality, then, for \Ptwo, action $a_2$ dominates action $a_1$.
For this case, all strategies $P_{A_1} \in \simplex{\set{A}_1}$, together with strategy $P_{A_2} \in \simplex{\set{A}_2}$ such that $P_{A_2}(a_1) = 0$, form an NE. 
Then there are infinitely many NEs, which contradicts the assumption of the lemma.
Alternatively, if both the inequalities \eqref{EqWait5} and  \eqref{EqWait6} hold with equality, all pair of strategies $(P_{A_1}, P_{A_2}) \in \simplex{\set{A}_1} \times \simplex{\set{A}_2}$ form an NE. 
Then there are infinitely many NEs, which contradicts the assumption of the lemma and completes the proof.

The proof of Case V is as follows. 
If $(P_{A_1}^\star, P_{A_2}^\star)$ forms the NE, then from Definition \ref{Def:Equ_1}, it holds that for all $Q \in \simplex{\set{A}_1}$ such that $Q(a_1)=\alpha$, 
\begin{IEEEeqnarray}{c}\label{EqBor21}
u_{2,2} \geq (\alpha, 1- \alpha) \matx{u} (0,1)^{\sf T};
\end{IEEEeqnarray}
and for all $Q \in \simplex{\set{A}_2}$ such that $Q(a_1)=\beta$, 
\begin{IEEEeqnarray}{c}\label{EqBor22}
u_{2,2} \leq (0, 1) \matx{u} (\beta, 1-\beta)^{\sf T}. 
\end{IEEEeqnarray}
Setting $\alpha$ in \eqref{EqBor21} and $\beta$ in \eqref{EqBor22} to one yields
\begin{IEEEeqnarray}{c}\label{EqCof25}
u_{2,1} \geq u_{2,2} \geq u_{1,2}. 
\end{IEEEeqnarray}
Similarly, if $(Q_{A_1}^\star, Q_{A_2}^\star)$ forms the NE, then from Definition \ref{Def:Equ_1}, it holds that for all $Q \in \simplex{\set{A}_1}$ such that $Q(a_1)=\alpha$, 
\begin{IEEEeqnarray}{c}\label{EqBor23}
u_{2,1} \geq (\alpha, 1- \alpha) \matx{u} (1,0)^{\sf T};
\end{IEEEeqnarray}
and for all $Q \in \simplex{\set{A}_2}$ such that $Q(a_1)=\beta$, 
\begin{IEEEeqnarray}{c}\label{EqBor24}
u_{2,1} \leq (0, 1) \matx{u} (\beta, 1-\beta)^{\sf T}.
\end{IEEEeqnarray}
Setting $\alpha$ in \eqref{EqBor23} to one and $\beta$ in \eqref{EqBor24} to zero yields
\begin{IEEEeqnarray}{c}\label{EqCof26}
u_{2,2} \geq u_{2,1} \geq u_{1,1}. 
\end{IEEEeqnarray}
Hence, the inequalities in \eqref{EqCof25} and \eqref{EqCof26} yields 
\begin{IEEEeqnarray}{l}
u_{2,2} = u_{2,1} \geq u_{1,2} \quad \textnormal{and} \label{EqWait7}\\
u_{2,2} = u_{2,1} \geq u_{1,1}. \label{EqWait8}
\end{IEEEeqnarray}
If at most one of the inequalities in \eqref{EqWait7} and  \eqref{EqWait8} holds with equality, then, for \Pone, action $a_2$ dominates action $a_1$.
For this case, all strategies $P_{A_1} \in \simplex{\set{A}_1}$, together with strategy $P_{A_2} \in \simplex{\set{A}_2}$ such that $P_{A_2}(a_1) = 0$, form an NE. 
Then there are infinitely many NEs, which contradicts the assumption of the lemma.
Alternatively, if both the inequalities \eqref{EqWait7} and  \eqref{EqWait8} hold with equality, all pair of strategies $(P_{A_1}, P_{A_2}) \in \simplex{\set{A}_1} \times \simplex{\set{A}_2}$ form an NE. 
Then there are infinitely many NEs, which contradicts the assumption of the lemma and completes the proof.

If Case VI holds, from Definition \ref{Def:Equ_1}, the pair of strategies $(P_{A_1}^\star, P_{A_2}^\star)$ forming the NE implies that for all $\beta \in [0,1]$ and $\alpha \in [0,1]$, it holds that 
\begin{IEEEeqnarray}{c}\label{Equ:r6}
(0, 1) \matx{u} (1,0)^{\sf T} \geq (\alpha, 1 - \alpha) \matx{u} (1,0)^{\sf T}; 
\end{IEEEeqnarray}
and 
\begin{IEEEeqnarray}{c}\label{Equ:r7}
(0, 1) \matx{u} (1,0)^{\sf T} \leq (0, 1) \matx{u} (\beta,1-\beta)^{\sf T}. 
\end{IEEEeqnarray}
Setting $\alpha$ in (\ref{Equ:r6}) to $1$ and $\beta$ in (\ref{Equ:r7}) to zero yields 
\begin{IEEEeqnarray}{c}\label{Equ:r8}
u_{1,1} \leq u_{2,1} \leq u_{2,2}. 
\end{IEEEeqnarray}
Similarly, from Definition \ref{Def:Equ_1}, the pair of strategies  $(Q_{A_1}^\star, Q_{A_2}^\star)$ forming the NE implies that for all $\beta \in [0,1]$ and $\alpha \in [0,1]$, it holds that 
\begin{IEEEeqnarray}{c}\label{Equ:r9}
( 1, 0) \matx{u} (0, 1)^{\sf T} \leq (1, 0) \matx{u} (\beta,1-\beta)^{\sf T}; 
\end{IEEEeqnarray}
and 
\begin{IEEEeqnarray}{c}\label{Equ:r10}
(1, 0) \matx{u} (0, 1)^{\sf T} \geq (\alpha, 1- \alpha) \matx{u} (0, 1)^{\sf T}. 
\end{IEEEeqnarray}
Setting $\beta$ in (\ref{Equ:r9}) to $1$ and $\alpha$ in (\ref{Equ:r10}) to zero yields 
\begin{IEEEeqnarray}{c}\label{Equ:r11}
u_{2,2} \leq u_{1,2} \leq u_{1,1}. 
\end{IEEEeqnarray}
Combing (\ref{Equ:r8}) with (\ref{Equ:r11}) yields that 
\begin{IEEEeqnarray}{c}
u_{1,1} = u_{1,2} = u_{2,1} = u_{2,2}. 
\end{IEEEeqnarray}
Then, for all strategies $Q \in \Delta (\set{A}_1)$ and  $P \in \Delta (\set{A}_2)$, it holds that
\begin{IEEEeqnarray}{c}
( Q(a_1), 1- Q(a_1))\matx{u} ( P(a_1), 1- P(a_1))^{\sf T} = u_{1,1} = u_{1,2} = u_{2,1} = u_{2,2}, 
\end{IEEEeqnarray}
which implies that there exist infinitely many NEs. 
This contradicts the setting that there exists a finite number of NEs in the game $\game{G}$.

This completes the proof. 

\section{Proof of Theorem \ref{TheoNumberNE}}\label{Sec:App5}

The number of NEs in the game $\game{G}$ in \eqref{EqTheGame} is either infinite or finite. 
First, consider the case in which there are finite number NEs.
From \cite[Theorem 4.49]{Maschler_2013_game}, it holds that there exists at least one NE, denoted by $(P_{A_1}^{\star}, P_{A_2}^{\star})$, in the game $\game{G}$. 
Note that from Lemma \ref{Lemma:infinite}, $(P_{A_1}^{\star}, P_{A_2}^{\star})$ satisfies 
\begin{IEEEeqnarray}{l}
P_{A_1}^{\star}(a_1) \in \{ 0,1\} \quad \textnormal{and} \quad P_{A_2}^{\star}(a_1) \in \{ 0,1\}; \quad \textnormal{or} \label{Eqbottle1}\\
P_{A_1}^{\star}(a_1) \in ( 0,1) \quad \textnormal{and} \quad P_{A_2}^{\star}(a_1) \in ( 0,1). \label{Eqbottle2}
\end{IEEEeqnarray}
If $(P_{A_1}^{\star}, P_{A_2}^{\star})$ satisfies \eqref{Eqbottle1}, then $(P_{A_1}^{\star}, P_{A_2}^{\star})$ is the unique NE, see Lemma \ref{Lemma:strictlymixed_unique}. 
If $(P_{A_1}^{\star}, P_{A_2}^{\star})$ satisfies \eqref{Eqbottle2}, then $(P_{A_1}^{\star}, P_{A_2}^{\star})$ is the unique NE, see Lemma \ref{Lemma:pure_unique}.
As a result, if the zero-sum game $\game{G}$ in \eqref{EqTheGame} exhibits finite number of NEs, then there exist only one NE in the game $\game{G}$.

This completes the proof. 

\section{Proof of Lemma \ref{LemmaBR1}}\label{AppLemmaBR1}
Note that for all $P_{A_1} \in \simplex{\set{A}_1}$ and all $P_{A_2} \in \simplex{\set{A}_2}$, the payoff in \eqref{Eqv} satisfies 
\begin{IEEEeqnarray}{rl}
& u(P_{A_1}, P_{A_2}) \IEEEnonumber \\
&= u_{1,1} P_{A_1}(a_1) P_{A_2}(a_1) +  u_{1,2} P_{A_1}(a_1) P_{A_2}(a_2)  +  u_{2,1} P_{A_1}(a_2) P_{A_2}(a_1) +u_{2,2} P_{A_1}(a_2) P_{A_2}(a_2) \IEEEeqnarraynumspace \\
& = u_{1,1} P_{A_1}(a_1) P_{A_2}(a_1)  + u_{1,2} P_{A_1}(a_1) \bigl(1- P_{A_2}(a_1)\bigr) + u_{2,1} \bigl(1-P_{A_1}(a_1)\bigr) P_{A_2}(a_1) \IEEEnonumber \\
& \quad + u_{2,2} \bigl(1-P_{A_1}(a_1)\bigr) \bigl(1-P_{A_2}(a_1)\bigr)\\
& = (u_{1,1} -u_{1,2} -u_{2,1}+u_{2,2}) P_{A_1}(a_1) P_{A_2}(a_1)+( u_{1,2}- u_{2,2})P_{A_1}(a_1) + (u_{2,1}-u_{2,2})P_{A_2}(a_1)\IEEEnonumber \\
& \quad  +u_{2,2}.
\end{IEEEeqnarray}
As a result, the best response of \Pone to a strategy $P_{A_2} \in \simplex{\set{A}_2}$ is given by 
\begin{IEEEeqnarray}{rl}
& \BR_1(P_{A_2}) = \IEEEnonumber \\
& \ \arg \max_{P\in \simplex{\set{A}_1} }  \bigl((u_{1,1} -u_{1,2} -u_{2,1}+u_{2,2}) P_{A_2}(a_1)+( u_{1,2}- u_{2,2}) \bigl)P(a_1) + (u_{2,1}-u_{2,2})P_{A_2}(a_1) +u_{2,2}. \Tsupersqueezeequ \IEEEeqnarraynumspace \label{EqBR1}
\end{IEEEeqnarray}

Assume that $\delta = u_{1,1} -u_{1,2} -u_{2,1}+u_{2,2} \neq 0$. 
Further assume that $\delta > 0$. 
Hence, if $P_{A_2}(a_1) <p^{(2)}$, the payoff in \eqref{EqBR1} satisfies 
\begin{IEEEeqnarray}{ll}
&\bigl((u_{1,1} -u_{1,2} -u_{2,1}+u_{2,2}) P_{A_2}(a_1)+( u_{1,2}- u_{2,2}) \bigl)P(a_1) + (u_{2,1}-u_{2,2})P_{A_2}(a_1) +u_{2,2} \IEEEeqnarraynumspace \IEEEnonumber \\
= & (u_{1,1} -u_{1,2} -u_{2,1}+u_{2,2})\left( P_{A_2}(a_1) - p^{(2)}\right) P(a_1) + (u_{2,1}-u_{2,2})P_{A_2}(a_1) +u_{2,2} \label{EqHand}\\
\leq & (u_{2,1}-u_{2,2})P_{A_2}(a_1) +u_{2,2}, 
\end{IEEEeqnarray}
where the inequality holds with equality if and only if 
\begin{IEEEeqnarray}{c}
P(a_1) = 0. 
\end{IEEEeqnarray}
Hence, it holds that 
\begin{IEEEeqnarray}{c}
\BR_1\left( P_{A_2} \right) =  \{ P \in \Delta(\set{A}_1): P(a_1) = 0 \}. 
\end{IEEEeqnarray}
If $P_{A_2}(a_1) >p^{(2)}$, the payoff in \eqref{EqBR1} satisfies 
\begin{IEEEeqnarray}{ll}
&\bigl((u_{1,1} -u_{1,2} -u_{2,1}+u_{2,2}) P_{A_2}(a_1)+( u_{1,2}- u_{2,2}) \bigl)P(a_1) + (u_{2,1}-u_{2,2})P_{A_2}(a_1) +u_{2,2} \IEEEeqnarraynumspace \IEEEnonumber \\
= & (u_{1,1} -u_{1,2} -u_{2,1}+u_{2,2})\left( P_{A_2}(a_1) - p^{(2)}\right) P(a_1) + (u_{2,1}-u_{2,2})P_{A_2}(a_1) +u_{2,2} \\
\leq & (u_{1,1}-u_{1,2})P_{A_2}(a_1) +u_{1,2}, 
\end{IEEEeqnarray}
where the inequality holds with equality if and only if 
\begin{IEEEeqnarray}{c}
P(a_1) = 1. 
\end{IEEEeqnarray}
Hence, it holds that 
\begin{IEEEeqnarray}{c}
\BR_1\left( P_{A_2} \right) =  \{ P \in \Delta(\set{A}_1): P(a_1) = 1 \}. 
\end{IEEEeqnarray}
If $P_{A_2}(a_1) =p^{(2)}$, for all $P \in \simplex{\set{A}_1}$,  the payoff in \eqref{EqBR1}  satisfies 
\begin{IEEEeqnarray}{ll}
&\bigl((u_{1,1} -u_{1,2} -u_{2,1}+u_{2,2}) P_{A_2}(a_1)+( u_{1,2}- u_{2,2}) \bigl)P(a_1) + (u_{2,1}-u_{2,2})P_{A_2}(a_1) +u_{2,2} \IEEEeqnarraynumspace \IEEEnonumber \\
= & (u_{1,1} -u_{1,2} -u_{2,1}+u_{2,2})\left( P_{A_2}(a_1) - p^{(2)}\right) P(a_1) + (u_{2,1}-u_{2,2})P_{A_2}(a_1) +u_{2,2} \\
= & (u_{2,1}-u_{2,2})P_{A_2}(a_1) +u_{2,2}, 
\end{IEEEeqnarray}
Hence, it holds that 
\begin{IEEEeqnarray}{c}
\BR_1\left( P_{A_2} \right) =  \{ P \in \Delta(\set{A}_1): P(a_1) = \beta, \beta  \in [0,1]  \}. 
\end{IEEEeqnarray}
Alternatively, assume that $\delta = u_{1,1} -u_{1,2} -u_{2,1}+u_{2,2} < 0$. 
Hence, if $P_{A_2}(a_1) >p^{(2)}$, the payoff in \eqref{EqBR1} satisfies 
\begin{IEEEeqnarray}{ll}
&\bigl((u_{1,1} -u_{1,2} -u_{2,1}+u_{2,2}) P_{A_2}(a_1)+( u_{1,2}- u_{2,2}) \bigl)P(a_1) + (u_{2,1}-u_{2,2})P_{A_2}(a_1) +u_{2,2} \IEEEeqnarraynumspace \IEEEnonumber \\
= & (u_{1,1} -u_{1,2} -u_{2,1}+u_{2,2})\left( P_{A_2}(a_1) - p^{(2)}\right) P(a_1) + (u_{2,1}-u_{2,2})P_{A_2}(a_1) +u_{2,2} \\
\leq & (u_{2,1}-u_{2,2})P_{A_2}(a_1) +u_{2,2}, 
\end{IEEEeqnarray}
where the inequality holds with equality if and only if 
\begin{IEEEeqnarray}{c}
P(a_1) = 0. 
\end{IEEEeqnarray}
Hence, it holds that 
\begin{IEEEeqnarray}{c}
\BR_1\left( P_{A_2} \right) =  \{ P \in \Delta(\set{A}_1): P(a_1) = 0 \}. 
\end{IEEEeqnarray}
Hence, if $P_{A_2}(a_1) < p^{(2)}$, the payoff in \eqref{EqBR1} satisfies 
\begin{IEEEeqnarray}{ll}
&\bigl((u_{1,1} -u_{1,2} -u_{2,1}+u_{2,2}) P_{A_2}(a_1)+( u_{1,2}- u_{2,2}) \bigl)P(a_1) + (u_{2,1}-u_{2,2})P_{A_2}(a_1) +u_{2,2} \IEEEeqnarraynumspace \IEEEnonumber \\
= & (u_{1,1} -u_{1,2} -u_{2,1}+u_{2,2})\left( P_{A_2}(a_1) - p^{(2)}\right) P(a_1) + (u_{2,1}-u_{2,2})P_{A_2}(a_1) +u_{2,2} \\
\leq & (u_{1,1}-u_{1,2})P_{A_2}(a_1) +u_{1,2}, 
\end{IEEEeqnarray}
where the inequality holds with equality if and only if 
\begin{IEEEeqnarray}{c}
P(a_1) = 1. 
\end{IEEEeqnarray}
Hence, it holds that 
\begin{IEEEeqnarray}{c}
\BR_1\left( P_{A_2} \right) =  \{ P \in \Delta(\set{A}_1): P(a_1) = 1 \}. 
\end{IEEEeqnarray}
Hence, if $P_{A_2}(a_1) =p^{(2)}$,  for all $P \in \simplex{\set{A}_1}$, the payoff in \eqref{EqBR1} satisfies 
\begin{IEEEeqnarray}{ll}
&\bigl((u_{1,1} -u_{1,2} -u_{2,1}+u_{2,2}) P_{A_2}(a_1)+( u_{1,2}- u_{2,2}) \bigl)P(a_1) + (u_{2,1}-u_{2,2})P_{A_2}(a_1) +u_{2,2} \IEEEeqnarraynumspace \IEEEnonumber \\
= & (u_{1,1} -u_{1,2} -u_{2,1}+u_{2,2})\left( P_{A_2}(a_1) - p^{(2)}\right) P(a_1) + (u_{2,1}-u_{2,2})P_{A_2}(a_1) +u_{2,2} \\
= & (u_{2,1}-u_{2,2})P_{A_2}(a_1) +u_{2,2}, 
\end{IEEEeqnarray}
Hence, it holds that 
\begin{IEEEeqnarray}{c}
\BR_1\left( P_{A_2} \right) =  \{ P \in \Delta(\set{A}_1): P(a_1) = \beta, \beta  \in [0,1]  \}. 
\end{IEEEeqnarray}
In a nutshell, if $\delta = u_{1,1} -u_{1,2} -u_{2,1}+u_{2,2} \neq 0$, it holds that 
\begin{IEEEeqnarray}{c}
\BR_1\left( P_{A_2} \right)= 
\left\{ 
 \begin{array}{cl}
 \{ P_{A_1} \in \Delta(\set{A}_1): P_{A_1}(a_1) = 0 \}, & \makecell[t]{ \textnormal{if } \delta > 0 \textnormal{ and } P_{A_2}(a_1) <p^{(2)}, \textnormal{ or  } \\
\hspace{-1em} \delta < 0  \textnormal{ and }  P_{A_2}(a_1) > p^{(2)} } \\
\{ P_{A_1} \in \Delta(\set{A}_1): P_{A_1}(a_1) = 1 \}, &  \makecell[t]{ \textnormal{if } \delta > 0 \textnormal{ and } P_{A_2}(a_1) > p^{(2)}, \textnormal{ or  } \\
\hspace{-1em} \delta< 0  \textnormal{ and }  P_{A_2}(a_1) < p^{(2)}} \\
\{ P_{A_1} \in \Delta(\set{A}_1): P_{A_1}(a_1) = \beta, \beta  \in [0,1] \}, & 
\makecell[t]{ \textnormal{if } P_{A_2}(a_1) = p^{(2)}. }
 \end{array}
 \right. \IEEEeqnarraynumspace \squeezeequ 
\end{IEEEeqnarray}

Then consider the case in which $\delta = u_{1,1} -u_{1,2} -u_{2,1}+u_{2,2} =  0$. 
From \eqref{EqBR1}, the best response of \Pone to a strategy $P_{A_2} \in \simplex{\set{A}_2}$ is given by 
\begin{IEEEeqnarray}{c}
 \BR_1(P_{A_2}) =  \arg \max_{P\in \simplex{\set{A}_1} }  ( u_{1,2}- u_{2,2}) P(a_1) + (u_{2,1}-u_{2,2})P_{A_2}(a_1) +u_{2,2}.  
\end{IEEEeqnarray}
Hence, if $\delta =  0$, the best responses $\BR_1\left( P_{A_2} \right)$  satisfies
\begin{IEEEeqnarray}{rl}
\BR_1\left( P_{A_2} \right)= 
&\left\{ 
 \begin{array}{cl}
 \{ P_{A_1} \in \Delta(\set{A}_1): P_{A_1}(a_1) = 0 \}, &   \textnormal{if } u_{1,2} < u_{2,2} \\
\{ P_{A_1} \in \Delta(\set{A}_1): P_{A_1}(a_1) = 1 \}, &   \textnormal{if } u_{1,2} > u_{2,2} \\
\{ P_{A_1} \in \Delta(\set{A}_1): P_{A_1}(a_1) = \beta, \beta  \in [0,1] \}, & 
\textnormal{if } u_{1,2} = u_{2,2}. 
 \end{array} \label{Equ:BR_12}
 \right. \IEEEeqnarraynumspace \squeezeequ
\end{IEEEeqnarray}

This completes the proof. 

\section{Proof of Lemma \ref{LemmaBR2}}\label{AppLemmaBR2}
Note that for all $P_{A_1} \in \simplex{\set{A}_1}$ and all $P_{A_2} \in \simplex{\set{A}_2}$, the payoff in \eqref{Eqv} satisfies 
\begin{IEEEeqnarray}{rl}
& u(P_{A_1}, P_{A_2}) \IEEEnonumber \\
&= u_{1,1} P_{A_1}(a_1) P_{A_2}(a_1) +  u_{1,2} P_{A_1}(a_1) P_{A_2}(a_2)  +  u_{2,1} P_{A_1}(a_2) P_{A_2}(a_1) +u_{2,2} P_{A_1}(a_2) P_{A_2}(a_2) \IEEEeqnarraynumspace \\
& = u_{1,1} P_{A_1}(a_1) P_{A_2}(a_1)  + u_{1,2} P_{A_1}(a_1) \bigl(1- P_{A_2}(a_1)\bigr) + u_{2,1} \bigl(1-P_{A_1}(a_1)\bigr) P_{A_2}(a_1) \IEEEnonumber \\
& \quad + u_{2,2} \bigl(1-P_{A_1}(a_1)\bigr) \bigl(1-P_{A_2}(a_1)\bigr)\\
& = (u_{1,1} -u_{1,2} -u_{2,1}+u_{2,2}) P_{A_1}(a_1) P_{A_2}(a_1)+( u_{1,2}- u_{2,2})P_{A_1}(a_1) + (u_{2,1}-u_{2,2})P_{A_2}(a_1)\IEEEnonumber \\
& \quad  +u_{2,2}. 
\end{IEEEeqnarray}
As a result, the best response of \Ptwo to a strategy $P_{A_1} \in \simplex{\set{A}_1}$ is given by 
\begin{IEEEeqnarray}{rl}
& \BR_2(P_{A_1}) = \IEEEnonumber \\
& \ \arg \min_{P\in \simplex{\set{A}_2} }  \bigl((u_{1,1} -u_{1,2} -u_{2,1}+u_{2,2}) P_{A_1}(a_1)+( u_{2,1}- u_{2,2}) \bigr)P(a_1) + (u_{1,2}-u_{2,2})P_{A_1}(a_1) +u_{2,2}. \Tsupersqueezeequ \IEEEeqnarraynumspace \label{EqBR2}
\end{IEEEeqnarray}

Assume that $\delta = u_{1,1} -u_{1,2} -u_{2,1}+u_{2,2} \neq 0$. 
Further assume that $\delta > 0$. 
Hence, if $P_{A_1}(a_1) <p^{(1)}$, the payoff in \eqref{EqBR2} satisfies 
\begin{IEEEeqnarray}{ll}
&\bigl((u_{1,1} -u_{1,2} -u_{2,1}+u_{2,2}) P_{A_1}(a_1)+( u_{2,1}- u_{2,2}) \bigr)P(a_1) + (u_{1,2}-u_{2,2})P_{A_1}(a_1) +u_{2,2}.  \IEEEeqnarraynumspace \IEEEnonumber \\
= & (u_{1,1} -u_{1,2} -u_{2,1}+u_{2,2})\left( P_{A_1}(a_1) - p^{(1)}\right) P(a_1)+  (u_{1,2}-u_{2,2})P_{A_1}(a_1) +u_{2,2} \\
\leq & (u_{1,1}-u_{2,1})P_{A_1}(a_1) +u_{2,1}, 
\end{IEEEeqnarray}
where the inequality holds with equality if and only if 
\begin{IEEEeqnarray}{c}
P(a_1) = 1. 
\end{IEEEeqnarray}
Hence, it holds that 
\begin{IEEEeqnarray}{c}
\BR_2\left( P_{A_1} \right) =  \{ P \in \Delta(\set{A}_1): P(a_1) = 1 \}. 
\end{IEEEeqnarray}
If $P_{A_1}(a_1) >p^{(1)}$, the payoff in \eqref{EqBR2} satisfies 
\begin{IEEEeqnarray}{ll}
&\bigl((u_{1,1} -u_{1,2} -u_{2,1}+u_{2,2}) P_{A_1}(a_1)+( u_{2,1}- u_{2,2}) \bigr)P(a_1) + (u_{1,2}-u_{2,2})P_{A_1}(a_1) +u_{2,2}.  \IEEEeqnarraynumspace \IEEEnonumber \\
= & (u_{1,1} -u_{1,2} -u_{2,1}+u_{2,2})\left( P_{A_1}(a_1) - p^{(1)}\right) P(a_1)+  (u_{1,2}-u_{2,2})P_{A_1}(a_1) +u_{2,2} \\
\leq &  (u_{1,2}-u_{2,2})P_{A_1}(a_1) +u_{2,2}, 
\end{IEEEeqnarray}
where the inequality holds with equality if and only if 
\begin{IEEEeqnarray}{c}
P(a_1) = 0. 
\end{IEEEeqnarray}
Hence, it holds that 
\begin{IEEEeqnarray}{c}
\BR_2\left( P_{A_1} \right) =  \{ P \in \Delta(\set{A}_1): P(a_1) = 0 \}. 
\end{IEEEeqnarray}
If $P_{A_1}(a_1) =p^{(1)}$, for all $P \in \simplex{\set{A}_2}$, the payoff in \eqref{EqBR2} satisfies 
\begin{IEEEeqnarray}{ll}
&\bigl((u_{1,1} -u_{1,2} -u_{2,1}+u_{2,2}) P_{A_1}(a_1)+( u_{2,1}- u_{2,2}) \bigr)P(a_1) + (u_{1,2}-u_{2,2})P_{A_1}(a_1) +u_{2,2}.  \IEEEeqnarraynumspace \IEEEnonumber \\
= & (u_{1,1} -u_{1,2} -u_{2,1}+u_{2,2})\left( P_{A_1}(a_1) - p^{(1)}\right) P(a_1)+  (u_{1,2}-u_{2,2})P_{A_1}(a_1) +u_{2,2} \\
= &  (u_{1,2}-u_{2,2})P_{A_1}(a_1) +u_{2,2}.
\end{IEEEeqnarray}
Hence, it holds that 
\begin{IEEEeqnarray}{c}
\BR_2\left( P_{A_1} \right) =  \{ P \in \Delta(\set{A}_1): P(a_1) = \beta, \beta \in [0,1] \}. 
\end{IEEEeqnarray}

Alternatively, assume that $\delta = u_{1,1} -u_{1,2} -u_{2,1}+u_{2,2} < 0$. 
Hence, if $P_{A_1}(a_1) >p^{(1)}$, the payoff in \eqref{EqBR2} satisfies 
\begin{IEEEeqnarray}{ll}
&\bigl((u_{1,1} -u_{1,2} -u_{2,1}+u_{2,2}) P_{A_1}(a_1)+( u_{2,1}- u_{2,2}) \bigr)P(a_1) + (u_{1,2}-u_{2,2})P_{A_1}(a_1) +u_{2,2}.  \IEEEeqnarraynumspace \IEEEnonumber \\
= & (u_{1,1} -u_{1,2} -u_{2,1}+u_{2,2})\left( P_{A_1}(a_1) - p^{(1)}\right) P(a_1)+  (u_{1,2}-u_{2,2})P_{A_1}(a_1) +u_{2,2} \\
\leq & (u_{1,1}-u_{2,1})P_{A_1}(a_1) +u_{2,1}, 
\end{IEEEeqnarray}
where the inequality holds with equality if and only if 
\begin{IEEEeqnarray}{c}
P(a_1) = 1. 
\end{IEEEeqnarray}
Hence, it holds that 
\begin{IEEEeqnarray}{c}
\BR_2\left( P_{A_1} \right) =  \{ P \in \Delta(\set{A}_1): P(a_1) = 1 \}. 
\end{IEEEeqnarray}
If $P_{A_1}(a_1) <p^{(1)}$, the payoff in \eqref{EqBR2} satisfies 
\begin{IEEEeqnarray}{ll}
&\bigl((u_{1,1} -u_{1,2} -u_{2,1}+u_{2,2}) P_{A_1}(a_1)+( u_{2,1}- u_{2,2}) \bigr)P(a_1) + (u_{1,2}-u_{2,2})P_{A_1}(a_1) +u_{2,2}.  \IEEEeqnarraynumspace \IEEEnonumber \\
= & (u_{1,1} -u_{1,2} -u_{2,1}+u_{2,2})\left( P_{A_1}(a_1) - p^{(1)}\right) P(a_1)+  (u_{1,2}-u_{2,2})P_{A_1}(a_1) +u_{2,2} \\
\leq &  (u_{1,2}-u_{2,2})P_{A_1}(a_1) +u_{2,2}, 
\end{IEEEeqnarray}
where the inequality holds with equality if and only if 
\begin{IEEEeqnarray}{c}
P(a_1) = 0. 
\end{IEEEeqnarray}
Hence, it holds that 
\begin{IEEEeqnarray}{c}
\BR_2\left( P_{A_1} \right) =  \{ P \in \Delta(\set{A}_1): P(a_1) = 0 \}. 
\end{IEEEeqnarray}
If $P_{A_1}(a_1) =p^{(1)}$, for all $P \in \simplex{\set{A}_2}$, the payoff in \eqref{EqBR2} satisfies 
\begin{IEEEeqnarray}{ll}
&\bigl((u_{1,1} -u_{1,2} -u_{2,1}+u_{2,2}) P_{A_1}(a_1)+( u_{2,1}- u_{2,2}) \bigr)P(a_1) + (u_{1,2}-u_{2,2})P_{A_1}(a_1) +u_{2,2}.  \IEEEeqnarraynumspace \IEEEnonumber \\
= & (u_{1,1} -u_{1,2} -u_{2,1}+u_{2,2})\left( P_{A_1}(a_1) - p^{(1)}\right) P(a_1)+  (u_{1,2}-u_{2,2})P_{A_1}(a_1) +u_{2,2} \\
= &  (u_{1,2}-u_{2,2})P_{A_1}(a_1) +u_{2,2}.
\end{IEEEeqnarray}
Hence, it holds that 
\begin{IEEEeqnarray}{c}
\BR_2\left( P_{A_1} \right) =  \{ P \in \Delta(\set{A}_1): P(a_1) = \beta, \beta \in [0,1] \}. 
\end{IEEEeqnarray}

In a nutshell, if $\delta \neq 0$, the best response $\BR_2\left( P_{A_1} \right)$ satisfies 
\begin{IEEEeqnarray}{c}
\BR_2\left( P_{A_1} \right)= 
\left\{ 
 \begin{array}{cl}
 \{ P_{A_2} \in \Delta(\set{A}_2): P_{A_2}(a_1) = 0 \}, & \makecell[t]{ \textnormal{if } \delta> 0 \textnormal{ and } P_{A_1}(a_1) >p^{(1)}, \textnormal{ or  } \\
\hspace{-1em} \delta < 0  \textnormal{ and }  P_{A_1}(a_1) <p^{(1)} } \\
\{ P_{A_2} \in \Delta(\set{A}_2): P_{A_2}(a_1) = 1 \}, &  \makecell[t]{ \textnormal{if }\delta > 0 \textnormal{ and } P_{A_1}(a_1) < p^{(1)}, \textnormal{ or  } \\
\hspace{-1em} \delta < 0  \textnormal{ and }  P_{A_1}(a_1) >p^{(1)}} \\
\{ P_{A_2} \in \Delta(\set{A}_2): P_{A_2}(a_1) = \beta, \beta  \in [0,1] \}, & 
\makecell[t]{ \textnormal{if } P_{A_1}(a_1) = p^{(1)} .}
 \end{array}
 \right. \IEEEeqnarraynumspace \squeezeequ 
\end{IEEEeqnarray}

Then, consider the case in which $\delta = u_{1,1} -u_{1,2} -u_{2,1}+u_{2,2} =  0$. 
From \eqref{EqBR2}, the best response of \Ptwo to a strategy $P_{A_1} \in \simplex{\set{A}_1}$ satisfies 
\begin{IEEEeqnarray}{c}
 \BR_2(P_{A_1}) =  \arg \min_{P\in \simplex{\set{A}_2} }  ( u_{2,1}- u_{2,2}) P(a_1) + (u_{1,2}-u_{2,2})P_{A_1}(a_1) +u_{2,2}.\end{IEEEeqnarray}
Hence, if $\delta=  0$, the best responses $\BR_2\left( P_{A_1} \right)$  satisfies
\begin{IEEEeqnarray}{rl}
\BR_2\left( P_{A_1} \right)= 
&\left\{ 
 \begin{array}{cl}
 \{ P_{A_2} \in \Delta(\set{A}_2): P_{A_2}(a_1) = 0 \}, & \textnormal{if } u_{2,1} > u_{2,2}\\
\{ P_{A_2} \in \Delta(\set{A}_2): P_{A_2}(a_1) = 1 \}, &  \textnormal{if } u_{2,1} < u_{2,2}\\
\{ P_{A_2} \in \Delta(\set{A}_2): P_{A_2}(a_1) = \beta, \beta  \in [0,1] \}, & 
\textnormal{if } u_{2,1} = u_{2,2}.
 \end{array}
 \right. \IEEEeqnarraynumspace \squeezeequ 
\end{IEEEeqnarray}

This completes the proof.

\section{Proof of Theorem \ref{TheoNE}}\label{Sec:App6}

The proof is divided into two parts. 
The first part provides conditions on the entries of the matrix $\matx{u}$ in (\ref{Equ:u}) such that $\left( P_{A_1}^\star, P_{A_2}^\star \right)$ is the unique NE and satisfies 
\begin{IEEEeqnarray}{c}\label{EqCup00}
P_{A_1}^\star(a_1) \in (0,1) \quad \textnormal{and} \quad P_{A_2}^\star(a_1) \in (0,1).
\end{IEEEeqnarray}
The second part provides conditions on the entries of the matrix $\matx{u}$ in (\ref{Equ:u}) such that $\left( P_{A_1}^\star, P_{A_2}^\star \right)$ is the unique NE and satisfies 
\begin{IEEEeqnarray}{c}\label{EqCup0}
P_{A_1}^\star(a_1) \in \{0,1\} \quad \textnormal{and} \quad P_{A_2}^\star(a_1) \in \{0,1\}.
\end{IEEEeqnarray}

The first part is as follows. 
First, consider the case in which $u_{1,1} -u_{1,2} -u_{2,1} +u_{2,2} \neq 0$.
If the pair of strategies $\left( P_{A_1}^\star, P_{A_2}^\star \right) \in \simplex{\set{A}_1}\times\simplex{\set{A}_2}$ forms an NE in the game $\game{G}$, then  it holds that 
\begin{IEEEeqnarray}{c}\label{EqPot1}
P_{A_1}^\star \in \BR_1( P_{A_2}^\star) \quad \textnormal{and} \quad P_{A_2}^\star \in \BR_2( P_{A_1}^\star).
\end{IEEEeqnarray}
Moreover, if 
\begin{IEEEeqnarray}{c}\label{Equ:cup1}
P_{A_1}^\star(a_1) = 1 - P_{A_1}^\star(a_2) \in \left(0,1 \right) \ \textnormal{and} \ \  P_{A_2}^\star(a_1) = 1 - P_{A_2}^\star(a_2) \in \left(0,1 \right),
\end{IEEEeqnarray}
then from Lemma \ref{LemmaBR1}, Lemma \ref{LemmaBR2}, and \eqref{EqPot1}, 
 it follows that  $p^{(1)}$ in \eqref{Equ:P_{A_1}} and $p^{(2)}$  in \eqref{Equ:P_{A_2}} satisfy
\begin{IEEEeqnarray}{l}
P_{A_1}^\star (a_1) = p^{(1)} = \frac{u_{2,2} - u_{2,1}}{u_{1,1} -u_{1,2} -u_{2,1}+u_{2,2}} \in (0,1) \quad \textnormal{and} \\
P_{A_2}^\star (a_1)  = p^{(2)} =  \frac{u_{2,2} - u_{1,2}}{u_{1,1} -u_{1,2} -u_{2,1}+u_{2,2}}\in (0,1). \IEEEeqnarraynumspace 
\end{IEEEeqnarray}

Note that $ 1> \frac{ u_{2,2} - u_{1,2}}{u_{1,1} - u_{1,2} - u_{2,1} + u_{2,2}}> 0$ is equivalent to  either 
\begin{IEEEeqnarray}{c} \label{Equ:con1}
u_{2,2} - u_{1,2} > 0,  u_{1,1} - u_{1,2} - u_{2,1} + u_{2,2}>0, \ \textnormal{and} \  u_{1,1} - u_{1,2} - u_{2,1} + u_{2,2} > u_{2,2} - u_{1,2};  \IEEEeqnarraynumspace
\end{IEEEeqnarray}
or 
\begin{IEEEeqnarray}{c}\label{Equ:con2}
u_{2,2} - u_{1,2} < 0,  u_{1,1} - u_{1,2} - u_{2,1} + u_{2,2}<0, \ \textnormal{and}  \ u_{1,1} - u_{1,2} - u_{2,1} + u_{2,2} < u_{2,2} - u_{1,2}. \IEEEeqnarraynumspace
\end{IEEEeqnarray}
In (\ref{Equ:con1}), inequalities $u_{2,2} - u_{1,2} > 0$ and $u_{1,1} - u_{1,2} - u_{2,1} + u_{2,2} > u_{2,2} - u_{1,2}$ guarantee that $u_{1,1} - u_{1,2} - u_{2,1} + u_{2,2}>0$.  
As a result, the condition in (\ref{Equ:con1}) can be simplified to 
\begin{IEEEeqnarray}{c}\label{Equ:con11}
u_{2,2} - u_{1,2} > 0  \ \textnormal{and} \ u_{1,1} -u_{2,1}> 0.
\end{IEEEeqnarray}
In  (\ref{Equ:con2}), inequalities $u_{2,2} - u_{1,2} < 0$ and $u_{1,1} - u_{1,2} - u_{2,1} + u_{2,2} < u_{2,2} - u_{1,2}$ guarantee that $u_{1,1} - u_{1,2} - u_{2,1} + u_{2,2}<0$. 
So the second condition in (\ref{Equ:con2}) can be simplified as 
\begin{IEEEeqnarray}{c}\label{Equ:con21}
u_{2,2} - u_{1,2} < 0 \  \textnormal{and}  \ u_{1,1} -u_{2,1}< 0. 
\end{IEEEeqnarray}

Furthermore, $ 1 > \frac{ u_{2,2} - u_{2,1}}{u_{1,1} - u_{1,2} - u_{2,1} + u_{2,2}} > 0$ is equivalent to either 
\begin{IEEEeqnarray}{c}\label{Equ:con3}
u_{2,2} - u_{2,1} > 0, u_{1,1} - u_{1,2} - u_{2,1} + u_{2,2}>0, \ \textnormal{and} \  u_{1,1} - u_{1,2} - u_{2,1} + u_{2,2} > u_{2,2} - u_{2,1}; \IEEEeqnarraynumspace
\end{IEEEeqnarray}
or 
\begin{IEEEeqnarray}{c}\label{Equ:con4}
u_{2,2} - u_{2,1} < 0,  u_{1,1} - u_{1,2} - u_{2,1} + u_{2,2}<0, \ \textnormal{and} \ u_{1,1} - u_{1,2} - u_{2,1} + u_{2,2} < u_{2,2} - u_{2,1}.  \IEEEeqnarraynumspace
\end{IEEEeqnarray}
In (\ref{Equ:con3}), inequalities $u_{2,2} - u_{2,1} > 0$ and $u_{1,1} - u_{1,2} - u_{2,1} + u_{2,2} > u_{2,2} - u_{2,1}$  guarantee that $u_{1,1} - u_{1,2} - u_{2,1} + u_{2,2}>0$. 
As a result, the first condition in (\ref{Equ:con3}) can be simplified to
\begin{IEEEeqnarray}{c}\label{Equ:con31}
u_{2,2} - u_{2,1} > 0 \ \textnormal{and}  \ u_{1,1} -u_{1,2} > 0.
\end{IEEEeqnarray}
In (\ref{Equ:con4}), inequalities $u_{2,2} - u_{2,1} < 0$ and $u_{1,1} - u_{1,2} - u_{2,1} + u_{2,2} < u_{2,2} - u_{2,1}$  guarantee that $u_{1,1} - u_{1,2} - u_{2,1} + u_{2,2}<0$. 
So the second condition in (\ref{Equ:con4}) can be simplified to
\begin{IEEEeqnarray}{c}\label{Equ:con41}
u_{2,2} - u_{2,1} < 0 \  \textnormal{and} \ u_{1,1} -u_{1,2}< 0. 
\end{IEEEeqnarray}

Now, consider the case in which $u_{1,1} - u_{1,2} -u_{2,1} +u_{2,2} = 0$. 
From Lemma \ref{LemmaBR1}, Lemma \ref{LemmaBR2} and \eqref{Equ:cup1}, if  \eqref{EqPot1} holds, then it follows that 
\begin{IEEEeqnarray}{c}
u_{1,2} = u_{2,1} = u_{2,2}. 
\end{IEEEeqnarray}
Given the fact that $u_{1,1} - u_{1,2} -u_{2,1} +u_{2,2} = 0$, it holds that 
\begin{IEEEeqnarray}{c}
u_{1,1} = u_{1,2} = u_{2,1} = u_{2,2}. 
\end{IEEEeqnarray}
Hence, from Lemma \ref{LemmaBR1} and Lemma \ref{LemmaBR2}, all pairs of strategies $(P_1, P_2) \in \simplex{\set{A}_1} \times \simplex{\set{A}_2}$  satisfy \eqref{EqPot1}. 
Hence, there is nothing to prove for this case, as there is no unique NE. 

This proves that if there exists a unique NE satisfying \eqref{EqUniqueMixCon}, one of the conditions in (\ref{Equ:con11}), (\ref{Equ:con21}), (\ref{Equ:con31}) and (\ref{Equ:con41}) holds. 

The converse is as follows. 
If the inequalities in \eqref{Equ:con41} and \eqref{Equ:con31} hold, $p^{(1)} \in (0,1)$. 
If the inequalities in \eqref{Equ:con11} and \eqref{Equ:con21} hold, $p^{(2)} \in (0,1)$. 
Hence, from Lemma \ref{LemmaBR1} and Lemma \ref{LemmaBR2}, the pair of strategies $(P_{A_1}, P_{A_2}) \in \simplex{\set{A}_1} \times \simplex{\set{A}_1}$ such that 
\begin{IEEEeqnarray}{c}
P_{A_1}(a_1) = p^{(1)} \in (0,1) \quad \textnormal{and} \quad P_{A_2}(a_1) = p^{(2)} \in (0,1) 
\end{IEEEeqnarray}
satisfies \eqref{EqPot1}. 
Given the fact that $|\set{P}| = 1$, then from Lemma \ref{Lemma:strictlymixed_unique} and from Theorem \ref{TheoNumberNE}, the pair of strategies $(P_{A_1}, P_{A_2}) $ forms the unique NE.

Combining the conditions in (\ref{Equ:con11}), (\ref{Equ:con21}), (\ref{Equ:con31}) and (\ref{Equ:con41}) yields the conclusion that there exists a unique NE satisfies \eqref{Equ:cup1} if and only if $(u_{11} -u_{1,2})(u_{2,2} -u_{2,1}) > 0$ and $(u_{1,1} -u_{2,1})(u_{2,2} -u_{1,2}) > 0$.

Furthermore, it holds that 
\begin{IEEEeqnarray}{rl}
&u(P_{A_1}^\star, P_{A_2}^\star) \IEEEnonumber\\
&= P_{A_1}^\star(a_1)P_{A_2}^\star(a_1) u_{1,1} + P_{A_1}^\star(a_1)P_{A_2}^\star(a_2) u_{1,2}+ P_{A_1}^\star(a_2)P_{A_2}^\star(a_1) u_{2,1} + P_{A_1}^\star(a_2)P_{A_2}^\star(a_2) u_{2,2} \IEEEeqnarraynumspace \\
& =  \left((u_{1,1} -u_{1,2}-u_{2,1}+u_{2,2}) P_{A_2}^\star(a_1) + (u_{1,2}-u_{2,2}) \right) P_{A_1}^\star(a_1)+ (u_{2,1}-u_{2,2}) P_{A_2}^\star(a_1) +u_{2,2} \squeezeequ\\
&= \frac{u_{1,1} u_{2,2} - u_{1,2} u_{2,1}}{u_{1,1} - u_{1,2} - u_{2,1} + u_{2,2}},
\end{IEEEeqnarray}
which completes the proof. 

The second part is as follows. 
Given the fact that there exists at most four NEs satisfying \eqref{EqCup0}, the following proof considers the following cases.

\noindent Case I: the NE $(P_{A_1}^\star, P_{A_2}^\star)$ satisfies 
\begin{IEEEeqnarray}{c}\label{EqVec1}
P_{A_1}^\star(a_{1})  =  1 - P_{A_1}^\star(a_{2})  = 1 \quad \textnormal{and} \quad 
P_{A_2}^\star(a_{1})    =  1 - P_{A_2}^\star(a_{2}) =1.
\end{IEEEeqnarray} 

\noindent Case II: the NE $(P_{A_1}^\star, P_{A_2}^\star)$ satisfies 
\begin{IEEEeqnarray}{c}\label{EqVec2}
P_{A_1}^\star(a_{1})  =  1 - P_{A_1}^\star(a_{2})  = 1,  \quad  \textnormal{and} \quad 
P_{A_2}^\star(a_{1})    =  1 - P_{A_2}^\star(a_{2}) =0.
\end{IEEEeqnarray} 

\noindent Case III: the NE $(P_{A_1}^\star, P_{A_2}^\star)$ satisfies 
\begin{IEEEeqnarray}{c}\label{EqVec3}
P_{A_1}^\star(a_{1})  =  1 - P_{A_1}^\star(a_{2})  = 0,  \quad  \textnormal{and} \quad 
P_{A_2}^\star(a_{1})    =  1 - P_{A_2}^\star(a_{2}) =1. 
\end{IEEEeqnarray} 
\noindent Case IV:  the NE $(P_{A_1}^\star, P_{A_2}^\star)$ satisfies 
\begin{IEEEeqnarray}{c}\label{EqVec4}
P_{A_1}^\star(a_{1})  =  1 - P_{A_1}^\star(a_{2})  = 0,  \quad  \textnormal{and} \quad 
P_{A_2}^\star(a_{1})    =  1 - P_{A_2}^\star(a_{2}) =0.
\end{IEEEeqnarray} 

The proof of Case I is as follows. 
If $\left( P_{A_1}^\star, P_{A_2}^\star \right)$ is the unique NE, from Definition \ref{Def:Equ_1}, for all $\alpha \in [0,1)$ and $\beta \in [0,1)$, it holds that 
\begin{IEEEeqnarray}{rl}
&u(P_{A_1}^\star, P_{A_2}^\star) = ( 1, 0) \matx{u} (1, 0)^{\sf T} = u_{1,1} > (\alpha, 1-\alpha) \matx{u} (1,0)^{\sf T} = \alpha u_{1,1} + (1-\alpha) u_{2,1}\quad \textnormal{and} \label{Equ:pen1} \IEEEeqnarraynumspace \\ 
&u(P_{A_1}^\star, P_{A_2}^\star) = ( 1, 0) \matx{u} (1, 0)^{\sf T} = u_{1,1} < (1, 0) \matx{u} (\beta,1 - \beta)^{\sf T} = \beta u_{1,1} + (1-\beta) u_{1,2}. \label{Equ:pen2}
\end{IEEEeqnarray}
The strict inequality follows from the assumption that the NE in unique. 
Note that the inequalities in (\ref{Equ:pen1}) and (\ref{Equ:pen2}) are equivalent to 
\begin{IEEEeqnarray}{c}\label{Eqpen01}
u_{1,1} > u_{2,1} \quad \textnormal{and} \quad u_{1,2} > u_{1,1},
\end{IEEEeqnarray}
respectively. 
As a result, the inequalities in (\ref{Equ:pen1}) and (\ref{Equ:pen2}) are satisfied when
\begin{IEEEeqnarray}{c}\label{EqVec5}
u_{1,2} > u_{1,1} > u_{2,1}, 
\end{IEEEeqnarray}
which is the condition in (\ref{Equ:pen01}). 

The converse is as follows. 
First, consider the case in which $u_{1,1} -u_{1,2} -u_{2,1}+u_{2,2} > 0$. 
Under the assumption that $u_{1,1} -u_{1,2} -u_{2,1}+u_{2,2} > 0$, if the inequalities in \eqref{EqVec5} hold, then it holds that 
\begin{IEEEeqnarray}{c}
u_{1,1} -u_{1,2} < 0 \quad \textnormal{and} \quad u_{2,2} -u_{2,1} > 0, 
\end{IEEEeqnarray}
which, from  \eqref{Equ:P_{A_1}}, implies that $p^{(1)}>1$. 
Furthermore, if the inequalities in \eqref{EqVec5} hold, then it holds that 
\begin{IEEEeqnarray}{c}
u_{1,1} -u_{2,1} > 0. 
\end{IEEEeqnarray}
Under the assumption that $u_{1,1} -u_{1,2} -u_{2,1}+u_{2,2} > 0$, if $u_{2,2} -u_{1,2} \leq 0$, then from  \eqref{Equ:P_{A_2}}, it holds that that $p^{(2)} \leq 0$. 
And if $u_{2,2} -u_{1,2} > 0$, from  \eqref{Equ:P_{A_2}}, it holds that that $p^{(2)} \in (0,1)$. 
Hence, in either case, it holds that $p^{(2)}<1$. 
Then from Lemma \ref{LemmaBR1} and Lemma \ref{LemmaBR2}, if $p^{(1)}>1$ and $p^{(2)}<1$, the pair of strategies $(P_{A_1}, P_{A_2}) \in \simplex{\set{A}_1} \times \simplex{\set{A}_1}$ such that 
\begin{IEEEeqnarray}{c}
P_{A_1}(a_1) = 1 \quad \textnormal{and} \quad P_{A_2}(a_1) = 1 
\end{IEEEeqnarray}
satisfies 
\begin{IEEEeqnarray}{c}
P_{A_1} \in \BR_1( P_{A_2}) \quad \textnormal{and} \quad P_{A_2} \in \BR_2( P_{A_1}).
\end{IEEEeqnarray}
Given the fact that $|\set{P}| = 1$, then from Lemma \ref{Lemma:strictlymixed_unique} and from Theorem \ref{TheoNumberNE}, the pair of strategies $(P_{A_1}, P_{A_2}) $ forms the unique NE. 

Then, consider the case in which $u_{1,1} -u_{1,2} -u_{2,1}+u_{2,2} < 0$. 
Under the assumption that $u_{1,1} -u_{1,2} -u_{2,1}+u_{2,2} < 0$, if the inequalities in \eqref{EqVec5} hold, then it holds that 
\begin{IEEEeqnarray}{c}
u_{1,1} -u_{2,1} > 0 \quad \textnormal{and} \quad u_{2,2} -u_{1,2} < 0, 
\end{IEEEeqnarray}
which, from  \eqref{Equ:P_{A_2}}, implies that $p^{(2)}>1$. 
Furthermore, if the inequalities in \eqref{EqVec5} hold, then it holds that 
\begin{IEEEeqnarray}{c}
u_{1,1} -u_{1,2} < 0. 
\end{IEEEeqnarray}
Under the assumption that $u_{1,1} -u_{1,2} -u_{2,1}+u_{2,2} < 0$, if $u_{2,2} -u_{2,1} \leq 0$, then from  \eqref{Equ:P_{A_1}}, it holds that that $p^{(1)} \leq 0$. 
And if $u_{2,2} -u_{2,1} > 0$, from  \eqref{Equ:P_{A_2}}, it holds that that $p^{(1)} <0$. 
Hence, in either case, it holds that $p^{(1)}\leq 0$. 
Then from Lemma \ref{LemmaBR1} and Lemma \ref{LemmaBR2},  if $p^{(2)}>1$ and $p^{(1)}\leq0$, the pair of strategies $(P_{A_1}, P_{A_2}) \in \simplex{\set{A}_1} \times \simplex{\set{A}_1}$ such that 
\begin{IEEEeqnarray}{c}
P_{A_1}(a_1) = 1 \quad \textnormal{and} \quad P_{A_2}(a_1) = 1 
\end{IEEEeqnarray}
satisfies 
\begin{IEEEeqnarray}{c}
P_{A_1} \in \BR_1( P_{A_2}) \quad \textnormal{and} \quad P_{A_2} \in \BR_2( P_{A_1}).
\end{IEEEeqnarray}
Given the fact that $|\set{P}| = 1$, then from Lemma \ref{Lemma:strictlymixed_unique} and from Theorem \ref{TheoNumberNE}, the pair of strategies $(P_{A_1}, P_{A_2}) $ forms the unique NE.

Finally, consider the case in which $u_{1,1} -u_{1,2} -u_{2,1}+u_{2,2} = 0$. 
Under the assumption that $u_{1,1} -u_{1,2} -u_{2,1}+u_{2,2} = 0$, it follows that 
\begin{IEEEeqnarray}{c}\label{EqWhen1}
u_{2,2} = u_{1,2} +u_{2,1} -u_{1,1}. 
\end{IEEEeqnarray}
If the inequalities in \eqref{EqVec5}  and the equality in \eqref{EqWhen1} hold,
\begin{IEEEeqnarray}{c}
u_{1,2} > u_{2,2} \quad \textnormal{and} \quad u_{2,2} >u_{2,1}.  
\end{IEEEeqnarray}
Hence, from Lemma \ref{LemmaBR1} and Lemma \ref{LemmaBR2},  the pair of strategies $(P_{A_1}, P_{A_2}) \in \simplex{\set{A}_1} \times \simplex{\set{A}_1}$ such that 
\begin{IEEEeqnarray}{c}
P_{A_1}(a_1) = 1 \quad \textnormal{and} \quad P_{A_2}(a_1) = 1 
\end{IEEEeqnarray}
satisfies 
\begin{IEEEeqnarray}{c}
P_{A_1} \in \BR_1( P_{A_2}) \quad \textnormal{and} \quad P_{A_2} \in \BR_2( P_{A_1})
\end{IEEEeqnarray}
and forms the unique NE. 
This completes the converse part.

Furthermore, the NE satisfying \eqref{EqVec1} implies that
\begin{IEEEeqnarray}{c}
u(P_{A_1}^\star, P_{A_2}^\star) = u_{1,1}, 
\end{IEEEeqnarray}
which completes the proof. 

The proof of Case II is as follows. 
If $\left( P_{A_1}^\star, P_{A_2}^\star \right)$ is the unique NE, from Definition \ref{Def:Equ_1}, for all $\alpha \in [0,1)$ and $\beta \in (0,1]$, it holds that
\begin{IEEEeqnarray}{rl}
&u(P_{A_1}^\star, P_{A_2}^\star) = ( 1, 0) \matx{u} (0, 1)^{\sf T} = u_{1,2} > (\alpha, 1-\alpha) \matx{u} (0, 1)^{\sf T} = \alpha u_{1,2} + (1-\alpha) u_{2,2}  \quad \textnormal{and} \label{Equ:pen3} \IEEEeqnarraynumspace \\ 
&u(P_{A_1}^\star, P_{A_2}^\star) = ( 1, 0) \matx{u} (0, 1)^{\sf T} = u_{1,2} < (1, 0) \matx{u} (\beta,1 - \beta)^{\sf T} = \beta u_{1,1} + (1-\beta) u_{1,2}. \label{Equ:pen4}
\end{IEEEeqnarray}
The strict inequality follows from the assumption that the NE in unique. 
Note that the inequalities in (\ref{Equ:pen3}) and (\ref{Equ:pen4}) are equivalent to 
\begin{IEEEeqnarray}{c}\label{Eqpen02}
u_{1,2} > u_{2,2} \quad \textnormal{and} \quad u_{1,1} > u_{1,2},
\end{IEEEeqnarray}
respectively. 
As a result, the inequalities in (\ref{Equ:pen3}) and (\ref{Equ:pen4}) are satisfied when 
\begin{IEEEeqnarray}{c}\label{EqVect1}
u_{1,1} > u_{1,2} > u_{2,2}, 
\end{IEEEeqnarray}
which is the condition in (\ref{Equ:pen02}). 

The converse is as follows. 
First, consider the case in which $u_{1,1} -u_{1,2} -u_{2,1}+u_{2,2} > 0$. 
Under the assumption that $u_{1,1} -u_{1,2} -u_{2,1}+u_{2,2} > 0$, if the inequalities in \eqref{EqVect1} hold, then it holds that 
\begin{IEEEeqnarray}{c}
u_{2,2} -u_{1,2} < 0 \quad \textnormal{and} \quad u_{1,1} -u_{2,1} > 0, 
\end{IEEEeqnarray}
which, from  \eqref{Equ:P_{A_2}}, implies that $p^{(2)}<0$. 
Furthermore, if the inequalities in \eqref{EqVect1} hold, then it holds that 
\begin{IEEEeqnarray}{c}
u_{1,1} -u_{1,2} > 0. 
\end{IEEEeqnarray}
Under the assumption that $u_{1,1} -u_{1,2} -u_{2,1}+u_{2,2} > 0$, if $u_{2,2} -u_{2,1} > 0$, then from  \eqref{Equ:P_{A_2}}, it holds that that $p^{(1)} \in (0,1)$. 
And if $u_{2,2} -u_{1,2} \leq 0$, from  \eqref{Equ:P_{A_2}}, it holds that that $p^{(1)} \leq 0$. 
Hence, in either case, it holds that $p^{(1)}<1$. 
Then from Lemma \ref{LemmaBR1} and Lemma \ref{LemmaBR2}, if $p^{(2)}<0$ and $p^{(1)}<1$,  
the pair of strategies $(P_{A_1}, P_{A_2}) \in \simplex{\set{A}_1} \times \simplex{\set{A}_1}$ such that 
\begin{IEEEeqnarray}{c}
P_{A_1}(a_1) = 1 \quad \textnormal{and} \quad P_{A_2}(a_1) = 0
\end{IEEEeqnarray}
satisfies 
\begin{IEEEeqnarray}{c}
P_{A_1} \in \BR_1( P_{A_2}) \quad \textnormal{and} \quad P_{A_2} \in \BR_2( P_{A_1}).
\end{IEEEeqnarray}
Given the fact that $|\set{P}| = 1$, then from Lemma \ref{Lemma:strictlymixed_unique} and from Theorem \ref{TheoNumberNE}, the pair of strategies $(P_{A_1}, P_{A_2}) $ forms the unique NE.

Then, consider the case in which $u_{1,1} -u_{1,2} -u_{2,1}+u_{2,2} < 0$. 
Under the assumption that $u_{1,1} -u_{1,2} -u_{2,1}+u_{2,2} < 0$, if the inequalities in \eqref{EqVect1} hold, then it holds that 
\begin{IEEEeqnarray}{c}
u_{1,1} -u_{1,2} > 0 \quad \textnormal{and} \quad u_{2,2} -u_{2,1} < 0, 
\end{IEEEeqnarray}
which, from  \eqref{Equ:P_{A_1}}, implies that $p^{(1)}>1$. 
Furthermore, if the inequalities in \eqref{EqVect1} hold, then it holds that 
\begin{IEEEeqnarray}{c}
u_{2,2} -u_{1,2} < 0. 
\end{IEEEeqnarray}
Under the assumption that $u_{1,1} -u_{1,2} -u_{2,1}+u_{2,2} < 0$, if $u_{1,1} -u_{2,1} \leq 0$, then from  \eqref{Equ:P_{A_2}}, it holds that that $ 0< p^{(2)} \leq 1$. 
And if $u_{1,1} -u_{2,1} > 0$, from  \eqref{Equ:P_{A_2}}, it holds that that $p^{(2)}>1$. 
Hence, in either case, it holds that $p^{(2)}>1$. 
Then from Lemma \ref{LemmaBR1} and Lemma \ref{LemmaBR2}, if $p^{(1)}>1$ and $p^{(2)}>0$, 
the pair of strategies $(P_{A_1}, P_{A_2}) \in \simplex{\set{A}_1} \times \simplex{\set{A}_1}$ such that 
\begin{IEEEeqnarray}{c}
P_{A_1}(a_1) = 1 \quad \textnormal{and} \quad P_{A_2}(a_1) = 0
\end{IEEEeqnarray}
satisfies 
\begin{IEEEeqnarray}{c}
P_{A_1} \in \BR_1( P_{A_2}) \quad \textnormal{and} \quad P_{A_2} \in \BR_2( P_{A_1}).
\end{IEEEeqnarray}
Given the fact that $|\set{P}| = 1$, then from Lemma \ref{Lemma:strictlymixed_unique} and from Theorem \ref{TheoNumberNE}, the pair of strategies $(P_{A_1}, P_{A_2}) $ forms the unique NE.

Finally, consider the case in which $u_{1,1} -u_{1,2} -u_{2,1}+u_{2,2} = 0$. 
Under the assumption that $u_{1,1} -u_{1,2} -u_{2,1}+u_{2,2} = 0$, it follows that 
\begin{IEEEeqnarray}{c}\label{EqWhen2}
u_{2,1} = u_{1,1} -u_{1,2} +u_{2,2}. 
\end{IEEEeqnarray}
If the inequalities in \eqref{EqVect1} and the equality in \eqref{EqWhen2} hold,
\begin{IEEEeqnarray}{c}
u_{2,1} > u_{2,2} \quad \textnormal{and} \quad u_{1,2} >u_{2,2}.  
\end{IEEEeqnarray}
Hence, from Lemma \ref{LemmaBR1} and Lemma \ref{LemmaBR2},  the pair of strategies $(P_{A_1}, P_{A_2}) \in \simplex{\set{A}_1} \times \simplex{\set{A}_1}$ such that 
\begin{IEEEeqnarray}{c}
P_{A_1}(a_1) = 1 \quad \textnormal{and} \quad P_{A_2}(a_1) = 0
\end{IEEEeqnarray}
satisfies 
\begin{IEEEeqnarray}{c}
P_{A_1} \in \BR_1( P_{A_2}) \quad \textnormal{and} \quad P_{A_2} \in \BR_2( P_{A_1})
\end{IEEEeqnarray}
and forms the unique NE. 
This completes the converse part.

Furthermore, the NE satisfying \eqref{EqVec2} implies that
\begin{IEEEeqnarray}{c}
u(P_{A_1}^\star, P_{A_2}^\star) = u_{1,2},
\end{IEEEeqnarray}
which completes the proof. 

The proof of Case III is as follows. 
If $\left( P_{A_1}^\star, P_{A_2}^\star \right)$ is the unique NE, from Definition \ref{Def:Equ_1}, for all $\alpha \in (0,1]$ and $\beta \in [0,1)$, it holds that
\begin{IEEEeqnarray}{rl}
&u(P_{A_1}^\star, P_{A_2}^\star)  = ( 0, 1) \matx{u} (1, 0)^{\sf T} = u_{2,1}> (\alpha, 1-\alpha) \matx{u} (1, 0)^{\sf T}  = \alpha u_{1,1} + (1-\alpha) u_{2,1}\quad \textnormal{and} \label{Equ:pen5} \IEEEeqnarraynumspace \\ 
&u(P_{A_1}^\star, P_{A_2}^\star)  = ( 0, 1) \matx{u} (1, 0)^{\sf T} = u_{2,1} < (0, 1) \matx{u} (\beta,1 - \beta)^{\sf T} = \beta u_{2,1} + (1-\beta)u_{2,2}. \label{Equ:pen6}
\end{IEEEeqnarray}
The strict inequality follows from the assumption that the NE in unique. 
Note that the inequalities in (\ref{Equ:pen5}) and (\ref{Equ:pen6}) are equivalent to 
\begin{IEEEeqnarray}{c}\label{Eqpen03}
u_{2,1} > u_{1,1} \quad \textnormal{and} \quad u_{2,2} > u_{2,1},
\end{IEEEeqnarray}
respectively. 
As a result, the inequalities in (\ref{Equ:pen5}) and (\ref{Equ:pen6}) are satisfied when 
\begin{IEEEeqnarray}{c}\label{EqVect2}
u_{2,2} > u_{2,1} > u_{1,1}, 
\end{IEEEeqnarray}
which is the condition in (\ref{Equ:pen03}). 

The converse is as follows. 
First, consider the case in which $u_{1,1} -u_{1,2} -u_{2,1}+u_{2,2} > 0$. 
Under the assumption that $u_{1,1} -u_{1,2} -u_{2,1}+u_{2,2} > 0$, if the inequalities in \eqref{EqVect2} hold, then it holds that 
\begin{IEEEeqnarray}{c}
u_{2,2} -u_{2,1} > 0 \quad \textnormal{and} \quad u_{1,1} -u_{2,1} < 0, 
\end{IEEEeqnarray}
which, from  \eqref{Equ:P_{A_1}}, implies that $p^{(1)}>1$. 
Furthermore, if the inequalities in \eqref{EqVect2} hold, then it holds that 
\begin{IEEEeqnarray}{c}
u_{1,1} -u_{2,1} < 0. 
\end{IEEEeqnarray}
Under the assumption that $u_{1,1} -u_{1,2} -u_{2,1}+u_{2,2} > 0$, it holds that $u_{2,2} -u_{1,2}> 0$, 
which, from  \eqref{Equ:P_{A_2}}, it holds that that $p^{(2)} >1$. 
Then from Lemma \ref{LemmaBR1} and Lemma \ref{LemmaBR2}, if $p^{(1)}>1$ and $p^{(2)}>1$, the pair of strategies $(P_{A_1}, P_{A_2}) \in \simplex{\set{A}_1} \times \simplex{\set{A}_1}$ such that 
\begin{IEEEeqnarray}{c}
P_{A_1}(a_1) = 0 \quad \textnormal{and} \quad P_{A_2}(a_1) = 1 
\end{IEEEeqnarray}
satisfies 
\begin{IEEEeqnarray}{c}
P_{A_1} \in \BR_1( P_{A_2}) \quad \textnormal{and} \quad P_{A_2} \in \BR_2( P_{A_1}).
\end{IEEEeqnarray}
Given the fact that $|\set{P}| = 1$, then from Lemma \ref{Lemma:strictlymixed_unique} and from Theorem \ref{TheoNumberNE}, the pair of strategies $(P_{A_1}, P_{A_2}) $ forms the unique NE. 

Then, consider the case in which $u_{1,1} -u_{1,2} -u_{2,1}+u_{2,2} < 0$. 
Under the assumption that $u_{1,1} -u_{1,2} -u_{2,1}+u_{2,2} < 0$, if the inequalities in \eqref{EqVect2} hold, then it holds that 
\begin{IEEEeqnarray}{c}
u_{2,2} -u_{2,1} > 0 \quad \textnormal{and} \quad u_{1,1} -u_{1,2} < 0, 
\end{IEEEeqnarray}
which, from  \eqref{Equ:P_{A_1}}, implies that $p^{(1)}<0$. 
Furthermore, if the inequalities in \eqref{EqVect2} hold, then it holds that 
\begin{IEEEeqnarray}{c}
u_{1,1} -u_{2,1} < 0. 
\end{IEEEeqnarray}
Under the assumption that $u_{1,1} -u_{1,2} -u_{2,1}+u_{2,2} < 0$, if $u_{2,2} -u_{1,2} < 0$, then from  \eqref{Equ:P_{A_2}}, it holds that that $ 0< p^{(2)} < 1$. 
And if $u_{2,2} -u_{1,2} \geq 0$, from  \eqref{Equ:P_{A_2}}, it holds that that $p^{(2)}\leq 0$. 
Hence, in either case, it holds that $p^{(2)}<1$. 
Then from Lemma \ref{LemmaBR1} and Lemma \ref{LemmaBR2}, if $p^{(1)}>1$ and $p^{(2)}<1$, the pair of strategies $(P_{A_1}, P_{A_2}) \in \simplex{\set{A}_1} \times \simplex{\set{A}_1}$ such that 
\begin{IEEEeqnarray}{c}
P_{A_1}(a_1) = 0 \quad \textnormal{and} \quad P_{A_2}(a_1) = 1 
\end{IEEEeqnarray}
satisfies 
\begin{IEEEeqnarray}{c}
P_{A_1} \in \BR_1( P_{A_2}) \quad \textnormal{and} \quad P_{A_2} \in \BR_2( P_{A_1}).
\end{IEEEeqnarray}
Given the fact that $|\set{P}| = 1$, then from Lemma \ref{Lemma:strictlymixed_unique} and from Theorem \ref{TheoNumberNE}, the pair of strategies $(P_{A_1}, P_{A_2}) $ forms the unique NE.

Finally, consider the case in which $u_{1,1} -u_{1,2} -u_{2,1}+u_{2,2} = 0$. 
Under the assumption that $u_{1,1} -u_{1,2} -u_{2,1}+u_{2,2} = 0$, it follows that 
\begin{IEEEeqnarray}{c}\label{EqWhen3}
u_{1,2} = u_{1,1} -u_{2,1} +u_{2,2}. 
\end{IEEEeqnarray}
If the inequalities in \eqref{EqVect3} and the equality in \eqref{EqWhen3} hold,
\begin{IEEEeqnarray}{c}
u_{2,1} < u_{2,2} \quad \textnormal{and} \quad u_{1,2} <u_{2,2}.  
\end{IEEEeqnarray}
Then from Lemma \eqref{LemmaBR1} and Lemma \eqref{LemmaBR2}, the pair of strategies $(P_{A_1}, P_{A_2}) \in \simplex{\set{A}_1} \times \simplex{\set{A}_1}$ such that 
\begin{IEEEeqnarray}{c}
P_{A_1}(a_1) = 0 \quad \textnormal{and} \quad P_{A_2}(a_1) = 1 
\end{IEEEeqnarray}
satisfies 
\begin{IEEEeqnarray}{c}
P_{A_1} \in \BR_1( P_{A_2}) \quad \textnormal{and} \quad P_{A_2} \in \BR_2( P_{A_1})
\end{IEEEeqnarray}
and forms the unique NE. 
This completes the converse part.

Furthermore, the NE satisfying \eqref{EqVec3} implies that
\begin{IEEEeqnarray}{c}
u(P_{A_1}^\star, P_{A_2}^\star) = u_{2,1},
\end{IEEEeqnarray}
which completes the proof. 

The proof of Case IV is as follows. 
If $\left( P_{A_1}^\star, P_{A_2}^\star \right)$ is the unique NE, from Definition \ref{Def:Equ_1}, for all $\alpha \in (0,1]$ and $\beta \in (0,1]$, it holds that 
\begin{IEEEeqnarray}{rl}
&u(P_{A_1}^\star, P_{A_2}^\star)  = ( 0, 1) \matx{u} (0,1)^{\sf T}  = u_{2,2}> (\alpha, 1-\alpha) \matx{u} (0, 1)^{\sf T} = \alpha u_{1,2} + (1-\alpha) u_{2,2}   \quad \textnormal{and} \label{Equ:pen7} \IEEEeqnarraynumspace \\ 
&u(P_{A_1}^\star, P_{A_2}^\star)  = ( 0, 1) \matx{u} (0, 1)^{\sf T}  = u_{2,2}< (0, 1) \matx{u} (\beta,1 - \beta)^{\sf T} = \beta u_{2,1} + (1-\beta) u_{2,2}, \label{Equ:pen8}
\end{IEEEeqnarray}
The strict inequality follows from the assumption that the NE in unique. 
Note that the inequalities in (\ref{Equ:pen7}) and (\ref{Equ:pen8}) are equivalent to 
\begin{IEEEeqnarray}{c}\label{Eqpen04}
u_{2,2} > u_{1,2} \quad \textnormal{and} \quad u_{2,1} > u_{2,2},
\end{IEEEeqnarray}
respectively. 
As a result, the inequalities in (\ref{Equ:pen7}) and (\ref{Equ:pen8}) are satisfied when
\begin{IEEEeqnarray}{c}\label{EqVect3} 
u_{2,1} > u_{2,2} > u_{1,2}, 
\end{IEEEeqnarray}
which is the condition in (\ref{Equ:pen04}). 

The converse is as follows. 
First, consider the case in which $u_{1,1} -u_{1,2} -u_{2,1}+u_{2,2} > 0$. 
Under the assumption that $u_{1,1} -u_{1,2} -u_{2,1}+u_{2,2} > 0$, if the inequalities in \eqref{EqVect3} hold, then it holds that 
\begin{IEEEeqnarray}{c}
u_{2,2} -u_{2,1} < 0 \quad \textnormal{and} \quad u_{1,1} -u_{1,2} > 0, 
\end{IEEEeqnarray}
which, from  \eqref{Equ:P_{A_1}}, implies that $p^{(1)}<0$. 
Furthermore, if the inequalities in \eqref{EqVect3} hold, then it holds that 
\begin{IEEEeqnarray}{c}
u_{2,2} -u_{1,2} > 0. 
\end{IEEEeqnarray}
Under the assumption that $u_{1,1} -u_{1,2} -u_{2,1}+u_{2,2} > 0$, if $u_{1,1} - u_{2,1} \geq 0$, from  \eqref{Equ:P_{A_2}}, it holds that $p^{(2)} \geq 1$. 
If $u_{1,1} - u_{2,1} < 0$, from  \eqref{Equ:P_{A_2}}, it holds that $p^{(2)} > 1$. 
Hence, in either case, it holds that $p^{(2)}\geq 1$. 
Then from Lemma \ref{LemmaBR1} and Lemma \ref{LemmaBR2}, if $p^{(1)}<0$ and $p^{(2)}\geq 1$, the pair of strategies $(P_{A_1}, P_{A_2}) \in \simplex{\set{A}_1} \times \simplex{\set{A}_1}$ such that 
\begin{IEEEeqnarray}{c}
P_{A_1}(a_1) = 0 \quad \textnormal{and} \quad P_{A_2}(a_1) = 0
\end{IEEEeqnarray}
satisfies 
\begin{IEEEeqnarray}{c}
P_{A_1} \in \BR_1( P_{A_2}) \quad \textnormal{and} \quad P_{A_2} \in \BR_2( P_{A_1}).
\end{IEEEeqnarray}
Given the fact that $|\set{P}| = 1$, then from Lemma \ref{Lemma:strictlymixed_unique} and from Theorem \ref{TheoNumberNE}, the pair of strategies $(P_{A_1}, P_{A_2}) $ forms the unique NE.

Then, consider the case in which $u_{1,1} -u_{1,2} -u_{2,1}+u_{2,2} < 0$. 
Under the assumption that $u_{1,1} -u_{1,2} -u_{2,1}+u_{2,2} < 0$, if the inequalities in \eqref{EqVect3} hold, then it holds that 
\begin{IEEEeqnarray}{c}
u_{2,2} -u_{1,2} > 0 \quad \textnormal{and} \quad u_{1,1} -u_{2,1} < 0, 
\end{IEEEeqnarray}
which, from  \eqref{Equ:P_{A_2}}, implies that $p^{(2)}<0$. 
Furthermore, if the inequalities in \eqref{EqVect3} hold, then it holds that 
\begin{IEEEeqnarray}{c}
u_{2,2} -u_{2,1} < 0. 
\end{IEEEeqnarray}
Under the assumption that $u_{1,1} -u_{1,2} -u_{2,1}+u_{2,2} < 0$, if $u_{1,1} -u_{1,2} \leq 0$, from  \eqref{Equ:P_{A_1}}, it holds that $0<p^{(1)}\leq 1$. 
If  $u_{1,1} -u_{1,2} > 0$, from  \eqref{Equ:P_{A_1}}, it holds that $p^{(1)}>1$. 
Hence, in either case, it holds that $p^{(1)}>0$. 
Then from Lemma \ref{LemmaBR1} and Lemma \ref{LemmaBR2}, if $p^{(2)}<0$ and $p^{(1)}>0$,  the pair of strategies $(P_{A_1}, P_{A_2}) \in \simplex{\set{A}_1} \times \simplex{\set{A}_1}$ such that 
\begin{IEEEeqnarray}{c}
P_{A_1}(a_1) = 0 \quad \textnormal{and} \quad P_{A_2}(a_1) = 0
\end{IEEEeqnarray}
satisfies 
\begin{IEEEeqnarray}{c}
P_{A_1} \in \BR_1( P_{A_2}^{\star}) \quad \textnormal{and} \quad P_{A_2} \in \BR_2( P_{A_1}^{\star}).
\end{IEEEeqnarray}
Given the fact that $|\set{P}| = 1$, then from Lemma \ref{Lemma:strictlymixed_unique} and from Theorem \ref{TheoNumberNE}, the pair of strategies $(P_{A_1}, P_{A_2}) $ forms the unique NE.

Finally, consider the case in which $u_{1,1} -u_{1,2} -u_{2,1}+u_{2,2} = 0$. 
Under the assumption that $u_{1,1} -u_{1,2} -u_{2,1}+u_{2,2} = 0$, it follows that 
\begin{IEEEeqnarray}{c}\label{EqWhen4}
u_{1,2} = u_{1,1} -u_{2,1} +u_{2,2}. 
\end{IEEEeqnarray}
If the inequalities in \eqref{EqVect3} and the equality in \eqref{EqWhen4} hold,
\begin{IEEEeqnarray}{c}
u_{1,2} < u_{2,2} \quad \textnormal{and} \quad u_{2,1} >u_{2,2}.  
\end{IEEEeqnarray}
Then from Lemma \ref{LemmaBR1} and Lemma \ref{LemmaBR2},  the pair of strategies $(P_{A_1}, P_{A_2}) \in \simplex{\set{A}_1} \times \simplex{\set{A}_1}$ such that 
\begin{IEEEeqnarray}{c}
P_{A_1}(a_1) = 0 \quad \textnormal{and} \quad P_{A_2}(a_1) = 0
\end{IEEEeqnarray}
satisfies 
\begin{IEEEeqnarray}{c}
P_{A_1} \in \BR_1( P_{A_2}) \quad \textnormal{and} \quad P_{A_2} \in \BR_2( P_{A_1})
\end{IEEEeqnarray}
and forms the unique NE. This completes the converse part. 

Furthermore, the NE satisfying \eqref{EqVec4} implies that
\begin{IEEEeqnarray}{c}
u(P_{A_1}^\star, P_{A_2}^\star) = u_{2,2}, 
\end{IEEEeqnarray}
which completes the proof. 

This completes the whole proof. 

\section{Proof of Theorem \ref{TheInfNE1}}\label{Sec:App7}
The proof is divided into two parts. 
The first part considers the case in which 
\begin{IEEEeqnarray}{c}
\set{P} = \left\{ \left( P_{1}, P_{2}\right) \in \simplex{\set{A}_1} \times \simplex{\set{A}_2}: P_{1} \in  \simplex{\set{A}_1} \ \textnormal{and} \ P_{2} \in \simplex{\set{A}_2} \right\}. 
\end{IEEEeqnarray}
The second part considers the case in which
\begin{IEEEeqnarray}{l}
\set{P} = \left\{ \left( P_{1}, P_{2}\right) \in \simplex{\set{A}_1} \times \simplex{\set{A}_2}: P_{1}(a_1) \in  \{0,1\} \ \textnormal{and} \ P_{2} \in \simplex{\set{A}_2} \right\} \quad \textnormal{or} \\
\set{P} = \left\{ \left( P_{1}, P_{2}\right) \in \simplex{\set{A}_1} \times \simplex{\set{A}_2}: P_{1} \in  \simplex{\set{A}_1} \ \textnormal{and} \ P_{2}(a_1)\in  \{ 0,1\}\right\}. 
\end{IEEEeqnarray}

The first part is as follows. 
If the pair of strategies $(P_{A_1}^\star, P_{A_2}^\star) \in \simplex{\set{A}_1} \times \simplex{\set{A}_2}$ satisfies 
\begin{IEEEeqnarray}{c}
P_{A_1}^\star(a_{1})  =  1 - P_{A_1}^\star(a_{2})  = \alpha \quad \textnormal{and} \quad 
P_{A_2}^\star(a_{1})    =  1 - P_{A_2}^\star(a_{2}) = \beta
\end{IEEEeqnarray} 
with $\alpha \in [0,1]$ and $\beta \in [0,1]$. 
From Definition \ref{Def:Equ_1}, it holds that for all strategies $Q_{A_1}\in \Delta(\set{A}_1)$, with $Q_{A_1}(a_1) = \alpha' \in [0,1]$, the following inequality holds: 
\begin{IEEEeqnarray}{rl}
&u(P_{A_1}^\star, P_{A_2}^\star) = ( \alpha, 1 -\alpha) \matx{u} ( \beta, 1- \beta)^{\sf T} \geq ( \alpha', 1 -\alpha') \matx{u} (\beta, 1 - \beta)^{\sf T} = (Q_{A_1}, P_{A_2}^\star), 
\end{IEEEeqnarray}
which is equivalent to 
\begin{IEEEeqnarray}{rl}
&\alpha \beta u_{1,1} + \alpha(1-\beta)u_{1,2} + (1-\alpha) \beta u_{2,1} + (1-\alpha)(1-\beta)u_{2,2} \IEEEnonumber \\
& \qquad  \qquad  \qquad \geq \alpha' \beta u_{1,1} + \alpha'(1-\beta)u_{1,2} + (1-\alpha') \beta u_{2,1} + (1-\alpha')(1-\beta)u_{2,2}. 
\end{IEEEeqnarray}
The above inequality can be simplified as 
\begin{IEEEeqnarray}{c}
(\alpha - \alpha') \beta u_{1,1} + (\alpha - \alpha')(1-\beta)u_{1,2} + (\alpha' - \alpha) \beta u_{2,1} +  (\alpha' - \alpha)(1- \beta)u_{2,2} \geq 0, 
\end{IEEEeqnarray}
which is equivalent to 
\begin{IEEEeqnarray}{ll}
0 &\leq (\alpha - \alpha') \beta (u_{1,1} - u_{2,1}) + (\alpha - \alpha')(1-\beta)(u_{1,2} - u_{2,2})    \\
& =  (\alpha - \alpha') \bigl(\beta (u_{1,1} - u_{2,1}) + (1-\beta)(u_{1,2} - u_{2,2})\bigr) 
\end{IEEEeqnarray}
As a result, the above inequality holds for all $\alpha' \in [0,1]$, $\alpha \in [0,1]$, and $\beta \in [0,1]$, if and only if it holds that
\begin{IEEEeqnarray}{c}\label{Equ:filter_1}
u_{1,1} = u_{2,1} \quad \textnormal{and} \quad u_{1,2} = u_{2,2}. 
\end{IEEEeqnarray}

Similarly, from Definition \ref{Def:Equ_1}, it holds that for all strategies $Q_{A_2}^\star \in \Delta(\set{A}_2)$ with $Q_{A_2}(a_1) = \beta' \in [0,1]$, the following inequality holds: 
\begin{IEEEeqnarray}{rl}
&u(P_{A_1}^\star, P_{A_2}^\star) = ( \alpha, 1 -\alpha) \matx{u} ( \beta, 1- \beta)^{\sf T} \leq ( \alpha, 1 -\alpha) \matx{u} (\beta', 1 - \beta')^{\sf T} = u(P_{A_1}^\star, Q_{A_2}), 
\end{IEEEeqnarray}
which is equivalent to 
\begin{IEEEeqnarray}{rl}
&\alpha \beta u_{1,1} + \alpha(1-\beta)u_{1,2} + (1-\alpha) \beta u_{2,1} + (1-\alpha)(1-\beta)u_{2,2} \IEEEnonumber \\
& \qquad  \qquad  \qquad \leq \alpha \beta' u_{1,1} + \alpha(1-\beta')u_{1,2} + (1-\alpha) \beta' u_{2,1} + (1-\alpha)(1-\beta')u_{2,2}. 
\end{IEEEeqnarray}
The above inequality can be simplified as 
\begin{IEEEeqnarray}{c}
(\beta- \beta') \alpha u_{1,1} + (\beta' - \beta) \alpha u_{1,2} + (1 - \alpha) (\beta -\beta') u_{2,1} +  (\beta' - \beta)(1- \alpha)u_{2,2} \leq 0, 
\end{IEEEeqnarray}
which is equivalent to 
\begin{IEEEeqnarray}{cl}
0 & \geq (\beta- \beta') \alpha (u_{1,1} - u_{1,2}) + (1 - \alpha) (\beta -\beta') (u_{2,1} -u_{2,2})  \\
& = (\beta- \beta') \bigl(\alpha (u_{1,1} - u_{1,2}) + (1 - \alpha)  (u_{2,1} -u_{2,2})\bigr) 
\end{IEEEeqnarray}
As a result, the above inequality holds for all  $\alpha \in [0,1]$, $\beta \in [0,1]$, and $\beta' \in [0,1]$,  if and only if it holds that
\begin{IEEEeqnarray}{c}\label{Equ:filter_2}
u_{1,1} = u_{1,2} \quad \textnormal{and} \quad u_{2,1} = u_{2,2}. 
\end{IEEEeqnarray}

Combining the inequalities in \eqref{Equ:filter_1} and \eqref{Equ:filter_2} yields
\begin{IEEEeqnarray}{c}\label{Equ:fig_5}
u_{1,1} = u_{1,2} = u_{2,1} = u_{2,2},
\end{IEEEeqnarray}
which is the condition in \eqref{Equ:mouse05}. 

The converse follows trivially from the fact that if the equalities in \eqref{Equ:fig_5} holds then it holds all pairs of strategies $(P_{A_1}, P_{A_2}) \in \simplex{\set{A}_1} \times \simplex{\set{A}_2}$ are NEs.

This completes the first part of the proof. 

The second part is as follows. 
Given the fact the game $\game{G}$ in \eqref{EqTheGame} is a $2 \times 2$ game, there are at most four cases when the strategy of only one of the players is pure. 
The four cases are: 

\noindent Case I: the pair of strategies $(P_{A_1}^\star, P_{A_2}^\star)$ satisfies 
\begin{IEEEeqnarray}{c}\label{Equ:fig_1}
P_{A_1}^\star(a_{1})  =  1 - P_{A_1}^\star(a_{2})  = 1 \quad \textnormal{and} \quad 
P_{A_2}^\star(a_{1})    =  1 - P_{A_2}^\star(a_{2}) = \beta
\end{IEEEeqnarray} 
with $\beta \in [0,1]$.

\noindent Case II: the pair of strategies $(P_{A_1}^\star, P_{A_2}^\star)$ satisfies 
\begin{IEEEeqnarray}{c}\label{Equ:fig_2}
P_{A_1}^\star(a_{1})  =  1 - P_{A_1}^\star(a_{2})  = 0 \quad \textnormal{and} \quad 
P_{A_2}^\star(a_{1})    =  1 - P_{A_2}^\star(a_{2}) = \beta
\end{IEEEeqnarray} 
with $\beta \in [0,1]$.

\noindent Case III: the pair of strategies $(P_{A_1}^\star, P_{A_2}^\star)$ satisfies 
\begin{IEEEeqnarray}{c}\label{Equ:fig_3}
P_{A_1}^\star(a_{1})  =  1 - P_{A_1}^\star(a_{2})  = \alpha \quad \textnormal{and} \quad 
P_{A_2}^\star(a_{1})    =  1 - P_{A_2}^\star(a_{2}) = 1
\end{IEEEeqnarray} 
with $\alpha \in [0,1]$. 

\noindent Case IV:  the pair of strategies $(P_{A_1}^\star, P_{A_2}^\star)$ satisfies 
\begin{IEEEeqnarray}{c}\label{Equ:fig_4}
P_{A_1}^\star(a_{1})  =  1 - P_{A_1}^\star(a_{2})  = \alpha \quad \textnormal{and} \quad 
P_{A_2}^\star(a_{1})    =  1 - P_{A_2}^\star(a_{2}) = 0
\end{IEEEeqnarray} 
with $\alpha \in [0,1]$. 

The proof of Case I is as follows. 
If the pair of strategies $(P_{A_1}^\star, P_{A_2}^\star)$ satisfies \eqref{Equ:fig_1}, 
from Definition \ref{Def:Equ_1}, for all strategies $P_{A_2}' \in \Delta(\set{A}_2)$ such that $P_{A_2}'(a_1) = \beta' $, with $\beta' \in [0,1]$, it holds that 
\begin{IEEEeqnarray}{ll}
u(P_{A_1}^\star, P_{A_2}^\star) = ( 1, 0) \matx{u} (\beta, 1-\beta)^{\sf T} &= \beta u_{1,1} + (1-\beta)u_{1,2} \IEEEnonumber \squeezeequ \\
& \leq \beta' u_{1,1} + (1-\beta')u_{1,2}  = (1, 0) \matx{u} (\beta', 1-\beta')^{\sf T} = u(P_{A_1}^\star, P_{A_2}'), \squeezeequ \IEEEeqnarraynumspace
\end{IEEEeqnarray}
which implies that, for all $\beta' \in [0,1]$, it holds that 
\begin{IEEEeqnarray}{rl}\label{Equ:temP_{A_1}}
&( \beta' -\beta) u_{1,1} \geq ( \beta' -\beta) u_{1,2} . 
\end{IEEEeqnarray}
The inequality in \eqref{Equ:temP_{A_1}} can be written as
\begin{IEEEeqnarray}{rl}\label{EqVec6}
&( \beta' -\beta) (u_{1,1}-u_{1,2}) \geq 0. 
\end{IEEEeqnarray}
As a result, the above inequality holds if 
\begin{IEEEeqnarray}{c}\label{Equ:mouse1}
u_{1,1} = u_{1,2}. 
\end{IEEEeqnarray}

Moreover, for all strategies $P_{A_1}' \in \Delta(\set{A}_1)$ such that $P_{A_1}'(a_1)= \alpha'$, with $ \alpha' \in [0,1]$, it holds that 
\begin{IEEEeqnarray}{rl}\label{Equ:mouse2}
&u(P_{A_1}^\star, P_{A_2}^\star) = ( 1, 0) \matx{u} (\beta, 1-\beta)^{\sf T} \geq (\alpha', 1-\alpha') \matx{u} (\beta, 1-\beta)^{\sf T} = u(P_{A_1}', P_{A_2}^\star), 
\end{IEEEeqnarray}
which is equivalent to 
\begin{IEEEeqnarray}{c}
u_{1,1}\beta + u_{1,2}(1-\beta) \geq \alpha'\beta u_{1,1} + \alpha'(1-\beta) u_{1,2} + (1-\alpha')\beta u_{2,1} + (1-\alpha')(1-\beta) u_{2,2}. 
\end{IEEEeqnarray}
From (\ref{Equ:mouse1}), the above inequality is simplified to 
\begin{IEEEeqnarray}{c}
u_{1,1} = u_{1,2} \geq \alpha' u_{1,1} + (1-\alpha')\beta u_{2,1} + (1-\alpha')(1-\beta) u_{2,2},
\end{IEEEeqnarray}
which is equivalent to 
\begin{IEEEeqnarray}{c}\label{Equ:mouse2_1}
u_{1,1} = u_{1,2}\geq  \beta u_{2,1} + (1-\beta) u_{2,2}.
\end{IEEEeqnarray}

If $u_{2,1} = u_{2,2}$, then for all $\beta \in [0,1]$, the inequality in (\ref{Equ:mouse2_1}) implies that 
\begin{IEEEeqnarray}{c}\label{Equ:cotton_1}
u_{1,1} = u_{1,2} \geq u_{2,1} = u_{2,2}.
\end{IEEEeqnarray}
However, if the inequality in \eqref{Equ:cotton_1} holds with equality, i.e. $u_{1,1} = u_{1,2} = u_{2,1} = u_{2,2}$, then all strategies $P_{A_1}'' \in \Delta(\set{A}_1)$, together with $P_{A_2}^\star$ in (\ref{Equ:fig_1}), form an NE, see \eqref{Equ:fig_5}. 
This contradicts the assumption that $P_{A_1}^\star(a_1) = 1$. 
As a result,  the inequality in \eqref{Equ:cotton_1} is considered with strict inequality. That is, 
\begin{IEEEeqnarray}{c}\label{Equ:cotton_2}
u_{1,1} = u_{1,2} > u_{2,1} = u_{2,2}.
\end{IEEEeqnarray}
If $u_{2,1} \neq u_{2,2}$, for all $\beta \in [0,1]$, the inequality in \eqref{Equ:mouse2_1} implies that 
\begin{IEEEeqnarray}{c}\label{Equ:cotton_3}
u_{1,1} = u_{1,2} \geq \textnormal{max} \{ u_{2,1}, u_{2,2}\} > \textnormal{min} \{ u_{2,1}, u_{2,2}\}.
\end{IEEEeqnarray}

Combining the inequalities in \eqref{Equ:cotton_2} and \eqref{Equ:cotton_3} yields the conditions in \eqref{Equ:mouse01} and \eqref{Equ:mouse011}.

The converse is as follows. 
If the conditions  in \eqref{Equ:mouse01} hold, for all $P_{A_2} \in \simplex{\set{A}_2}$ with $P_{A_2}(a_1) = \beta \in [0,1]$, it holds that 
\begin{IEEEeqnarray}{c}
(1,0) \matx{u} (\beta, 1-\beta)^{\sf T} > (0,1) \matx{u} (\beta, 1-\beta)^{\sf T},
\end{IEEEeqnarray}
which implies that, for \Pone, the action $a_2$ is dominated by action $a_1$. 
Furthermore, if the conditions in \eqref{Equ:mouse011} hold, for all $P_{A_2} \in \simplex{\set{A}_2}$ with $P_{A_2}(a_1) = \beta \in [0,1]$, it holds that 
\begin{IEEEeqnarray}{c}
(1,0) \matx{u} (\beta, 1-\beta)^{\sf T} \geq (0,1) \matx{u} (\beta, 1-\beta)^{\sf T}; 
\end{IEEEeqnarray}
and for all $P_{A_2} \in \simplex{\set{A}_2}$ with $P_{A_2}(a_1) = \beta \in (0,1)$, it holds that 
\begin{IEEEeqnarray}{c}
(1,0) \matx{u} (\beta, 1-\beta)^{\sf T} > (0,1) \matx{u} (\beta, 1-\beta)^{\sf T}. 
\end{IEEEeqnarray}
Hence, if the conditions in \eqref{Equ:mouse011} hold,  for \Pone, the action $a_2$ is dominated by action $a_1$. 
As a result, it holds that $P_{A_1}^\star(a_1) = 1$. 
Given the fact that $P_{A_1}^\star(a_1) = 1$, if either the conditions in \eqref{Equ:mouse01} or the conditions in  \eqref{Equ:mouse011} hold, \Ptwo is indifferent between action $a_1$ and action $a_2$. 
This implies that 
\begin{IEEEeqnarray}{c}
P_{A_2}^{\star}(a_1) = \beta, 
\end{IEEEeqnarray}
with $\beta \in [0,1]$, which completes the proof of the converse part. 

This completes the proof of Case I.

The proof of Case II is as follows. 
If the pair of strategies $(P_{A_1}^\star, P_{A_2}^\star)$ satisfies \eqref{Equ:fig_2}, 
from Definition \ref{Def:Equ_1}, for all strategies $P_{A_2}' \in \Delta(\set{A}_2)$ such that $P_{A_2}'(a_1) = \beta' $, with $\beta' \in [0,1]$, it holds that 
\begin{IEEEeqnarray}{ll}
u(P_{A_1}^\star, P_{A_2}^\star) = (0, 1) \matx{u} (\beta, 1-\beta)^{\sf T} &= \beta u_{2,1} + (1-\beta)u_{2,2} \IEEEnonumber \\
& \leq \beta' u_{2,1} + (1-\beta')u_{2,2}  = (0, 1) \matx{u} (\beta', 1-\beta')^{\sf T} = u(P_{A_1}^\star, P_{A_2}'), \IEEEeqnarraynumspace \squeezeequ
\end{IEEEeqnarray}
which implies that, for all $\beta' \in [0,1]$, it holds that 
\begin{IEEEeqnarray}{rl}\label{Equ:temP_{A_2}}
&( \beta' -\beta) u_{2,1} \geq ( \beta' -\beta) u_{2,2}. 
\end{IEEEeqnarray}
The inequality in \eqref{Equ:temP_{A_2}} can be written as
\begin{IEEEeqnarray}{rl}\label{EqVec7}
& ( \beta' -\beta) (u_{2,1} -  u_{2,2}) \geq 0. 
\end{IEEEeqnarray}
As a result, the above inequality holds if it holds that 
\begin{IEEEeqnarray}{c}\label{Equ:mouse3}
u_{2,1} = u_{2,2}. 
\end{IEEEeqnarray}

Moreover, for all strategies $P_{A_1}' \in \Delta(\set{A}_1)$ such that $P_{A_1}'(a_1)= \alpha'$, with $\alpha' \in [0,1]$, it holds that 
\begin{IEEEeqnarray}{rl}\label{Equ:mouse4}
&u(P_{A_1}^\star, P_{A_2}^\star) = ( 0, 1) \matx{u} (\beta, 1-\beta)^{\sf T} \geq (\alpha', 1-\alpha') \matx{u} (\beta, 1-\beta)^{\sf T} = u(P_{A_1}', P_{A_2}^\star), 
\end{IEEEeqnarray}
which is equivalent to 
\begin{IEEEeqnarray}{c}
u_{2,1}\beta + u_{2,2}(1-\beta) \geq \alpha'\beta u_{1,1} + \alpha'(1-\beta) u_{1,2} + (1-\alpha')\beta u_{2,1} + (1-\alpha')(1-\beta) u_{2,2}. 
\end{IEEEeqnarray}
From (\ref{Equ:mouse3}), the above inequality is simplified to 
\begin{IEEEeqnarray}{c}
u_{2,1} = u_{2,2} \geq \alpha' \beta u_{1,1} + \alpha'(1-\beta) u_{1,2} + (1-\alpha') u_{2,2},
\end{IEEEeqnarray}
which is equivalent to 
\begin{IEEEeqnarray}{c}\label{Equ:mouse4_1}
u_{2,1} = u_{2,2}\geq  \beta u_{1,1} + (1-\beta) u_{1,2}.
\end{IEEEeqnarray}

If $u_{1,1} = u_{1,2}$, then for all $\beta \in [0,1]$, the inequality in (\ref{Equ:mouse4_1}) implies that 
\begin{IEEEeqnarray}{c}\label{Equ:cotton_30}
u_{2,1} = u_{2,2} \geq u_{1,1} = u_{1,2}.
\end{IEEEeqnarray}
However, if the inequality in \eqref{Equ:cotton_30} holds with equality, i.e.  $u_{1,1} = u_{1,2} = u_{2,1} = u_{2,2}$, then all strategies $P_{A_1}'' \in \Delta(\set{A}_1)$, together with $P_{A_2}^\star$ in (\ref{Equ:fig_2}), form an NE, see \eqref{Equ:fig_5}. 
This contradicts the assumption that $P_{A_1}^\star(a_1) = 0$.
As a result, the inequality in \eqref{Equ:cotton_30} is considered with strict inequality. That is, 
\begin{IEEEeqnarray}{c}\label{Equ:cotton_4}
u_{2,1} = u_{2,2} > u_{1,1} = u_{1,2}.
\end{IEEEeqnarray}
If $u_{1,1} \neq u_{1,2}$, for all $\beta \in [0,1]$, the inequality in \eqref{Equ:mouse4_1} implies that 
\begin{IEEEeqnarray}{c}\label{Equ:cotton_50}
u_{2,1} = u_{2,2} \geq \textnormal{max} \{ u_{1,1}, u_{1,2}\} > \textnormal{min} \{ u_{1,1}, u_{1,2}\}.
\end{IEEEeqnarray}

Combining the inequalities in \eqref{Equ:cotton_4} and \eqref{Equ:cotton_50} yields the conditions in \eqref{Equ:mouse02} and \eqref{Equ:mouse021}. 

The converse is as follows. 
If the conditions  in \eqref{Equ:mouse02} hold, for all $P_{A_2} \in \simplex{\set{A}_2}$ with $P_{A_2}(a_1) = \beta \in [0,1]$, it holds that 
\begin{IEEEeqnarray}{c}
(0,1) \matx{u} (\beta, 1-\beta)^{\sf T} > (1,0) \matx{u} (\beta, 1-\beta)^{\sf T},
\end{IEEEeqnarray}
which implies that, for \Pone, the action $a_1$ is dominated by action $a_2$. 
Furthermore, if the conditions in \eqref{Equ:mouse021} hold, for all $P_{A_2} \in \simplex{\set{A}_2}$ with $P_{A_2}(a_1) = \beta \in [0,1]$, it holds that 
\begin{IEEEeqnarray}{c}
(0,1) \matx{u} (\beta, 1-\beta)^{\sf T} \geq (1,0) \matx{u} (\beta, 1-\beta)^{\sf T}; 
\end{IEEEeqnarray}
and for all $P_{A_2} \in \simplex{\set{A}_2}$ with $P_{A_2}(a_1) = \beta \in (0,1)$, it holds that 
\begin{IEEEeqnarray}{c}
(0,1) \matx{u} (\beta, 1-\beta)^{\sf T} > (1,0) \matx{u} (\beta, 1-\beta)^{\sf T}. 
\end{IEEEeqnarray}
Hence, if the conditions in \eqref{Equ:mouse021} hold,  for \Pone, the action $a_1$ is dominated by action $a_2$. 
As a result, it holds that $P_{A_1}^\star(a_1) = 0$. 
Given the fact that $P_{A_1}^\star(a_1) = 0$, if either the conditions in \eqref{Equ:mouse02} or the conditions in  \eqref{Equ:mouse021} hold, \Ptwo is indifferent between action $a_1$ and action $a_2$. 
This implies that 
\begin{IEEEeqnarray}{c}
P_{A_2}^{\star}(a_1) = \beta, 
\end{IEEEeqnarray}
with $\beta \in [0,1]$, which completes the proof of the converse part. 

This completes the proof of Case II.

The proof of Case III is as follows. 
If the pair of strategies $(P_{A_1}^\star, P_{A_2}^\star)$ satisfies \eqref{Equ:fig_3}, 
From Definition \ref{Def:Equ_1}, for all strategies $P_{A_1}' \in \Delta(\set{A}_1)$ such that $P_{A_1}'(a_1) = \alpha' $, with $\alpha' \in [0,1]$, it holds that 
\begin{IEEEeqnarray}{ll}
u(P_{A_1}^\star, P_{A_2}^\star) = (\alpha, 1-\alpha) \matx{u} (1, 0)^{\sf T}& = \alpha u_{1,1} + (1-\alpha)u_{2,1} \IEEEnonumber \\
&  \geq \alpha' u_{1,1} + (1-\alpha')u_{2,1}  = (\alpha', 1 - \alpha') \matx{u} (1, 0)^{\sf T} = u(P_{A_1}', P_{A_2}^\star), \IEEEeqnarraynumspace \squeezeequ
\end{IEEEeqnarray}
which implies that, for all $\alpha' \in [0,1]$, it holds that 
\begin{IEEEeqnarray}{rl}\label{Equ:temp_3}
&( \alpha -\alpha') u_{1,1} \geq ( \alpha -\alpha') u_{2,1}. 
\end{IEEEeqnarray}
The inequality in \eqref{Equ:temp_3} can be written as
\begin{IEEEeqnarray}{rl}\label{EqVec8}
& ( \alpha -\alpha') (u_{1,1} -  u_{2,1}) \geq 0. 
\end{IEEEeqnarray}
As a result, the above inequality holds if it holds that 
\begin{IEEEeqnarray}{c}\label{Equ:mouse5}
u_{1,1} = u_{2,1}. 
\end{IEEEeqnarray}

Moreover, for all strategies $P_{A_2}' \in \Delta(\set{A}_2)$ such that $P_{A_2}'(a_1)= \beta'$, with $\beta' \in [0,1]$, it holds that 
\begin{IEEEeqnarray}{rl}\label{Equ:mouse6}
&u(P_{A_1}^\star, P_{A_2}^\star) = ( \alpha, 1 - \alpha) \matx{u} (1, 0)^{\sf T} \leq (\alpha, 1-\alpha) \matx{u} (\beta', 1-\beta')^{\sf T} = u(P_{A_1}^\star, P_{A_2}'), 
\end{IEEEeqnarray}
which is equivalent to 
\begin{IEEEeqnarray}{c}
u_{1,1}\alpha + u_{2,1}(1-\alpha) \leq \alpha\beta' u_{1,1} + \alpha(1-\beta') u_{1,2} + (1-\alpha)\beta' u_{2,1} + (1-\alpha)(1-\beta') u_{2,2}. 
\end{IEEEeqnarray}
From (\ref{Equ:mouse5}), the above inequality is simplified to 
\begin{IEEEeqnarray}{c}
u_{1,1} = u_{2,1} \leq \beta' u_{2,1} + \alpha(1-\beta') u_{1,2} + (1-\alpha)(1-\beta') u_{2,2},
\end{IEEEeqnarray}
which is equivalent to 
\begin{IEEEeqnarray}{c}\label{Equ:mouse6_1}
u_{1,1} = u_{2,1}\leq  \alpha u_{1,2} + (1-\alpha) u_{2,2}.
\end{IEEEeqnarray}

If $u_{1,2} = u_{2,2}$, then for all $\alpha \in [0,1]$, the inequality in (\ref{Equ:mouse6_1}) implies that 
\begin{IEEEeqnarray}{c}\label{Equ:cotton_5}
u_{1, 1} = u_{2,1} \leq u_{1,2} = u_{2,2}.
\end{IEEEeqnarray}
However, if the inequality in \eqref{Equ:cotton_5} holds with equality, i.e.  $u_{1,1} = u_{1,2} = u_{2,1} = u_{2,2}$, all strategies $P_{A_2}'' \in \Delta(\set{A}_2)$, together with $P_{A_1}^\star$ in (\ref{Equ:fig_3}), form an NE, see \eqref{Equ:fig_5}. 
This contradicts the assumption that $P_{A_2}^\star(a_1) =1$. 
As a result, the inequality in \eqref{Equ:cotton_5} is considered with strict inequality. That is, 
\begin{IEEEeqnarray}{c}\label{Equ:cotton_6}
u_{1,1} = u_{2,1} < u_{1,2} = u_{2,2}.
\end{IEEEeqnarray}
If $u_{1,2} \neq u_{2,2}$, for all $\alpha \in [0,1]$, the inequality in \eqref{Equ:mouse6_1} implies that 
\begin{IEEEeqnarray}{c}\label{Equ:cotton_7}
u_{1,1} = u_{2,1} \leq \textnormal{min} \{ u_{1,2}, u_{2,2}\} < \textnormal{max} \{ u_{1,2}, u_{2,2}\}.
\end{IEEEeqnarray}

Combining the inequalities in \eqref{Equ:cotton_6} and \eqref{Equ:cotton_7} yields the conditions in \eqref{Equ:mouse03} and \eqref{Equ:mouse031}.

The converse is as follows. 
If the conditions  in \eqref{Equ:mouse03} hold, for all $P_{A_1} \in \simplex{\set{A}_1}$ with $P_{A_1}(a_1) = \alpha \in [0,1]$, it holds that 
\begin{IEEEeqnarray}{c}
(\alpha, 1-\alpha) \matx{u} (1,0)^{\sf T} < (\alpha, 1-\alpha) \matx{u} (0, 1)^{\sf T},
\end{IEEEeqnarray}
which implies that, for \Ptwo, the action $a_2$ is dominated by action $a_1$. 
Furthermore, if the conditions in \eqref{Equ:mouse031} hold, for all $P_{A_1} \in \simplex{\set{A}_1}$ with $P_{A_1}(a_1) = \alpha \in [0,1]$, it holds that 
\begin{IEEEeqnarray}{c}
(\alpha, 1-\alpha) \matx{u} (1,0)^{\sf T} \leq (\alpha, 1-\alpha) \matx{u} (0, 1)^{\sf T};
\end{IEEEeqnarray}
and for all $P_{A_1} \in \simplex{\set{A}_1}$ with $P_{A_1}(a_1) =\alpha \in [0,1]$, it holds that 
\begin{IEEEeqnarray}{c}
(\alpha, 1-\alpha) \matx{u} (1,0)^{\sf T} < (\alpha, 1-\alpha) \matx{u} (0, 1)^{\sf T}.
\end{IEEEeqnarray}
Hence, if the conditions in \eqref{Equ:mouse021} hold,  for \Ptwo, the action $a_2$ is dominated by action $a_1$. 
As a result, it holds that $P_{A_2}^\star(a_1) = 1$. 
Given the fact that $P_{A_2}^\star(a_1) = 1$, if either the conditions in \eqref{Equ:mouse03} or the conditions in  \eqref{Equ:mouse031} hold, \Pone is indifferent between action $a_1$ and action $a_2$. 
This implies that 
\begin{IEEEeqnarray}{c}
P_{A_1}^{\star}(a_1) = \beta, 
\end{IEEEeqnarray}
with $\beta \in [0,1]$, which completes the proof of the converse part. 

This completes the proof of Case III.

The proof of Case IV is as follows. 
If the pair of strategies $(P_{A_1}^\star, P_{A_2}^\star)$ satisfies \eqref{Equ:fig_4}, 
from Definition \ref{Def:Equ_1}, for all strategies $P_{A_1}' \in \Delta(\set{A}_1)$ such that $P_{A_1}'(a_1) = \alpha' $, with $\alpha' \in [0,1]$, it holds that 
\begin{IEEEeqnarray}{ll}
u(P_{A_1}^\star, P_{A_2}^\star) = (\alpha, 1-\alpha) \matx{u} (0, 1)^{\sf T} &= \alpha u_{1,2} + (1-\alpha)u_{2,2} \IEEEnonumber \\
& \geq \alpha' u_{1,2} + (1-\alpha')u_{2,2}  = (\alpha', 1 - \alpha') \matx{u} (0, 1)^{\sf T} = u(P_{A_1}', P_{A_2}^\star), \IEEEeqnarraynumspace \squeezeequ
\end{IEEEeqnarray}
which implies that, for all $\alpha' \in [0,1]$, it holds that 
\begin{IEEEeqnarray}{rl}\label{Equ:temp_4}
&( \alpha -\alpha') u_{1,2} \geq ( \alpha -\alpha') u_{2,2} 
\end{IEEEeqnarray}
The inequality in \eqref{Equ:temp_4} can be written as 
\begin{IEEEeqnarray}{rl}\label{EqVec9}
& ( \alpha -\alpha') (u_{1,2} -  u_{2,2}) \geq 0. 
\end{IEEEeqnarray}
As a result, the above inequality holds if it holds that 
\begin{IEEEeqnarray}{c}\label{Equ:mouse8}
u_{1,2} = u_{2,2}. 
\end{IEEEeqnarray}

Moreover, for all strategies $P_{A_2}' \in \Delta(\set{A}_2)$ such that $P_{A_2}'(a_1)= \beta'$, with $\beta' \in [0,1]$, it holds that 
\begin{IEEEeqnarray}{rl}\label{Equ:mouse9}
&u(P_{A_1}^\star, P_{A_2}^\star) = ( \alpha, 1 - \alpha) \matx{u} (0, 1)^{\sf T} \leq (\alpha, 1-\alpha) \matx{u} (\beta', 1-\beta')^{\sf T} = u(P_{A_1}^\star, P_{A_2}'), 
\end{IEEEeqnarray}
which is equivalent to 
\begin{IEEEeqnarray}{c}
u_{1,2}\alpha + u_{2,2}(1-\alpha) \leq \alpha\beta' u_{1,1} + \alpha(1-\beta') u_{1,2} + (1-\alpha)\beta' u_{2,1} + (1-\alpha)(1-\beta') u_{2,2}. 
\end{IEEEeqnarray}
From (\ref{Equ:mouse8}), the above inequality is simplified to 
\begin{IEEEeqnarray}{c}
u_{1,2} = u_{2,2} \leq (1-\beta') u_{2,2} + \alpha \beta' u_{1,1} + (1-\alpha) \beta' u_{2,1}
\end{IEEEeqnarray}
which is equivalent to 
\begin{IEEEeqnarray}{c}\label{Equ:mouse9_1}
u_{1,2} = u_{2,2}\leq  \alpha u_{1,1} + (1-\alpha) u_{2,1}.
\end{IEEEeqnarray}

If $u_{1,1} = u_{2,1}$, then for all $\alpha \in [0,1]$, the inequality in (\ref{Equ:mouse9_1}) implies that 
\begin{IEEEeqnarray}{c}\label{Equ:cotton_8}
u_{1, 2} = u_{2,2} \leq u_{1,1} = u_{2,1}.
\end{IEEEeqnarray}
However, if the inequality in \eqref{Equ:cotton_8} holds with equality, i.e.  $u_{1,1} = u_{1,2} = u_{2,1} = u_{2,2}$, all strategies $P_{A_2}'' \in \Delta(\set{A}_2)$, together with $P_{A_1}^\star$ in (\ref{Equ:fig_4}), form an NE, see \eqref{Equ:fig_4}. 
This contradicts the assumption that $P_{A_2}^\star(a_1) = 0$. 
As a result, the inequality in \eqref{Equ:cotton_8} is considered with strict inequality. That is, 
\begin{IEEEeqnarray}{c}\label{Equ:cotton_9}
u_{1,2} = u_{2,2} < u_{1,1} = u_{2,1}.
\end{IEEEeqnarray}
If $u_{1,1} \neq u_{2,1}$, for all $\alpha \in [0,1]$, the inequality in \eqref{Equ:mouse9_1} implies that 
\begin{IEEEeqnarray}{c}\label{Equ:cotton_10}
u_{1,2} = u_{2,2} \leq \textnormal{min} \{ u_{1,1}, u_{2,1}\} < \textnormal{max} \{ u_{1,1}, u_{2,1}\}.
\end{IEEEeqnarray}

Combining the inequalities in \eqref{Equ:cotton_9} and \eqref{Equ:cotton_10} yields the conditions in \eqref{Equ:mouse04} and \eqref{Equ:mouse041}. 

The converse is as follows. 
If the conditions  in \eqref{Equ:mouse04} hold, for all $P_{A_1} \in \simplex{\set{A}_1}$ with $P_{A_1}(a_1) = \alpha \in [0,1]$, it holds that 
\begin{IEEEeqnarray}{c}
(\alpha, 1-\alpha) \matx{u} (0,1)^{\sf T} < (\alpha, 1-\alpha) \matx{u} (1, 0)^{\sf T},
\end{IEEEeqnarray}
which implies that, for \Ptwo, the action $a_1$ is dominated by action $a_2$. 
Furthermore, if the conditions in \eqref{Equ:mouse041} hold, for all $P_{A_1} \in \simplex{\set{A}_1}$ with $P_{A_1}(a_1) = \alpha \in [0,1]$, it holds that 
\begin{IEEEeqnarray}{c}
(\alpha, 1-\alpha) \matx{u} (0,1)^{\sf T} \leq (\alpha, 1-\alpha) \matx{u} (1, 0)^{\sf T};
\end{IEEEeqnarray}
and for all $P_{A_1} \in \simplex{\set{A}_1}$ with $P_{A_1}(a_1) =\alpha \in [0,1]$, it holds that 
\begin{IEEEeqnarray}{c}
(\alpha, 1-\alpha) \matx{u} (0,1)^{\sf T} < (\alpha, 1-\alpha) \matx{u} (1, 0)^{\sf T}.
\end{IEEEeqnarray}
Hence, if the conditions in \eqref{Equ:mouse041} hold,  for \Ptwo, the action $a_1$ is dominated by action $a_2$. 
As a result, it holds that $P_{A_2}^\star(a_1) = 0$. 
Given the fact that $P_{A_2}^\star(a_1) = 0$, if either the conditions in \eqref{Equ:mouse05} or the conditions in  \eqref{Equ:mouse041} hold, \Pone is indifferent between action $a_1$ and action $a_2$. 
This implies that 
\begin{IEEEeqnarray}{c}
P_{A_1}^{\star}(a_1) = \beta, 
\end{IEEEeqnarray}
with $\beta \in [0,1]$, which completes the proof of the converse part. 

This completes the proof of Case IV. 

This completes the whole proof.

\section{Proof of Theorem \ref{TheInfNE2}} \label{Sec:App7_1}

The proof of Case I is as follows. 
Note that the set $\set{P}$ satisfying \eqref{Equ:mous0112} implies that $u_{1,1} -u_{1,2} -u_{2,1}+u_{2,2} \neq 0$. 
To that end, only the case in which $u_{1,1} -u_{1,2} -u_{2,1}+u_{2,2} \neq 0$ is considered in the following proof. 
Given the fact that $0 <\frac{u_{2,2} -u_{1,2}}{u_{1,1} - u_{1,2} -u_{2,1}+u_{2,2}} < 1$, it holds that 
$p^{(2)}$ in \eqref{Equ:P_{A_2}} satisfies
\begin{IEEEeqnarray}{c}\label{EqWake5}
1 > p^{(2)} > 0.
\end{IEEEeqnarray}
Let a pair of strategies $(P_{A_1}^\star, P_{A_2}^\star) \in \simplex{\set{A}_1} \times \simplex{\set{A}_2}$ be such that $(P_{A_1}^\star, P_{A_2}^\star) \in \set{P}$. 
Then, it follows that 
\begin{IEEEeqnarray}{c}\label{EqRoll1}
P_{A_2}^\star \in \BR_2(P_{A_1}^\star). 
\end{IEEEeqnarray}
Given the fact that $\set{P}$ satisfies \eqref{Equ:mous0112}, from Lemma \ref{LemmaBR2} and \eqref{EqRoll1},
it follows that $p^{(1)}$ in \eqref{Equ:P_{A_1}} satisfies
\begin{IEEEeqnarray}{c}\label{EqWake4}
 p^{(1)} = 0.
\end{IEEEeqnarray}
Note that, from Lemma \ref{LemmaBR1}, only when $u_{1,1} -u_{1,2} -u_{2,1}+u_{2,2}>0$, the best response $\BR_1(P_{A_2}^\star)$ satisfies 
\begin{IEEEeqnarray}{c}
\BR_1(P_{A_2}^\star) = \{ P_{A_1} \in \simplex{\set{A}_1}: P_{A_1}(a_1) = 0 \}
\end{IEEEeqnarray}
for all $P_{A_2}^\star(a_1) \in \left[0, \frac{u_{2,2} -u_{1,2}}{u_{1,1} - u_{1,2} -u_{2,1}+u_{2,2}} \right]$.
As a result, only the case in which $u_{1,1} -u_{1,2} -u_{2,1}+u_{2,2}>0$ needs to be considered. 
Hence, if $u_{1,1} -u_{1,2} -u_{2,1}+u_{2,2}>0$, the equality in \eqref{EqWake4} and the inequality in \eqref{EqWake5} implies that 
\begin{IEEEeqnarray}{c}
u_{2,2} = u_{2,1}, \ u_{1,1} >u_{2,1},  \textnormal{ and } u_{2,2} > u_{1,2}, \label{EqWake6}
\end{IEEEeqnarray}
which proves the condition in \eqref{Equ:mous011}. 
Note that the condition in \eqref{Equ:mous011} guarantees that $u_{1,1} -u_{1,2} -u_{2,1}+u_{2,2}>0$. 
Hence, there is no need to include $u_{1,1} -u_{1,2} -u_{2,1}+u_{2,2}>0$ as a condition. 

The converse is as follows. 
If \eqref{Equ:mous011} holds, then it holds that $u_{1,1} -u_{1,2} -u_{2,1}+u_{2,2}>0$, the equality in \eqref{EqWake4} holds, and the inequalities \eqref{EqWake5} hold. 
Hence, from Lemma \ref{LemmaBR1} and Lemma \ref{LemmaBR2}, for a pair of strategies $(P_{A_1}, P_{A_2}) \in \simplex{\set{A}_1} \times \simplex{\set{A}_2}$, if 
$P_{A_1}(a_1) > 0$, it holds that 
\begin{IEEEeqnarray}{l}
\BR_{2}(P_{A_1}) = \{P_{A_2} \in \simplex{\set{A}_2}: P_{A_2}(a_1) = 0 \} 
\quad \textnormal{and} \quad \\
\BR_{1} \left( \BR_{2}(P_{A_1}) \right) =  \{P_{A_1} \in \simplex{\set{A}_1}: P_{A_1}(a_1) = 0 \}. 
\end{IEEEeqnarray}
As a result, for all $(P_{A_1}^{\star}, P_{A_2}^{\star}) \in \set{P}$, it holds that 
\begin{IEEEeqnarray}{c}
P_{A_1}^{\star}(a_1) = 0,
\end{IEEEeqnarray}
which, from Lemma \ref{LemmaBR1}, implies that 
\begin{IEEEeqnarray}{c}
P_{A_2}^{\star}(a_1) \in \left[0,  \frac{u_{2,2} -u_{1,2}}{u_{1,1} -u_{1,2} -u_{2,1}+u_{2,2}}\right]. 
\end{IEEEeqnarray}
This completes the proof.

Furthermore, it holds that
\begin{IEEEeqnarray}{c}
u(P_{A_1}^\star, P_{A_2}^{\star}) = u_{2,2} = u_{2,1},
\end{IEEEeqnarray}
which follows from the fact that $P_{A_1}^\star(a_1) = 0$, $P_{A_2}^{\star}(a_1)$ can be equal to zero and $u_{2,2} = u_{2,1}$. 
This completes the proof.

The proof of Case II is as follows. 
Note that the set $\set{P}$ satisfying \eqref{Equ:mous0122} implies that $u_{1,1} -u_{1,2} -u_{2,1}+u_{2,2} \neq 0$. 
To that end, only the case in which $u_{1,1} -u_{1,2} -u_{2,1}+u_{2,2} \neq 0$ is considered in the following proof. 
Given the fact that $0 <\frac{u_{2,2} -u_{1,2}}{u_{1,1} - u_{1,2} -u_{2,1}+u_{2,2}} < 1$, it holds that 
$p^{(2)}$ in \eqref{Equ:P_{A_2}} satisfies
\begin{IEEEeqnarray}{c}\label{EqWake51}
1 > p^{(2)} > 0.
\end{IEEEeqnarray}
Let a pair of strategies $(P_{A_1}^\star, P_{A_2}^\star) \in \simplex{\set{A}_1} \times \simplex{\set{A}_2}$ be such that $(P_{A_1}^\star, P_{A_2}^\star) \in \set{P}$. 
Then, it follows that 
\begin{IEEEeqnarray}{c}\label{EqRoll2}
P_{A_2}^\star \in \BR_2(P_{A_1}^\star). 
\end{IEEEeqnarray}
Given the fact that $\set{P}$ satisfies \eqref{Equ:mous0122}, from Lemma \ref{LemmaBR2} and \eqref{EqRoll2},
it follows that $p^{(1)}$ in \eqref{Equ:P_{A_1}} satisfies
\begin{IEEEeqnarray}{c}\label{EqWake41}
 p^{(1)} = 0.
\end{IEEEeqnarray}
Note that, from Lemma \ref{LemmaBR1}, only when $u_{1,1} -u_{1,2} -u_{2,1}+u_{2,2}<0$, the best response $\BR_1(P_{A_2}^\star)$ satisfies 
\begin{IEEEeqnarray}{c}
\BR_1(P_{A_2}^\star) = \{ P_{A_1} \in \simplex{\set{A}_1}: P_{A_1}(a_1) = 0 \}
\end{IEEEeqnarray}
for all $P_{A_2}^\star(a_1) \in \left[ \frac{u_{2,2} -u_{1,2}}{u_{1,1} - u_{1,2} -u_{2,1}+u_{2,2}}, 1 \right]$.
As a result, only the case in which $u_{1,1} -u_{1,2} -u_{2,1}+u_{2,2}<0$ needs to be considered. 
Hence, if $u_{1,1} -u_{1,2} -u_{2,1}+u_{2,2}<0$, the equality in \eqref{EqWake41} and the inequality in \eqref{EqWake51} hold if the entries of the payoff matrix $\matx{u}$ satisfy 
\begin{IEEEeqnarray}{c}
u_{2,2} = u_{2,1}, \ u_{1,1} <u_{2,1},  \textnormal{ and } u_{2,2} < u_{1,2}, \label{EqWake6}
\end{IEEEeqnarray}
which proves the condition in \eqref{Equ:mous012}. 
Note that the condition in \eqref{Equ:mous012} guarantees that $u_{1,1} -u_{1,2} -u_{2,1}+u_{2,2}<0$. 
Hence, there is no need to include $u_{1,1} -u_{1,2} -u_{2,1}+u_{2,2}<0$ as a condition.

The converse is as follows. 
If \eqref{Equ:mous012} holds, then it holds that $u_{1,1} -u_{1,2} -u_{2,1}+u_{2,2}<0$, the equality in \eqref{EqWake41} holds, and the inequalities \eqref{EqWake51} hold. 
Hence, from Lemma \ref{LemmaBR1} and Lemma \ref{LemmaBR2}, 
 for a pair of strategies $(P_{A_1}, P_{A_2}) \in \simplex{\set{A}_1} \times \simplex{\set{A}_2}$, if 
$P_{A_1}(a_1) > 0$, it holds that 
\begin{IEEEeqnarray}{l}
\BR_{2}(P_{A_1}) = \{P_{A_2} \in \simplex{\set{A}_2}: P_{A_2}(a_1) = 1 \} 
\quad \textnormal{and} \quad \\
\BR_{1} \left( \BR_{2}(P_{A_1}) \right) =  \{P_{A_1} \in \simplex{\set{A}_1}: P_{A_1}(a_1) = 0 \}. 
\end{IEEEeqnarray}
As a result, for all $(P_{A_1}^{\star}, P_{A_2}^{\star}) \in \set{P}$, it holds that 
\begin{IEEEeqnarray}{c}
P_{A_1}^{\star}(a_1) = 0,
\end{IEEEeqnarray}
which, from Lemma \ref{LemmaBR1}, implies that 
\begin{IEEEeqnarray}{c}
P_{A_2}^{\star}(a_1) \in \left[ \frac{u_{2,2} -u_{1,2}}{u_{1,1} -u_{1,2} -u_{2,1}+u_{2,2}}, 1\right]. 
\end{IEEEeqnarray}
This completes the proof.

Furthermore, it holds that
\begin{IEEEeqnarray}{c}
u(P_{A_1}^\star, P_{A_2}^{\star}) = u_{2,2} = u_{2,1},
\end{IEEEeqnarray}
which follows from the fact that $P_{A_1}^\star(a_1) = 0$, $P_{A_2}^{\star}(a_1)$ can be equal to one and $u_{2,2} = u_{2,1}$. 
This completes the proof. 

The proof of Case III is as follows.  
Note that the set $\set{P}$ satisfying \eqref{Equ:mous0212} implies that $u_{1,1} -u_{1,2} -u_{2,1}+u_{2,2} \neq 0$. 
To that end, only the case in which $u_{1,1} -u_{1,2} -u_{2,1}+u_{2,2} \neq 0$ is considered in the following proof. 
Given the fact that $0 <\frac{u_{2,2} -u_{1,2}}{u_{1,1} - u_{1,2} -u_{2,1}+u_{2,2}} < 1$, it holds that 
$p^{(2)}$ in \eqref{Equ:P_{A_2}} satisfies
\begin{IEEEeqnarray}{c}\label{EqWake15}
1 > p^{(2)} > 0.
\end{IEEEeqnarray}
Let a pair of strategies $(P_{A_1}^\star, P_{A_2}^\star) \in \simplex{\set{A}_1} \times \simplex{\set{A}_2}$ be such that $(P_{A_1}^\star, P_{A_2}^\star) \in \set{P}$. 
Then, it follows that 
\begin{IEEEeqnarray}{c}\label{EqRoll3}
P_{A_2}^\star \in \BR_2(P_{A_1}^\star). 
\end{IEEEeqnarray}
Given the fact that $\set{P}$ satisfies \eqref{Equ:mous0112}, from Lemma \ref{LemmaBR2} and \eqref{EqRoll3},
it follows that $p^{(1)}$ in \eqref{Equ:P_{A_1}} satisfies
\begin{IEEEeqnarray}{c}\label{EqWake14}
 p^{(1)} = 1.
\end{IEEEeqnarray}
Note that, from Lemma \ref{LemmaBR1}, only when $u_{1,1} -u_{1,2} -u_{2,1}+u_{2,2}>0$, the best response $\BR_1(P_{A_2}^\star)$ satisfies 
\begin{IEEEeqnarray}{c}
\BR_1(P_{A_2}^\star) = \{ P_{A_1} \in \simplex{\set{A}_1}: P_{A_1}(a_1) = 1 \}
\end{IEEEeqnarray}
for all $P_{A_2}^\star(a_1) \in \left[\frac{u_{2,2} -u_{1,2}}{u_{1,1} - u_{1,2} -u_{2,1}+u_{2,2}}, 1 \right]$.
As a result, only the case in which $u_{1,1} -u_{1,2} -u_{2,1}+u_{2,2}>0$ needs to be considered. 
Hence, if $u_{1,1} -u_{1,2} -u_{2,1}+u_{2,2}>0$, the equality in \eqref{EqWake14} and the inequality in \eqref{EqWake15} hold if the entries of the payoff matrix $\matx{u}$ satisfy 
\begin{IEEEeqnarray}{c}
u_{1,1} = u_{1,2}, \ u_{1,1} >u_{2,1},  \textnormal{ and } u_{2,2} > u_{1,2}, \label{EqWake16}
\end{IEEEeqnarray}
which proves the condition in \eqref{Equ:mous021}. 
Note that the condition in \eqref{Equ:mous021} guarantees that $u_{1,1} -u_{1,2} -u_{2,1}+u_{2,2}>0$. 
Hence, there is no need to include $u_{1,1} -u_{1,2} -u_{2,1}+u_{2,2}>0$ as a condition.

The converse is as follows. 
If \eqref{Equ:mous021} holds, then it holds that $u_{1,1} -u_{1,2} -u_{2,1}+u_{2,2}>0$, the equality in \eqref{EqWake14} holds, and the inequalities \eqref{EqWake15} hold. 
Hence, from Lemma \ref{LemmaBR1} and Lemma \ref{LemmaBR2}, 
 for a pair of strategies $(P_{A_1}, P_{A_2}) \in \simplex{\set{A}_1} \times \simplex{\set{A}_2}$, if 
$P_{A_1}(a_1) <1$, it holds that 
\begin{IEEEeqnarray}{l}
\BR_{2}(P_{A_1}) = \{P_{A_2} \in \simplex{\set{A}_2}: P_{A_2}(a_1) = 1 \} 
\quad \textnormal{and} \quad \\
\BR_{1} \left( \BR_{2}(P_{A_1}) \right) =  \{P_{A_1} \in \simplex{\set{A}_1}: P_{A_1}(a_1) = 1 \}. 
\end{IEEEeqnarray}
As a result, for all $(P_{A_1}^{\star}, P_{A_2}^{\star}) \in \set{P}$, it holds that 
\begin{IEEEeqnarray}{c}
P_{A_1}^{\star}(a_1) = 1,
\end{IEEEeqnarray}
which, from Lemma \ref{LemmaBR1}, implies that 
\begin{IEEEeqnarray}{c}
P_{A_2}^{\star}(a_1) \in \left[ \frac{u_{2,2} -u_{1,2}}{u_{1,1} -u_{1,2} -u_{2,1}+u_{2,2}}, 1\right]. 
\end{IEEEeqnarray}
This completes the proof.

Furthermore, it holds that
\begin{IEEEeqnarray}{c}
u(P_{A_1}^\star, P_{A_2}^{\star}) = u_{1,1} = u_{1,2},
\end{IEEEeqnarray}
which follows from the fact that $P_{A_1}^\star(a_1) = 1$, $P_{A_2}^{\star}(a_1)$ can be equal to one and $u_{1,1} = u_{1,2}$. 
This completes the proof.

The proof of Case IV is as follows. 
Note that the set $\set{P}$ satisfying \eqref{Equ:mous0222} implies that $u_{1,1} -u_{1,2} -u_{2,1}+u_{2,2} \neq 0$. 
To that end, only the case in which $u_{1,1} -u_{1,2} -u_{2,1}+u_{2,2} \neq 0$ is considered in the following proof. 
Given the fact that $0 <\frac{u_{2,2} -u_{1,2}}{u_{1,1} - u_{1,2} -u_{2,1}+u_{2,2}} < 1$, it holds that 
$p^{(2)}$ in \eqref{Equ:P_{A_2}} satisfies
\begin{IEEEeqnarray}{c}\label{EqWake151}
1 > p^{(2)} > 0.
\end{IEEEeqnarray}
Let a pair of strategies $(P_{A_1}^\star, P_{A_2}^\star) \in \simplex{\set{A}_1} \times \simplex{\set{A}_2}$ be such that $(P_{A_1}^\star, P_{A_2}^\star) \in \set{P}$. 
Then, it follows that 
\begin{IEEEeqnarray}{c}\label{EqRoll4}
P_{A_2}^\star \in \BR_2(P_{A_1}^\star). 
\end{IEEEeqnarray}
Given the fact that $\set{P}$ satisfies \eqref{Equ:mous0112}, from Lemma \ref{LemmaBR2} and \eqref{EqRoll3},
it follows that $p^{(1)}$ in \eqref{Equ:P_{A_1}} satisfies
\begin{IEEEeqnarray}{c}\label{EqWake141}
 p^{(1)} = 1.
\end{IEEEeqnarray}
Note that, from Lemma \ref{LemmaBR1}, only when $u_{1,1} -u_{1,2} -u_{2,1}+u_{2,2}<0$, the best response $\BR_1(P_{A_2}^\star)$ satisfies 
\begin{IEEEeqnarray}{c}
\BR_1(P_{A_2}^\star) = \{ P_{A_1} \in \simplex{\set{A}_1}: P_{A_1}(a_1) = 1 \}
\end{IEEEeqnarray}
for all $P_{A_2}^\star(a_1) \in \left[0, \frac{u_{2,2} -u_{1,2}}{u_{1,1} - u_{1,2} -u_{2,1}+u_{2,2}} \right]$.
As a result, only the case in which $u_{1,1} -u_{1,2} -u_{2,1}+u_{2,2}<0$ needs to be considered. 
Hence, if $u_{1,1} -u_{1,2} -u_{2,1}+u_{2,2}<0$, the equality in \eqref{EqWake141} and the inequality in \eqref{EqWake151} hold if the entries of the payoff matrix $\matx{u}$ satisfy 
\begin{IEEEeqnarray}{c}
u_{1,1} = u_{1,2}, \ u_{1,1} <u_{2,1},  \textnormal{ and } u_{2,2} < u_{1,2}.
\end{IEEEeqnarray}
which proves the condition in \eqref{Equ:mous022}. 
Note that if the inequalities in  \eqref{Equ:mous022} hold, then it holds that $u_{1,1} -u_{1,2} -u_{2,1}+u_{2,2}<0$. 
Hence, there is no need to include $u_{1,1} -u_{1,2} -u_{2,1}+u_{2,2}<0$ as a condition. 

The converse is as follows. 
If \eqref{Equ:mous022} holds, then it holds that $u_{1,1} -u_{1,2} -u_{2,1}+u_{2,2}<0$, the equality in \eqref{EqWake141} holds, and the inequalities \eqref{EqWake151} hold. 
Hence, from Lemma \ref{LemmaBR1} and Lemma \ref{LemmaBR2}, 
 for a pair of strategies $(P_{A_1}, P_{A_2}) \in \simplex{\set{A}_1} \times \simplex{\set{A}_2}$, if 
$P_{A_1}(a_1) <1$, it holds that 
\begin{IEEEeqnarray}{l}
\BR_{2}(P_{A_1}) = \{P_{A_2} \in \simplex{\set{A}_2}: P_{A_2}(a_1) = 0 \} 
\quad \textnormal{and} \quad \\
\BR_{1} \left( \BR_{2}(P_{A_1}) \right) =  \{P_{A_1} \in \simplex{\set{A}_1}: P_{A_1}(a_1) = 1 \}. 
\end{IEEEeqnarray}
As a result, for all $(P_{A_1}^{\star}, P_{A_2}^{\star}) \in \set{P}$, it holds that 
\begin{IEEEeqnarray}{c}
P_{A_1}^{\star}(a_1) = 1,
\end{IEEEeqnarray}
which, from Lemma \ref{LemmaBR1}, implies that 
\begin{IEEEeqnarray}{c}
P_{A_2}^{\star}(a_1) \in \left[ 0, \frac{u_{2,2} -u_{1,2}}{u_{1,1} -u_{1,2} -u_{2,1}+u_{2,2}}\right]. 
\end{IEEEeqnarray}
This completes the proof.

Furthermore, it holds that
\begin{IEEEeqnarray}{c}
u(P_{A_1}^\star, P_{A_2}^{\star}) = u_{1,1} = u_{1,2},
\end{IEEEeqnarray}
which follows from the fact that $P_{A_1}^\star(a_1) = 1$, $P_{A_2}^{\star}(a_1)$ can be equal to zero and $u_{1,1} = u_{1,2}$. 
This completes the proof.

The proof of Case V is as follows. 
Note that the set $\set{P}$ satisfying \eqref{Equ:mous0312} implies that $u_{1,1} -u_{1,2} -u_{2,1}+u_{2,2} \neq 0$. 
To that end, only the case in which $u_{1,1} -u_{1,2} -u_{2,1}+u_{2,2} \neq 0$ is considered in the following proof. 
Given the fact that $0 <\frac{u_{2,2} -u_{2,1}}{u_{1,1} - u_{1,2} -u_{2,1}+u_{2,2}} < 1$, it holds that 
$p^{(1)}$ in \eqref{Equ:P_{A_1}} satisfies
\begin{IEEEeqnarray}{c}\label{EqWake35}
1 > p^{(1)} > 0. 
\end{IEEEeqnarray}
Let a pair of strategies $(P_{A_1}^\star, P_{A_2}^\star) \in \simplex{\set{A}_1} \times \simplex{\set{A}_2}$ be such that $(P_{A_1}^\star, P_{A_2}^\star) \in \set{P}$. 
Then, it follows that 
\begin{IEEEeqnarray}{c}\label{EqRoll5}
P_{A_1}^\star \in \BR_1(P_{A_2}^\star). 
\end{IEEEeqnarray}
Given the fact that $\set{P}$ satisfies \eqref{Equ:mous0312}, from Lemma \ref{LemmaBR1} and \eqref{EqRoll5},
it follows that $p^{(2)}$ in \eqref{Equ:P_{A_2}} satisfies
\begin{IEEEeqnarray}{c}\label{EqWake34}
 p^{(2)} = 0,
\end{IEEEeqnarray}
Note that, from Lemma \ref{LemmaBR1}, only when $u_{1,1} -u_{1,2} -u_{2,1}+u_{2,2}>0$, the best response $\BR_1(P_{A_2}^\star)$ satisfies 
\begin{IEEEeqnarray}{c}
\BR_1(P_{A_2}^\star) = \{ P_{A_1} \in \simplex{\set{A}_1}: P_{A_1}(a_1) = 0 \}
\end{IEEEeqnarray}
for all $P_{A_2}^\star(a_1) \in \left[ \frac{u_{2,2} -u_{2,1}}{u_{1,1} - u_{1,2} -u_{2,1}+u_{2,2}}, 1\right]$.
As a result, only the case in which $u_{1,1} -u_{1,2} -u_{2,1}+u_{2,2}>0$ needs to be considered. 
Hence, if $u_{1,1} -u_{1,2} -u_{2,1}+u_{2,2}>0$, the equality in \eqref{EqWake34} and the inequality in \eqref{EqWake35} hold if the entries of the payoff matrix $\matx{u}$ satisfy 
\begin{IEEEeqnarray}{c}
u_{2,2} = u_{1,2}, \ u_{2,2} >u_{2,1},  \textnormal{ and } u_{1,1} > u_{1,2}., \label{EqWake36}
\end{IEEEeqnarray}
which proves the condition in \eqref{Equ:mous031}. 
Note that if the inequalities in  \eqref{Equ:mous031} hold, then it holds that $u_{1,1} -u_{1,2} -u_{2,1}+u_{2,2}>0$.
Hence, there is no need to include $u_{1,1} -u_{1,2} -u_{2,1}+u_{2,2}>0$ as a condition. 

The converse is as follows. 
If \eqref{Equ:mous031} holds, then it holds that $u_{1,1} -u_{1,2} -u_{2,1}+u_{2,2}>0$, the equality in \eqref{EqWake34} holds, and the inequalities \eqref{EqWake35} hold. 
Hence, from Lemma \ref{LemmaBR1} and Lemma \ref{LemmaBR2}, 
 for a pair of strategies $(P_{A_1}, P_{A_2}) \in \simplex{\set{A}_1} \times \simplex{\set{A}_2}$, if 
$P_{A_2}(a_1) >0$, it holds that 
\begin{IEEEeqnarray}{l}
\BR_{1}(P_{A_2}) = \{P_{A_1} \in \simplex{\set{A}_1}: P_{A_1}(a_1) = 1 \} 
\quad \textnormal{and} \quad \\
\BR_{2} \left( \BR_{1}(P_{A_1}) \right) =  \{P_{A_2} \in \simplex{\set{A}_2}: P_{A_2}(a_1) = 0 \}. 
\end{IEEEeqnarray}
As a result, for all $(P_{A_1}^{\star}, P_{A_2}^{\star}) \in \set{P}$, it holds that 
\begin{IEEEeqnarray}{c}
P_{A_2}^{\star}(a_1) = 0,
\end{IEEEeqnarray}
which, from Lemma \ref{LemmaBR2}, implies that 
\begin{IEEEeqnarray}{c}
P_{A_1}^{\star}(a_1) \in \left[ \frac{u_{2,2} -u_{2,1}}{u_{1,1} -u_{1,2} -u_{2,1}+u_{2,2}}, 1\right]. 
\end{IEEEeqnarray}
This completes the proof.

Furthermore, it holds that
\begin{IEEEeqnarray}{c}
u(P_{A_1}^\star, P_{A_2}^{\star}) = u_{1,2} = u_{2,2},
\end{IEEEeqnarray}
which follows from the fact that $P_{A_2}^\star(a_1) = 0$, $P_{A_2}^{\star}(a_1)$ can be equal to one and $u_{1,2} = u_{2,2}$. 
This completes the proof.

The proof of Case VI is as follows. 
Note that the set $\set{P}$ satisfying \eqref{Equ:mouse0322} implies that $u_{1,1} -u_{1,2} -u_{2,1}+u_{2,2} \neq 0$. 
To that end, only the case in which $u_{1,1} -u_{1,2} -u_{2,1}+u_{2,2} \neq 0$ is considered in the following proof. 
Given the fact that $0 <\frac{u_{2,2} -u_{2,1}}{u_{1,1} - u_{1,2} -u_{2,1}+u_{2,2}} < 1$, it holds that 
$p^{(1)}$ in \eqref{Equ:P_{A_1}} satisfies
\begin{IEEEeqnarray}{c}\label{EqWake351}
1 > p^{(1)} > 0. 
\end{IEEEeqnarray}
Let a pair of strategies $(P_{A_1}^\star, P_{A_2}^\star) \in \simplex{\set{A}_1} \times \simplex{\set{A}_2}$ be such that $(P_{A_1}^\star, P_{A_2}^\star) \in \set{P}$. 
Then, it follows that 
\begin{IEEEeqnarray}{c}\label{EqRoll6}
P_{A_1}^\star \in \BR_1(P_{A_2}^\star). 
\end{IEEEeqnarray}
Given the fact that $\set{P}$ satisfies \eqref{Equ:mouse0322}, from Lemma \ref{LemmaBR1} and \eqref{EqRoll6},
it follows that $p^{(2)}$ in \eqref{Equ:P_{A_2}} satisfies
\begin{IEEEeqnarray}{c}\label{EqWake341}
 p^{(2)} = 0,
\end{IEEEeqnarray}
Note that, from Lemma \ref{LemmaBR1}, only when $u_{1,1} -u_{1,2} -u_{2,1}+u_{2,2}<0$, the best response $\BR_1(P_{A_2}^\star)$ satisfies 
\begin{IEEEeqnarray}{c}
\BR_1(P_{A_2}^\star) = \{ P_{A_1} \in \simplex{\set{A}_1}: P_{A_1}(a_1) = 0 \}
\end{IEEEeqnarray}
for all $P_{A_2}^\star(a_1) \in \left[ 0,\frac{u_{2,2} -u_{2,1}}{u_{1,1} - u_{1,2} -u_{2,1}+u_{2,2}}\right]$.
As a result, only the case in which $u_{1,1} -u_{1,2} -u_{2,1}+u_{2,2}<0$ needs to be considered. 
Hence, if $u_{1,1} -u_{1,2} -u_{2,1}+u_{2,2}<0$, the equality in \eqref{EqWake341} and the inequality in \eqref{EqWake351} hold if the entries of the payoff matrix $\matx{u}$ satisfy 
\begin{IEEEeqnarray}{c}
u_{2,2} = u_{1,2}, \ u_{2,2} >u_{2,1},  \textnormal{ and } u_{1,1} > u_{1,2}., \label{EqWake36}
\end{IEEEeqnarray}
which proves the condition in \eqref{Equ:mous031}. 
Note that if the inequalities in  \eqref{Equ:mous031} hold, then it holds that $u_{1,1} -u_{1,2} -u_{2,1}+u_{2,2}<0$.
Hence, there is no need to include $u_{1,1} -u_{1,2} -u_{2,1}+u_{2,2}<0$ as a condition. 

The converse is as follows. 
If \eqref{Equ:mous031} holds, then it holds that $u_{1,1} -u_{1,2} -u_{2,1}+u_{2,2}<0$, the equality in \eqref{EqWake341} holds, and the inequalities \eqref{EqWake351} hold. 
Hence, from Lemma \ref{LemmaBR1} and Lemma \ref{LemmaBR2}, 
 for a pair of strategies $(P_{A_1}, P_{A_2}) \in \simplex{\set{A}_1} \times \simplex{\set{A}_2}$, if 
$P_{A_2}(a_1) >0$, it holds that 
\begin{IEEEeqnarray}{l}
\BR_{1}(P_{A_2}) = \{P_{A_1} \in \simplex{\set{A}_1}: P_{A_1}(a_1) = 0 \} 
\quad \textnormal{and} \quad \\
\BR_{2} \left( \BR_{1}(P_{A_1}) \right) =  \{P_{A_2} \in \simplex{\set{A}_2}: P_{A_2}(a_1) = 0 \}. 
\end{IEEEeqnarray}
As a result, for all $(P_{A_1}^{\star}, P_{A_2}^{\star}) \in \set{P}$, it holds that 
\begin{IEEEeqnarray}{c}
P_{A_2}^{\star}(a_1) = 0,
\end{IEEEeqnarray}
which, from Lemma \ref{LemmaBR2}, implies that 
\begin{IEEEeqnarray}{c}
P_{A_1}^{\star}(a_1) \in \left[ 0, \frac{u_{2,2} -u_{2,1}}{u_{1,1} -u_{1,2} -u_{2,1}+u_{2,2}}\right]. 
\end{IEEEeqnarray}
This completes the proof.

The proof of Case VII is as follows. 
Note that the set $\set{P}$ satisfying \eqref{Equ:mouse0411} implies that $u_{1,1} -u_{1,2} -u_{2,1}+u_{2,2} \neq 0$. 
To that end, only the case in which $u_{1,1} -u_{1,2} -u_{2,1}+u_{2,2} \neq 0$ is considered in the following proof. 
Given the fact that $0 <\frac{u_{2,2} -u_{2,1}}{u_{1,1} - u_{1,2} -u_{2,1}+u_{2,2}} < 1$, it holds that 
$p^{(1)}$ in \eqref{Equ:P_{A_1}} satisfies
\begin{IEEEeqnarray}{c}\label{EqWake45}
1 > p^{(1)} > 0. 
\end{IEEEeqnarray}
Let a pair of strategies $(P_{A_1}^\star, P_{A_2}^\star) \in \simplex{\set{A}_1} \times \simplex{\set{A}_2}$ be such that $(P_{A_1}^\star, P_{A_2}^\star) \in \set{P}$. 
Then, it follows that 
\begin{IEEEeqnarray}{c}\label{EqRoll7}
P_{A_1}^\star \in \BR_1(P_{A_2}^\star). 
\end{IEEEeqnarray}
Given the fact that $\set{P}$ satisfies \eqref{Equ:mouse0411}, from Lemma \ref{LemmaBR1} and \eqref{EqRoll7},
it follows that $p^{(2)}$ in \eqref{Equ:P_{A_2}} satisfies
\begin{IEEEeqnarray}{c}\label{EqWake44}
 p^{(2)} = 1.
\end{IEEEeqnarray}
Note that, from Lemma \ref{LemmaBR1}, only when $u_{1,1} -u_{1,2} -u_{2,1}+u_{2,2}>0$, the best response $\BR_1(P_{A_2}^\star)$ satisfies 
\begin{IEEEeqnarray}{c}
\BR_1(P_{A_2}^\star) = \{ P_{A_1} \in \simplex{\set{A}_1}: P_{A_1}(a_1) = 1 \}
\end{IEEEeqnarray}
for all $P_{A_2}^\star(a_1) \in \left[ 0,\frac{u_{2,2} -u_{2,1}}{u_{1,1} - u_{1,2} -u_{2,1}+u_{2,2}}\right]$.
As a result, only the case in which $u_{1,1} -u_{1,2} -u_{2,1}+u_{2,2}>0$ needs to be considered. 
Hence, if $u_{1,1} -u_{1,2} -u_{2,1}+u_{2,2}>0$, the equality in \eqref{EqWake44} and the inequality in \eqref{EqWake45} hold if the entries of the payoff matrix $\matx{u}$ satisfy 
\begin{IEEEeqnarray}{c}
u_{1,1} = u_{2,1}, \ u_{2,2} >u_{2,1},  \textnormal{ and } u_{1,1} > u_{1,2}., \label{EqWake46}
\end{IEEEeqnarray}
which proves the condition in \eqref{Equ:mouse041}. 
Note that if the inequalities in  \eqref{Equ:mouse041} hold, then it holds that $u_{1,1} -u_{1,2} -u_{2,1}+u_{2,2}>0$.
Hence, there is no need to include $u_{1,1} -u_{1,2} -u_{2,1}+u_{2,2}>0$ as a condition. 

The converse is as follows. 
If \eqref{Equ:mouse041} holds, then it holds that $u_{1,1} -u_{1,2} -u_{2,1}+u_{2,2}>0$, the equality in \eqref{EqWake44} holds, and the inequalities \eqref{EqWake45} hold. 
Hence, from Lemma \ref{LemmaBR1} and Lemma \ref{LemmaBR2}, 
 for a pair of strategies $(P_{A_1}, P_{A_2}) \in \simplex{\set{A}_1} \times \simplex{\set{A}_2}$, if 
$P_{A_2}(a_1) <1$, it holds that 
\begin{IEEEeqnarray}{l}
\BR_{1}(P_{A_2}) = \{P_{A_1} \in \simplex{\set{A}_1}: P_{A_1}(a_1) = 0 \} 
\quad \textnormal{and} \quad \\
\BR_{2} \left( \BR_{1}(P_{A_1}) \right) =  \{P_{A_2} \in \simplex{\set{A}_2}: P_{A_2}(a_1) = 1 \}. 
\end{IEEEeqnarray}
As a result, for all $(P_{A_1}^{\star}, P_{A_2}^{\star}) \in \set{P}$, it holds that 
\begin{IEEEeqnarray}{c}
P_{A_2}^{\star}(a_1) = 1,
\end{IEEEeqnarray}
which, from Lemma \ref{LemmaBR2}, implies that 
\begin{IEEEeqnarray}{c}
P_{A_1}^{\star}(a_1) \in \left[ 0, \frac{u_{2,2} -u_{2,1}}{u_{1,1} -u_{1,2} -u_{2,1}+u_{2,2}}\right]. 
\end{IEEEeqnarray}
This completes the proof.

The proof of Case VIII is as follows. 
Note that the set $\set{P}$ satisfying \eqref{Equ:mouse0421} implies that $u_{1,1} -u_{1,2} -u_{2,1}+u_{2,2} \neq 0$. 
To that end, only the case in which $u_{1,1} -u_{1,2} -u_{2,1}+u_{2,2} \neq 0$ is considered in the following proof. 
Given the fact that $0 <\frac{u_{2,2} -u_{2,1}}{u_{1,1} - u_{1,2} -u_{2,1}+u_{2,2}} < 1$, it holds that 
$p^{(1)}$ in \eqref{Equ:P_{A_1}} satisfies
\begin{IEEEeqnarray}{c}\label{EqWake045}
1 > p^{(1)} > 0. 
\end{IEEEeqnarray}
Let a pair of strategies $(P_{A_1}^\star, P_{A_2}^\star) \in \simplex{\set{A}_1} \times \simplex{\set{A}_2}$ be such that $(P_{A_1}^\star, P_{A_2}^\star) \in \set{P}$. 
Then, it follows that 
\begin{IEEEeqnarray}{c}\label{EqRoll8}
P_{A_1}^\star \in \BR_1(P_{A_2}^\star). 
\end{IEEEeqnarray}
Given the fact that $\set{P}$ satisfies \eqref{Equ:mouse0421}, from Lemma \ref{LemmaBR1} and \eqref{EqRoll8},
it follows that $p^{(2)}$ in \eqref{Equ:P_{A_2}} satisfies
\begin{IEEEeqnarray}{c}\label{EqWake044}
 p^{(2)} = 1.
\end{IEEEeqnarray}
Note that, from Lemma \ref{LemmaBR1}, only when $u_{1,1} -u_{1,2} -u_{2,1}+u_{2,2}<0$, the best response $\BR_1(P_{A_2}^\star)$ satisfies 
\begin{IEEEeqnarray}{c}
\BR_1(P_{A_2}^\star) = \{ P_{A_1} \in \simplex{\set{A}_1}: P_{A_1}(a_1) = 1 \}
\end{IEEEeqnarray}
for all $P_{A_2}^\star(a_1) \in \left[ \frac{u_{2,2} -u_{2,1}}{u_{1,1} - u_{1,2} -u_{2,1}+u_{2,2}}, 1\right]$.
As a result, only the case in which $u_{1,1} -u_{1,2} -u_{2,1}+u_{2,2}<0$ needs to be considered. 
Hence, if $u_{1,1} -u_{1,2} -u_{2,1}+u_{2,2}<0$, the equality in \eqref{EqWake044} and the inequality in \eqref{EqWake045} hold if the entries of the payoff matrix $\matx{u}$ satisfy 
\begin{IEEEeqnarray}{c}
u_{1,1} = u_{2,1}, \ u_{2,2} >u_{2,1},  \textnormal{ and } u_{1,1} > u_{1,2}., \label{EqWake46}
\end{IEEEeqnarray}
which proves the condition in \eqref{Equ:mouse042}. 
Note that if the inequalities in  \eqref{Equ:mouse042} hold, then it holds that $u_{1,1} -u_{1,2} -u_{2,1}+u_{2,2}<0$.
Hence, there is no need to include $u_{1,1} -u_{1,2} -u_{2,1}+u_{2,2}<0$ as a condition. 

The converse is as follows. 
If \eqref{Equ:mouse042} holds, then it holds that $u_{1,1} -u_{1,2} -u_{2,1}+u_{2,2}<0$, the equality in \eqref{EqWake044} holds, and the inequalities \eqref{EqWake045} hold. 
Hence, from Lemma \ref{LemmaBR1} and Lemma \ref{LemmaBR2},  for a pair of strategies $(P_{A_1}, P_{A_2}) \in \simplex{\set{A}_1} \times \simplex{\set{A}_2}$, if 
$P_{A_2}(a_1) <1$, it holds that 
\begin{IEEEeqnarray}{l}
\BR_{1}(P_{A_2}) = \{P_{A_1} \in \simplex{\set{A}_1}: P_{A_1}(a_1) = 1 \} 
\quad \textnormal{and} \quad \\
\BR_{2} \left( \BR_{1}(P_{A_1}) \right) =  \{P_{A_2} \in \simplex{\set{A}_2}: P_{A_2}(a_1) = 1 \}. 
\end{IEEEeqnarray}
As a result, for all $(P_{A_1}^{\star}, P_{A_2}^{\star}) \in \set{P}$, it holds that 
\begin{IEEEeqnarray}{c}
P_{A_2}^{\star}(a_1) = 1,
\end{IEEEeqnarray}
which, from Lemma \ref{LemmaBR2}, implies that 
\begin{IEEEeqnarray}{c}
P_{A_1}^{\star}(a_1) \in \left[ \frac{u_{2,2} -u_{2,1}}{u_{1,1} -u_{1,2} -u_{2,1}+u_{2,2}}, 1\right]. 
\end{IEEEeqnarray}
This completes the proof.

Furthermore, it holds that
\begin{IEEEeqnarray}{c}
u(P_{A_1}^\star, P_{A_2}^{\star}) = u_{1,1} = u_{2,1},
\end{IEEEeqnarray}
which follows from the fact that $P_{A_2}^\star(a_1) = 1$, $P_{A_2}^{\star}(a_1)$ can be equal to one, and $u_{1,1} = u_{2,1}$. 
This completes the proof. 

This completes the whole proof.

\section{Proof of Theorem \ref{TheoremNE}}\label{AppTheoremNE}

Note that Theorem \ref{TheoNE}, Theorem \ref{TheInfNE1}, and Theorem \ref{TheInfNE2} cover all the cases of the NEs in $2 \times 2$ zero sum games. 
Hence, the proof is separated into 2 parts. 
The first part proves that if the entries in the payoff matrix $\matx{u}$ satisfy \eqref{EqMixedAssumption}, then the NE is unique and is formed by \eqref{EqPA1StarExample} and \eqref{EqPA1StarExample}. 
The second part proves that if the entries in the payoff matrix $\matx{u}$ satisfy \eqref{EqNotMixedAssumption}, then the value of the game satisfies \eqref{Equ:f_3} and \eqref{Equ:f_4}.

The first part follows from Theorem \ref{TheoNE} (case $(i)$). 

The second part is as follows. 
Note that the conditions in \eqref{EqMixedAssumption} and \eqref{EqNotMixedAssumption} form a partition of $\reals^{2 \times 2}$. 
If the conditions in \eqref{EqMixedAssumption} hold, then from Theorem \ref{TheoNE}, there is a unique NE in strictly mixed strategies. 
Alternatively, from Theorem \ref{TheoNumberNE}, if the entries of the payoff matrix $\matx{u}$ satisfy \eqref{EqNotMixedAssumption}, then there is a unique NE in pure strategies or infinitely many NEs.
 
Note that from Theorem \ref{TheoNE}, Theorem \ref{TheInfNE1}, and Theorem \ref{TheInfNE2}, if the entries of the payoff matrix $\matx{u}$ satisfy \eqref{EqNotMixedAssumption}, the set of NE(s) always include a pair of strategies $(P_1, P_2) \in \simplex{\set{A}_1 \times \set{A}_2}$ such that 
\begin{IEEEeqnarray}{c}
P_1(a_1) \in \{0,1 \} \quad \textnormal{and} \quad P_2(a_1) \in \{0,1 \}. 
\end{IEEEeqnarray}
Given the fact that all the NEs yield the same payoff, if the entries of the payoff matrix $\matx{u}$ satisfy \eqref{EqNotMixedAssumption}, the value of the game can be searched 
exclusively in pure strategies.

For the game in which both players only use pure strategies, the maxmin value of the game \cite[Section 4.12]{Maschler_2013_game} equals to 
\begin{IEEEeqnarray}{c}
\max_{\alpha \in \{0,1\}} \min_{\beta \in \{0,1\}} u_{1,1}\alpha \beta + u_{1,2}\alpha (1-\beta) + u_{2,1}(1-\alpha) \beta + u_{2,2}(1-\alpha)(1-\beta),
\end{IEEEeqnarray}
which can be rewritten as 
\begin{IEEEeqnarray}{c}\label{EqLight1}
\max_{\alpha \in \{0,1\}} \min_{\beta \in \{0,1\}} \bigl(u_{1,1} \beta + u_{1,2}(1-\beta) \bigr)\alpha + \bigl(u_{2,1} \beta + u_{2,2}(1-\beta) \bigr)(1-\alpha). 
\end{IEEEeqnarray}
Note that the maxmin value in \eqref{EqLight1} can be further expressed as 
\begin{IEEEeqnarray}{c}\label{EqLight2}
\max \Big\{ \min_{\beta \in \{0,1\}} u_{1,1} \beta + u_{1,2}(1-\beta) , \min_{\beta \in \{0,1\}} u_{2,1} \beta + u_{2,2}(1-\beta) \Big\}, 
\end{IEEEeqnarray}
which follows from the fact that $\alpha \in \{0,1\}$. 
Given the fact that $\beta$ only can be either zero or one, the maximum value in \eqref{EqLight2} also equals to 
\begin{IEEEeqnarray}{c}\label{EqLight3}
\max \Big\{ \min \big\{ u_{1,1}, u_{1,2}\big\}, \min \big\{ u_{2,1} , u_{2,2}\big\} \Big\}. 
\end{IEEEeqnarray}

Similarly, for the game in which both players only use pure strategies, the minmax value of the game \cite[Section 4.12]{Maschler_2013_game} equals to 
\begin{IEEEeqnarray}{c}
\min_{\beta \in \{0,1\}} \max_{\alpha \in \{0,1\}}  u_{1,1}\alpha \beta + u_{1,2}\alpha (1-\beta) + u_{2,1}(1-\alpha) \beta + u_{2,2}(1-\alpha)(1-\beta),
\end{IEEEeqnarray}
which can be rewritten as 
\begin{IEEEeqnarray}{c}\label{EqLight4}
\min_{\beta \in \{0,1\}} \max_{\alpha \in \{0,1\}}  \bigl(u_{1,1} \beta + u_{1,2}(1-\beta) \bigr)\alpha + \bigl(u_{2,1} \beta + u_{2,2}(1-\beta) \bigr)(1-\alpha). 
\end{IEEEeqnarray}
Note that the maxmin value in \eqref{EqLight4} can be further expressed as 
\begin{IEEEeqnarray}{c}\label{EqLight5}
\min \Big\{ \max_{\alpha \in \{0,1\}} u_{1,1} \alpha + u_{2,1}(1-\alpha) , \max_{\alpha \in \{0,1\}} u_{1,2} \alpha + u_{2,2}(1-\alpha) \Big\}, 
\end{IEEEeqnarray}
which follows from the fact that $\beta \in \{0,1\}$. 
Given the fact that $\alpha$ only can be either zero or one, the minmax value in \eqref{EqLight5} also equals to 
\begin{IEEEeqnarray}{c}\label{EqLight6}
\min \Big\{ \max \big\{ u_{1,1}, u_{2,1}\big\}, \min \big\{ u_{1,2} , u_{2,2}\big\} \Big\}. 
\end{IEEEeqnarray}

In a $2 \times 2$ zero sum game, the maximin value equals the minmax value, which also equals the value of the game \cite[Section 4.12]{Maschler_2013_game}. 
As a result, it holds that 
\begin{IEEEeqnarray}{rcl}
u(P_{A_1}^{\star},P_{A_2}^{\star}) & = & \min \lbrace \max\lbrace u_{1,1}, u_{2,1}\rbrace,  \max\lbrace u_{1,2}, u_{2,2}\rbrace \rbrace \squeezeequ\\
& = & \max \lbrace \min\lbrace u_{1,1}, u_{1,2}\rbrace,  \min\lbrace u_{2,1}, u_{2,2}\rbrace \rbrace, \IEEEeqnarraynumspace \squeezeequ
\end{IEEEeqnarray}
which completes the proof. 

\section{Proof of Lemma \ref{LemmaMix}}\label{AppLemmaMix}
Note that 
\begin{IEEEeqnarray}{rl}
\frac{u_{1,1}u_{2,2} - u_{1,2} u_{2,1}}{u_{1,1} - u_{1,2} -u_{2,1} +u_{2,2}} &=  \frac{u_{1,1}u_{2,2} - u_{1,1}u_{2,1} + u_{1,1}u_{2,1} -u_{1,2} u_{2,1} }{u_{1,1} - u_{1,2} -u_{2,1} +u_{2,2}} \\
& =  \frac{u_{1,1}\left(u_{2,2} -u_{2,1}\right) + u_{2,1} \left(u_{1,1}-u_{1,2}\right)}{u_{1,1} - u_{1,2} -u_{2,1} +u_{2,2}} \\
& =  u_{1,1}P_{A_1}^{\star}(a_1) + u_{2,1}P_{A_1}^{\star}(a_2) \label{EqFinish1} \\
& < \max \left\{ u_{1,1},  u_{2,1} \right\}, \label{EqFinish2}
\end{IEEEeqnarray}
where \eqref{EqFinish1} follows from Theorem \ref{TheoNE}; 
the strictly inequality in \eqref{EqFinish2} follows from the fact that $P_{A_1}^{\star}(a_1) \in (0,1)$. 

Similarly, it holds that 
\begin{IEEEeqnarray}{rl}
\frac{u_{1,1}u_{2,2} - u_{1,2} u_{2,1}}{u_{1,1} - u_{1,2} -u_{2,1} +u_{2,2}} &=  \frac{u_{1,1}u_{2,2} - u_{1,2}u_{2,2} + u_{1,2}u_{2,2}-u_{1,2} u_{2,1} }{u_{1,1} - u_{1,2} -u_{2,1} +u_{2,2}} \\
& =  \frac{u_{2,2}\left(u_{1,1} -u_{1,2}\right) + u_{1,2} \left(u_{2,2}-u_{2,1}\right)}{u_{1,1} - u_{1,2} -u_{2,1} +u_{2,2}} \\
& =  u_{2,2}P_{A_1}^{\star}(a_2) + u_{1,2}P_{A_1}^{\star}(a_1) \label{EqFinish3}\\
& < \max \left\{ u_{1,2},  u_{2,2} \right\}, \label{EqFinish4}
\end{IEEEeqnarray}
where \eqref{EqFinish3} follows from Theorem \ref{TheoNE}; 
the strictly inequality in \eqref{EqFinish4} follows from the fact that $P_{A_1}^{\star}(a_1) \in (0,1)$. 

As a result, it holds that 
\begin{IEEEeqnarray}{c}
\frac{u_{1,1}u_{2,2} - u_{1,2} u_{2,1}}{u_{1,1} - u_{1,2} -u_{2,1} +u_{2,2}} < \min \left\{ \max \left\{ u_{1,1},  u_{2,1} \right\}, \max \left\{ u_{1,2},  u_{2,2} \right\}\right\}
\end{IEEEeqnarray}

Note that 
\begin{IEEEeqnarray}{rl}
\frac{u_{1,1}u_{2,2} - u_{1,2} u_{2,1}}{u_{1,1} - u_{1,2} -u_{2,1} +u_{2,2}} &=  \frac{u_{1,1}u_{2,2} - u_{1,1}u_{1,2} + u_{1,1}u_{1,2} -u_{1,2} u_{2,1} }{u_{1,1} - u_{1,2} -u_{2,1} +u_{2,2}} \\
& =  \frac{u_{1,1}\left(u_{2,2} -u_{1,2}\right) + u_{1,2} \left(u_{1,1}-u_{2,1}\right)}{u_{1,1} - u_{1,2} -u_{2,1} +u_{2,2}} \\
& =  u_{1,1}P_{A_2}^{\star}(a_1) + u_{1,2}P_{A_2}^{\star}(a_2) \label{EqFinish5} \\
& > \min \left\{ u_{1,1},  u_{1,2} \right\}, \label{EqFinish6}
\end{IEEEeqnarray}
where \eqref{EqFinish5} follows from Theorem \ref{TheoNE}; 
the strictly inequality in \eqref{EqFinish6} follows from the fact that $P_{A_2}^{\star}(a_1) \in (0,1)$. 
Similarly, it holds that 
\begin{IEEEeqnarray}{rl}
\frac{u_{1,1}u_{2,2} - u_{1,2} u_{2,1}}{u_{1,1} - u_{1,2} -u_{2,1} +u_{2,2}} &=  \frac{u_{1,1}u_{2,2} - u_{2,1}u_{2,2} + u_{2,1}u_{2,2}-u_{1,2} u_{2,1} }{u_{1,1} - u_{1,2} -u_{2,1} +u_{2,2}} \\
& =  \frac{u_{2,2}\left(u_{1,1} -u_{2,1}\right) + u_{2,1} \left(u_{2,2}-u_{1,2}\right)}{u_{1,1} - u_{1,2} -u_{2,1} +u_{2,2}} \\
& =  u_{2,2}P_{A_2}^{\star}(a_2) + u_{2,1}P_{A_2}^{\star}(a_1) \label{EqFinish7}\\
& > \min \left\{ u_{2,2},  u_{2,1} \right\},\label{EqFinish8}
\end{IEEEeqnarray}
where \eqref{EqFinish7} follows from Theorem \ref{TheoNE}; 
the strictly inequality in \eqref{EqFinish8} follows from the fact that $P_{A_2}^{\star}(a_1) \in (0,1)$. 

As a result, it holds that 
\begin{IEEEeqnarray}{c}
\frac{u_{1,1}u_{2,2} - u_{1,2} u_{2,1}}{u_{1,1} - u_{1,2} -u_{2,1} +u_{2,2}} > \max \left\{ \min \left\{ u_{1,1},  u_{1,2} \right\}, \min \left\{ u_{2,2},  u_{2,1} \right\}\right\}, 
\end{IEEEeqnarray}
which completes the proof. 

\section{Proof of Lemma \ref{LemmaHatu}} \label{AppLemmaHatu}
Note that  if the entries of the payoff matrix $\matx{u}$ in \eqref{Equ:u} satisfy \eqref{EqMixedAssumption}, from Theorem \ref{TheoNE}, it holds that   
\begin{IEEEeqnarray}{c}
p^{(2)} = P_{A_2}^{\star}(a_1) =  \frac{u_{2,2} - u_{1,2}}{u_{1,1} -u_{1,2} -u_{2,1}+u_{2,2}}  \in (0,1), 
\end{IEEEeqnarray}
where $p^{(2)}$ is in \eqref{Equ:P_{A_2}}, and the pair of strategies $(P_{A_1}^{\star}, P_{A_2}^{\star})$ forms the unique NE.

If given a strategy $P \in \simplex{\set{A}_2}$, it holds that $\arg \max_{V \in \simplex{\set{A}_1}} u\left(V, P \right) = \{ Q\in \Delta(\set{A}_1): Q(a_1) = 0 \}$, then, from \eqref{Eqv}, it follows that
\begin{IEEEeqnarray}{c}\label{Equ:add_1}
 \hat{u}(P)= u_{2,1}P(a_1)+u_{2,2} P(a_2). 
\end{IEEEeqnarray}
If given a $P \in \simplex{\set{A}_2}$, it holds that $\arg \max_{V \in \simplex{\set{A}_1}} u\left(V, P \right) = \{ Q\in \Delta(\set{A}_1): Q(a_1) = 1 \}$, then, from \eqref{Eqv}, it follows that
\begin{IEEEeqnarray}{c}\label{Equ:add_2}
 \hat{u}(P)= u_{1,1}P(a_1)+u_{1,2} P(a_2). 
\end{IEEEeqnarray}
If  $P = P^{\star}_{A_2}$, from Lemma \ref{LemmaBR1}, it holds that $\arg \max_{V \in \simplex{\set{A}_1}} u\left(V, P \right) = \{ Q \in \Delta(\set{A}_1): Q(a_1) = \beta, \beta  \in [0,1] \}$, then, from (\ref{Eqv}) and \eqref{EqHand}, it follows that 
\begin{IEEEeqnarray}{rCl}
&&\hat{u}( P^{\star}_{A_2}) \nonumber \\
&=& \max_{Q \in \simplex{\set{A}_1}} (u_{1,1} -u_{1,2}-u_{2,1}+u_{2,2}) \left( P_{A_2}^{\star}(a_1) - P_{A_2}^{\star}(a_1)\right)  Q(a_1)+ (u_{2,1}-u_{2,2}) P_{A_2}^{\star}(a_1) +u_{2,2} \squeezeequ \IEEEeqnarraynumspace  \\
& = & u_{2,1} P_{A_2}^{\star}(a_1) +u_{2,2}P_{A_2}^{\star}(a_2). \label{EqHand1}
\end{IEEEeqnarray}
Plugging \eqref{EqPA2StarExample} into \eqref{EqHand1} yields
\begin{IEEEeqnarray}{rl}
u_{2,1}P^{\star}_{A_2}(a_1)+u_{2,2} P^{\star}_{A_2}(a_2)= &
(u_{2,1} -u_{2,2}) P^{\star}_{A_2}(a_1)  +u_{2,2}  \\
& = (u_{2,1} -u_{2,2})\frac{ u_{2,2} - u_{1,2}}{u_{1,1} - u_{1,2} - u_{2,1} + u_{2,2}} +u_{2,2}  \\
& =  \frac{(u_{2,1} - u_{2,2}) (u_{2,2} - u_{1,2}) + u_{2,2} (u_{1,1} - u_{1,2} - u_{2,1} + u_{2,2})}{u_{1,1} - u_{1,2} - u_{2,1} + u_{2,2}} \IEEEeqnarraynumspace \\
& =  \frac{u_{1,1} u_{2,2} - u_{1,2} u_{2,1}}{u_{1,1} - u_{1,2} - u_{2,1} + u_{2,2}}. \label{Equ:add_12} 
\end{IEEEeqnarray}
Note that plugging \eqref{EqPA2StarExample} into into ~\eqref{Equ:add_2} also yields 
\begin{IEEEeqnarray}{rl}
u_{1,1}P^{\star}_{A_2}(a_1)+u_{1,2} P^{\star}_{A_2}(a_2)= &
(u_{1,1} -u_{1,2}) P^{\star}_{A_2}(a_1)  +u_{1,2}  \\
& = (u_{1,1} -u_{1,2})\frac{ u_{2,2} - u_{1,2}}{u_{1,1} - u_{1,2} - u_{2,1} + u_{2,2}} +u_{1,2}  \\
& =  \frac{(u_{1,1} - u_{1,2}) (u_{2,2} - u_{1,2}) + u_{1,2} (u_{1,1} - u_{1,2} - u_{2,1} + u_{2,2})}{u_{1,1} - u_{1,2} - u_{2,1} + u_{2,2}} \IEEEeqnarraynumspace \\
& =  \frac{u_{1,1} u_{2,2} - u_{1,2} u_{2,1}}{u_{1,1} - u_{1,2} - u_{2,1} + u_{2,2}}, \label{Equ:add_11} 
\end{IEEEeqnarray}
which, with \eqref{Equ:add_12}, implies that the function $\hat{u}$ satisfies 
\begin{IEEEeqnarray}{c}
\hat{u}(P_{A_2}^{\star}) = u_{2,1}P^{\star}_{A_2}(a_1)+u_{2,2} P^{\star}_{A_2}(a_2) = u_{1,1}P^{\star}_{A_2}(a_1)+u_{1,2} P^{\star}_{A_2}(a_2). 
\end{IEEEeqnarray}
In a nutshell, from Lemma \ref{LemmaBR1}, under the assumption of the lemma, for all $P \in \simplex{\set{A}_2}$,  it holds that 
\begin{IEEEeqnarray}{c}
\hat{u}(P)= 
\quad  \left\{ 
\begin{array}{cl}
u_{2,1} P(a_1)+u_{2,2}P(a_2), &  \ \ \makecell[t]{ \textnormal{ if } P(a_1) < P^{\star}_{A_2}(a_1) \ \textnormal{and} \ u_{1,1} -u_{1,2}-u_{2,1}+u_{2,2}> 0 \quad \textnormal{or } \\ \!\!\!\! P(a_1) > P^{\star}_{A_2}(a_1) \ \textnormal{and} \ u_{1,1} -u_{1,2}-u_{2,1}+u_{2,2}< 0,} \\
u_{1,1}P(a_1)  +u_{1,2}P(a_2), & \ \ \makecell[t]{\textnormal{ if }  P(a_1) > P^{\star}_{A_2}(a_1) \ \textnormal{and} \ u_{1,1} -u_{1,2}-u_{2,1}+u_{2,2}> 0 \quad \textnormal{or } \\ \!\!\!\!  P(a_1) < P^{\star}_{A_2}(a_1) \ \textnormal{and} \ u_{1,1} -u_{1,2}-u_{2,1}+u_{2,2}< 0,} \\
u(P_{A_1}^{\star}, P_{A_2}^{\star}), &\ \  \textnormal{ if }   P(a_1) = P^{\star}_{A_2}(a_1).
\end{array}
\right.  \IEEEeqnarraynumspace 
\end{IEEEeqnarray}

The proof is completed by noticing that if the matrix $\matx{u}$ in~\eqref{Equ:u} satisfies~\eqref{EqMixedAssumption}, it holds that $u_{1,1} - u_{1,2} -u_{2,1} +u_{2,2} \neq 0$.

\section{Proof of Lemma \ref{LemmaMono}}\label{AppLemmaMono}
Two cases are considered. First, the case in which $u_{1,1} -u_{1,2}-u_{2,1}+u_{2,2}>0$; Second, the case in which $u_{1,1} -u_{1,2}-u_{2,1}+u_{2,2}\leq0$. 

Consider the case in which $u_{1,1} -u_{1,2}-u_{2,1}+u_{2,2}>0$. 
From Theorem \ref{TheoremNE}, if the entries of the matrix $\matx{u}$ in~\eqref{Equ:u} satisfy \eqref{EqMixedAssumption}, then one of the following conditions holds: 
\begin{IEEEeqnarray}{rl}
&u_{1,1} - u_{1,2}>0 \quad \textnormal{and} \quad u_{2,1} -u_{2,2}<0, \qquad \textnormal{or}\label{EqShu1}\\
&u_{1,1} - u_{1,2}<0 \quad \textnormal{and} \quad u_{2,1} -u_{2,2}>0.
\end{IEEEeqnarray}
Nonetheless, only the first condition yields $u_{1,1} -u_{1,2}-u_{2,1}+u_{2,2}>0$. 
Hence, from Lemma \ref{LemmaHatu} and \eqref{EqShu1}, if $0 \leq P(a_1) < Q(a_1) \leq P_{A_2}^\star(a_1)$, then it holds that 
\begin{IEEEeqnarray}{c}
\hat{u}(P) > \hat{u}(Q);  
\end{IEEEeqnarray}
and if $P_{A_2}^\star(a_1) \leq P(a_1) < Q(a_1) \leq 1$, then it holds that 
\begin{IEEEeqnarray}{c}
\hat{u}(P) <\hat{u}(Q). 
\end{IEEEeqnarray}

Alternatively, consider the case in which $u_{1,1} -u_{1,2}-u_{2,1}+u_{2,2} \leq 0$. 
From Theorem \ref{TheoremNE}, if the entries of the matrix $\matx{u}$ in~\eqref{Equ:u} satisfy \eqref{EqMixedAssumption}, then one of the following conditions holds: 
\begin{IEEEeqnarray}{rl}
&u_{1,1} - u_{1,2}>0 \quad \textnormal{and} \quad u_{2,1} -u_{2,2}<0, \qquad \textnormal{or}\\
&u_{1,1} - u_{1,2}<0 \quad \textnormal{and} \quad u_{2,1} -u_{2,2}>0.\label{Eqshu2}
\end{IEEEeqnarray}
Nonetheless, only the second condition yields $u_{1,1} -u_{1,2}-u_{2,1}+u_{2,2}\leq 0$. 
Hence, from Theorem \ref{TheoremNE} and \eqref{Eqshu2}, if $0 \leq P(a_1) < Q(a_1)  \leq P_{A_2}^\star(a_1)$, then it holds that 
\begin{IEEEeqnarray}{c}
\hat{u}(P) > \hat{u}(Q);  
\end{IEEEeqnarray}
and if $ P_{A_2}^\star(a_1) \leq P(a_1) < Q(a_1) \leq 1$, then it holds that 
\begin{IEEEeqnarray}{c}
\hat{u}(P) <\hat{u}(Q). 
\end{IEEEeqnarray}

This completes the proof. 

\section{Proof of Lemma \ref{LemmaNESE}}\label{AppLemmaNESE}
The proof follows from the fact that 
\begin{IEEEeqnarray}{c}
u(P_{A_1}^\star, P_{A_2}^\star) = \min_{P \in \simplex{\set{A}_2}} \max_{Q \in \simplex{\set{A}_1}}u(Q, P)  =  \min_{P \in \simplex{\set{A}_2}} \hat{u}( P). \IEEEeqnarraynumspace
\end{IEEEeqnarray}
This completes the proof.

\bibliographystyle{IEEEbib}
\bibliography{reference_SunPerlaza}

\begin{thebibliography}{1}

\bibitem{Maschler_2013_game}
M.~Maschler, E.~Solan, and S.~Zamir,
\newblock {\em Game Theory},
\newblock Cambridge University Press, 2013.

\end{thebibliography}

\end{document}